\documentclass[iop]{emulateapj}

\usepackage{natbib}
\usepackage{graphics}

\shorttitle{LMC INTEGRATED LIGHT FE ABUNDANCES}
\shortauthors{COLUCCI ET AL.}

\begin{document}
\newcommand{\msol}{M_\odot}
\newcommand{\etal}{et al.\ }
\newcommand{\kms}{km~s$^{-1}$ }
\newcommand{\rAA}{{\AA \enskip}}
\newcommand{\ew}{W_\lambda}
\newcommand{\mv}{$M_{V}^{\rm{tot}}$}
\newcommand{\bvo}{($B-V$)$_{\rm{0}}$} 
\defcitealias{2008ApJ...684..326M}{Paper I}
\defcitealias{milkyway}{Paper II}
\defcitealias{paper4}{Paper IV}


\title{ Globular Cluster Abundances from High-Resolution,
  Integrated-Light Spectroscopy. III.  The Large Magellanic Cloud: Fe and Ages\footnotemark[1]}

\footnotetext[1]{This paper includes data gathered with the 6.5 meter Magellan 
Telescopes located at Las Campanas Observatory, Chile.}

\author{Janet E. Colucci}
\affil{Department of Astronomy and Astrophysics, 1156  High Street, UCO/Lick Observatory, \\ University of California, Santa Cruz, CA 95064; jcolucci@ucolick.org}

\author{Rebecca A. Bernstein}
\affil{Department of Astronomy and Astrophysics, 1156  High Street, UCO/Lick Observatory,\\
 University of California, Santa Cruz, CA 95064; rab@ucolick.org}

\author{Scott  A. Cameron}
\affil{Science Department, 3000 College Heights Blvd., Cerro Coso Community College,  Ridgecrest, CA 93555; scameron@cerrocoso.edu}

\and

\author{Andrew McWilliam}
\affil{The Observatories of the Carnegie Institute of Washington, 813 Santa Barbara Street, Pasadena, CA 91101-1292; andy@ociw.edu}

\begin{abstract}

In this paper we refine our method for the abundance analysis of high resolution  spectroscopy of the integrated light of unresolved globular clusters (GCs).
This method was previously demonstrated for the analysis of old ($>$10 Gyr) Milky Way GCs.
 Here we extend the technique to young clusters using a training set of 9 GCs in  the Large Magellanic Cloud (LMC).
Depending on the signal-to-noise ratio of the data, we use 20-100 Fe lines per cluster to successfully constrain the ages of old clusters to within a $\sim$5 Gyr range, the ages of $\sim$2 Gyr clusters to a 1-2 Gyr range, and the ages of the youngest clusters (0.05-1 Gyr) to a $\sim$200 Myr range.    
We also demonstrate that we  can measure [Fe/H] in clusters with any age less than 12 Gyrs with similar or
only slightly larger uncertainties (0.1-0.25 dex) than those obtained
for old Milky Way GCs (0.1 dex);  the slightly larger uncertainties are
due to the rapid evolution in stellar populations at these ages.
In this paper, we present only Fe abundances and ages.   In the next paper in this series, we present our complete analysis of the $\sim 20$ elements for  which we are able to measure abundances.
For several of the clusters in this sample, there are no high resolution abundances in the literature from individual member stars;  our results are the first detailed chemical abundances available. 
 The spectra used in this paper were obtained at Las Campanas with the echelle on the du Pont Telescope and  with the MIKE spectrograph on the Magellan  Clay Telescope.

\end{abstract}

\keywords{galaxies: individual (LMC) --- galaxies: star clusters --- galaxies: abundances --- globular clusters: individual(NGC 2005, NGC 2019, NGC 1916, NGC 1978, NGC 1718, NGC 1866, NGC 1711, NGC 2100, NGC 2002) --- stars: abundances}

\section{Introduction}	
\label{sec:intro}
\setcounter{footnote}{1}

Globular clusters (GCs) are unique tools for probing the formation history and evolution of galaxies.   
The $\sim$150 GCs \citep{1996AJ....112.1487H} of the Milky Way (MW) have been studied extensively and have provided fundamental clues to our current picture of the formation of the MW, and by extension, other normal spiral galaxies.  Particularly interesting constraints have come from detailed chemical abundances of individual member stars in the MW GCs, which show
relative abundance patterns that can be used to constrain the early star formation history of the MW.

Because GCs are so luminous, extragalactic GCs can also be individually identified and studied as unresolved sources. Because they are spatially unresolved, however, extragalactic GC systems have typically been studied with low resolution spectroscopy and photometry. Detailed abundance work requires high resolution, high signal-to-noise ratio (S/N) spectra and so has been done solely on individual, resolved stars in the past, which limited such studies to stars in the MW and nearby dwarf galaxies.  In order to measure detailed chemical abundances of stellar systems beyond the MW, we have developed a new technique for analyzing high resolution spectra of the integrated light (IL) of GCs.  With this technique, we can measure both ages and detailed chemical abundances of $\sim$20 elements in a moderate quality (S/N$\sim$50), high resolution ($R\equiv \lambda/\Delta\lambda \sim 20,000$) spectrum of a typical GC.  Our technique now allows us to study the chemical evolutionary history of distant galaxies at an unprecedented level of detail that has previously only been possible within the MW.  We can now begin to constrain the star formation rate (SFR), stellar initial mass function (IMF), gas inflow/outflow, and chemical evolution history of galaxies within roughly 4 Mpc, well beyond the nearest neighbors of the MW.
 
\begin{deluxetable*}{rrrrrrrrrr}
\small
\tablecolumns{10}
\tablewidth{0pc}
\tablecaption{LMC Cluster Sample \label{tab:lmctable}}
\tablehead{
\colhead{Cluster} & \colhead{RA}   & \colhead{Dec}   &\colhead{V\tablenotemark{a}}&\colhead{$B-V$} &\colhead{$E(B-V)$}& \colhead{Age} &    \colhead{[Fe/H]}& \colhead{[$\alpha$/Fe]\tablenotemark{b}} &\colhead{References\tablenotemark{c}} \\ \colhead{}&\colhead{(2000)}&\colhead{(2000)}&&&&\colhead{(Gyrs)}  }

\startdata
\sidehead{Old ($>$ 5 Gyrs)}

\hline
\\
NGC 1916	&5 18 37.9	&$-$69 24 23	&10.38&	0.78$\pm$0.02	&0.13$\pm$0.02	&$>$10	&$-$2.08	&\nodata		 	& 1,13	\\
NGC 2019	&5 31 56.5	&$-$70 09 32	&10.95&	0.77$\pm$0.01	&0.06$\pm$0.02	&$>$10	&$-$1.37	&$+$0.20	&2, 14\\
NGC 2005	&5 30 10.4	&$-$69 45 09	&11.57&	0.73		&0.10$\pm$0.02	&$>$10	&$-$1.80	&$+$0.05	&2, 14 \\
\\
\sidehead{Intermediate Age (1-3 Gyrs)}

\hline
\\
NGC 1978	&5 28 45.0	&$-$66 14 14	&10.74&0.78$\pm$0.04	&0.09		&1.9		&$-$0.38	&$+$0.02 	& 3,7 \\
			&			&				&&	&			&		&$-$0.96	 &$+$0.38  	& 4 \\
NGC 1718	&4 52 25.0	&$-$67 03 06	&12.25&0.76$\pm$0.01	&0.10$\pm$0.03	&2.1		&$-$0.80	&\nodata	&  5,6 \\
\\
\sidehead{Young ($<$ 1 Gyr)}

\hline
\\
NGC 1866	&5 13 38.9	&$-$65 27 52
&9.89&0.26$\pm$0.02&0.06	&0.13	&$-$0.51	&$+$0.08  &
4,15\\
			&			&				&&	&			&		&$-$0.43	 &$+$0.00 	& 16 \\

NGC 1711	&4 50 37.0	&$-$69 59 06	&10.11&0.20$\pm$0.08	&0.09$\pm$0.05		&0.050	&$-$0.57	&\nodata & 8	\\
NGC 2002	&5 30 21.0	&$-$66 53 02	&10.1&0.00	&0.20		&0.018	&$-$2.2		&\nodata& 11, 12	\\
NGC 2100	&5 42 08.6	&$-$69 12 44	&9.60&0.16$\pm$0.02	&0.26$\pm$0.01		&0.015	&$-$0.32	&$-$0.06 & 9, 10, 13	\\

\enddata

\tablenotetext{a}{From  \cite{2008MNRAS.385.1535P} and \cite{1996ApJS..102...57B}}
\tablenotetext{b}{Mean of [Si/Fe], [Ca/Fe], and [Ti/Fe] abundances
  from 2, 3, 9, and 16 and mean of [O/Fe] from  4.}
\tablenotetext{c}{References for columns 4, 5, 6 and 7:
  1.\cite{1991AJ....101..515O}, 2.   \cite{2006ApJ...640..801J},
  3. \cite{2008AJ....136..375M}, 4. \cite{2000A&A...364L..19H},
  5. \cite{2006AJ....132.1630G}, 6. \cite{2007A&A...462..139K},
  7. \cite{2007AJ....133.2053M}, 8. \cite{2000A&A...360..133D},
  9. \cite{1994A&A...282..717J}, 10. \cite{1991ApJS...76..185E},
  11. \cite{2008MNRAS.386.1380K}, 12. \cite{2007ApJ...655..179W},
  13. \cite{2008MNRAS.385.1535P},
  14. \cite{1998MNRAS.300..665O},15. \cite{2001ApJ...560L.139W}, 16. \cite{mucc1866}  }

\end{deluxetable*}

Previous techniques utilizing IL spectra of unresolved, extragalactic clusters focused on estimating metallicities (e.g. [Fe/H]) and $\alpha$-element abundance ratios from low resolution spectra with line index systems like the ``Lick'' system \citep[e.g.][]{1985ApJS...57..711F}.   While very useful for obtaining information on the general chemical properties of GC systems, line indexes are  inaccurate, and even    state of the art line index or full spectrum fitting techniques for lower resolution will always be limited to measurements of at most $\sim$
5  elements \citep[e.g.][]{2008ApJS..177..446G}.  
In fact, line index systems were originally designed to target unresolved galaxies in which spectral lines of individual elements are not possible to resolve due to the high internal velocity dispersions (100$-$300 \kms) of the galaxies themselves \citep[e.g.][]{1976ApJ...204..668F}.  GCs, on the other hand, have small velocity dispersions of 2$-$25 \kms, so that weak lines are not blended and broadened;  features of many elements are easily identifiable in spectra of their integrated light.
The availability of over 20 individual elements makes our method particularly   powerful for chemical evolution studies compared to low and medium resolution \citep[see new ``LIS'' system of ][]{2010MNRAS.404.1639V} line index methods. 
 
Our IL abundance analysis technique  relies on the  fact that GCs are SSPs, and thus can be easily modeled \citep{2003SPIE.4841.1694B,2002IAUS..207..739B,2005astro.ph..7042B,2008ApJ...684..326M}.   Unlike line index systems \citep[see][]{2002A&A...395...45P}, this method is based on explicit measurements of individual absorption features  and is  analogous to standard analyses of  red giant branch (RGB) stars.  Thus, our abundance analysis does not rely on a calibration to local MW stars, and there are no systematic errors due to built in assumptions on the chemical nature of the target clusters.   Briefly, our IL abundance analysis utilizes the fact that stellar evolution is reasonably well understood \citep[e.g.][]{2005ARA&A..43..387G}, so that it is straightforward to use theoretical stellar isochrones to create arbitrary synthetic stellar populations of bound star clusters that are characterized only by their age and metallicity. 
 These synthetic GC populations can then be used to synthesize flux-weighted, IL spectra for comparison to the observed IL spectra of unresolved GCs.  
Of course there are some aspects of stellar evolution that  are not perfectly produced in the isochrones.  For example, it is well known that the properties of blue horizontal branch stars are not accurately produced from first principles  and that  parameters such as mass loss 
must be adjusted to match observations \citep[e.g.][]{2004ApJ...612..168P}.   Other sources of uncertainty in the isochrones include fractional contributions from blue stragglers, asymptotic giant branch stars, and young supergiants.  Fortunately, the presence of these types of stars can be inferred from our spectra themselves.   We discuss several of the potential diagnostic tools for evaluating the more uncertain phases of stellar evolution and their impact on our analysis in this paper. 

While we do so in a integrated light spectrum, our IL abundance analysis technique  utilizes the same standard stellar high resolution abundance analysis methods that have been in use for decades;  we use  equivalent widths (EWs) and line profile fitting  of individual spectral features and spectral line synthesis to derive abundances.   Clean, unblended features with well determined transition strengths can be readily measured in our spectra,  and the self-consistency and stability of the abundance solution for  many (50-150) individual transitions of Fe can be used to constrain the best-fitting synthetic GC population.  Using the stability of the Fe line solutions, the most appropriate stellar population (characterized by its age and abundance)  can be matched to any unresolved GC  using  the observed IL spectra alone.  This best-fitting stellar population can then be used to derive abundances of $\sim$20 different chemical elements.

In \cite{2008ApJ...684..326M} (hereafter \citetalias{2008ApJ...684..326M}),   \citet{milkyway} (hereafter \citetalias{milkyway} ), 
\cite{scottphd} and \cite{mythesis},
extensive testing on a ``training set'' of MW GCs of  known properties was presented with the development of our IL abundance analysis method.    These works focused on demonstrating the analysis method and the accuracy of the derived abundances for a set of ``typical'' MW GCs, which have old ages ($>$10 Gyr), span a range in metallicity from  [Fe/H]$\sim-2$ to $-0.4$, and have a spread in horizontal branch morphology and internal velocity dispersion ($\sigma_{V}$).  The results of \citetalias{2008ApJ...684..326M} and \citetalias{milkyway} showed that the  abundances measured using our IL analysis method have accuracies of $\leq$0.1 dex for nearly all species when compared to standard analysis of individual stars in the training set clusters, and that the statistical uncertainties in abundances for individual element species are $\sim$0.2 dex, which is comparable to the precision in abundance measurements in those GCs obtained from single member RGB stars ($\lesssim$0.15 dex). For old GCs, ages can be constrained  to a range of 5 Gyr, which is small enough that the adopted age affects the abundance results by $<$0.1 dex, which is less than the statistical error of the measured spectral lines.  

In this paper, the third of the series, we have two primary goals.  The first is to demonstrate that the IL abundance analysis method can be applied to clusters younger than 10 Gyr in age. Because there are few, if any, high mass, high surface brightness young star clusters in the MW, we have used clusters in the Large Magellanic Cloud (LMC)  for this purpose.
Although most young clusters are not massive enough to survive a Hubble time, they are still valuable tools for studying the chemical evolution of galaxies as a function of  time.
As part of further developing the IL method for young clusters, we also  present an additional technique for evaluating the effect of stochastic stellar population fluctuations on the measurement of age and [Fe/H] of  unresolved clusters.  Our second goal in this work  is to measure new detailed chemical abundances for  LMC clusters that have not been previously studied with high resolution spectroscopy.  In this paper we present new results for [Fe/H] in three, previously unstudied clusters;  results for approximately 20 additional elements in each cluster are presented in a following paper \cite{paper4}, hereafter ``Paper IV.''  

The properties of the LMC training set clusters  are discussed in detail in  \textsection~\ref{sec:trainingSet} . In \textsection~\ref{sec:obs} we discuss the observations and data reduction, and in \textsection~\ref{sec:analysis} we describe our analysis techniques as well as age and [Fe/H] results for each cluster.  In \textsection~\ref{sec:discussion} we compare the IL abundances for the training set clusters to  previous work using high resolution spectra of individual stars, and  to results from  photometry and  low resolution spectra. We also discuss the age-metallicity relation  we have found for the LMC clusters in this work.

\begin{figure*}
\centering
\includegraphics[angle=90,scale=1.1]{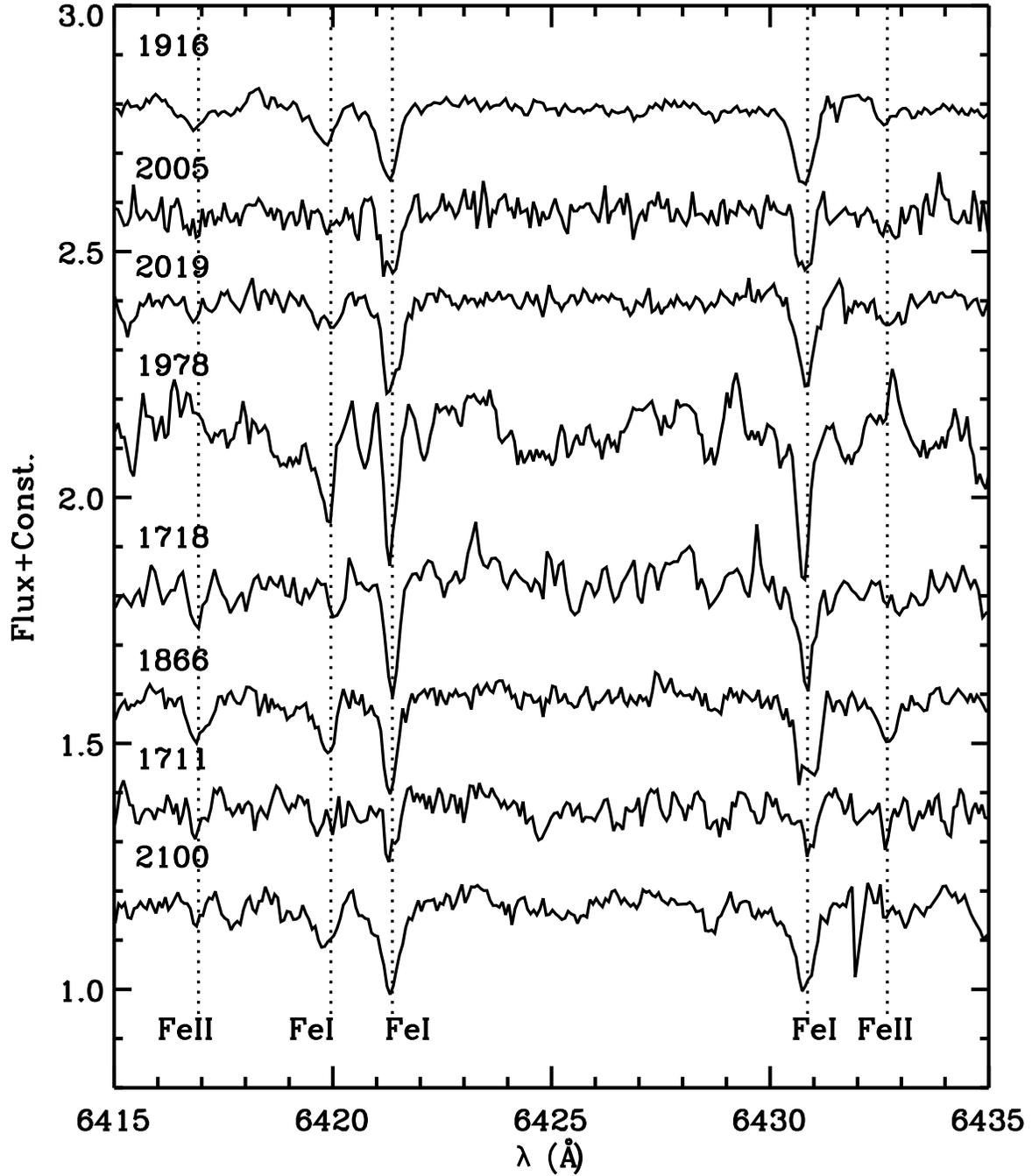}
\caption{Example LMC cluster spectra  shown decreasing in age (top to bottom).  Individual Fe I and Fe II lines that can be used  in   abundance analysis of one or more clusters are  marked by dotted lines. }
\label{fig:lmcspectra} 
\end{figure*}

\section{Globular Cluster Training Set}
\label{sec:trainingSet}

The goals of this paper  and \citetalias{paper4} are to refine the  integrated light method developed in \citetalias{2008ApJ...684..326M} and \citetalias{milkyway} on clusters in the LMC, and also to measure new abundances for  LMC clusters previously unstudied at high resolution. 
To this end, it is important to consider differences between the LMC training set and the MW training set along with the limitations of the LMC training set when compared to distant, unresolved clusters. 
Overall, the LMC training set data is unavoidably of lower quality  (i.e. lower S/N) than the data obtained for the  MW training set  because the clusters are generally less massive, and therefore less luminous, and they lie at much greater distances (D$\sim$50 kpc).   This makes the LMC training set data useful to evaluate the accuracy of the IL method when applied to lower S/N data of distant, unresolved clusters.

Our  training set includes seven MW clusters, as described in \citetalias{2008ApJ...684..326M} and \citetalias{milkyway}, as well as  nine LMC clusters (NGC 1916, NGC 2005, NGC 2019, NGC 1978, NGC 1718, NGC 1866, NGC 1711, NGC 2100, and NGC 2002).  The LMC training set includes clusters $<$10 Gyr old, which is an age range that cannot be probed using clusters in the MW alone.   In addition, the LMC training set provides old clusters with a range of abundance ratios \citep{2006ApJ...640..801J}. This is in contrast to MW clusters, which have very uniform abundance ratios \citep[e.g.][]{2005AJ....130.2140P}, and reflects the difference in star formation histories between galaxies with significantly different masses.
For this reason, we discuss the  LMC training set  as  three groups, which we divide according to ages previously determined using other techniques: old ($>$5 Gyrs), intermediate age (1-4 Gyrs) and young ($<$ 1 Gyr).   Coordinates, photometric properties, and previous abundance information that is available in the literature are given in Table~\ref{tab:lmctable}.

 As for the MW training set, we have only  observed the core regions of the LMC clusters (see \textsection~\ref{sec:obs}).    Because of this, and because the clusters are lower mass and lower density, incomplete sampling  is a potential complication for integrated light analysis that arises from only observing a fraction of the total population of the cluster.  
 Because our analysis method involves using theoretical color magnitude diagrams (CMDs), which accurately represent the full stellar population, incomplete sampling must be explicitly included in our analysis strategy.
 In this paper, we use the phrase  ``sampling uncertainties''  to refer to statistical fluctuations in the small numbers of bright stars present in the core region of a cluster at the time of our observations.  These statistical fluctuations  can potentially have a significant impact on the integrated properties of the cluster because the stars are very luminous.  
 In general, the importance of  sampling uncertainties scales with the total number of stars in the cluster, and therefore with the total luminosity  (\mv) of the cluster.  In our training set clusters in particular, we have effectively reduced the total luminosity of the clusters by observing only a fraction of the total flux, so we can expect that sampling uncertainties will be a greater issue than they would be for integrated light analysis of distant clusters in which a large well-populated fraction of the cluster will always be observed within one seeing disk.

  It is also important to note that sampling uncertainties are more important for clusters of younger ages, due to the contribution of very few, but very luminous giant stars in rapid phases of stellar evolution \citep[e.g. see][]{1999A&AS..136...65B}.   Young stellar populations may change significantly over only a $<$1 Gyr age range \citep[e.g.][]{1999A&AS..136...65B,2007A&A...462..107F,2003MNRAS.344.1000B}.  In general,  it is necessary to evaluate sampling uncertainties on younger ($<$5 Gyr) clusters because they will suffer  from greater intrinsic statistical fluctuations in the stellar populations than old clusters.  For the younger LMC training set clusters in particular, sampling uncertainties will be exacerbated due to observing only a fraction of the total cluster population. 

Because statistical fluctuations will always be an issue for young, rapidly evolving clusters, in this work we have extended the integrated light abundance analysis method developed in \citetalias{2008ApJ...684..326M} and \citetalias{milkyway} to include extensive tests designed to address sampling uncertainties in the LMC training set.  These tests will be discussed further in \textsection~\ref{sec:analysis}.  We  aim to evaluate  how sampling uncertainties  in the young LMC training set clusters  affect the constraints we can derive for the cluster ages and abundances.

Finally, we note that in this work on the LMC training set, we do not analyze the IL spectra using resolved star photometry as we did for the MW training set in \citetalias{2008ApJ...684..326M} and \citetalias{milkyway}.  This is mainly due to the fact that in the core regions we have observed, the quality of the available photometry  is not high enough   to make this analysis meaningful, with the possible exception of the deep Hubble Space Telescope (HST)  photometry  for NGC 1978, by \cite{2007AJ....133.2053M}.   At the distance of the LMC crowding in the cluster core regions always makes the photometry incomplete  for main sequence stars.  Moreover, as discussed in  \citetalias{milkyway}, even tests with the best photometry of Galactic GCs  \citep[e.g.][]{2002A&A...391..945P,2007AJ....133.1658S}  result in a higher scatter in our abundance solutions than tests performed with isochrones.  Thus, for clusters in the LMC training set, we limit ourselves to broad consistency checks with the available photometry, discussed in detail in \textsection~\ref{sec:photo}.

In summary, when compared to the MW training set, first we expect that the old LMC clusters will be more difficult to analyze  due to lower S/N data.  Second, we expect that the method strategy will be very similar to the MW training set strategy except for the addition of sampling uncertainty tests developed in this work.  We  will discuss the extent that these tests can  improve our abundance solutions by allowing for statistical variations in the stellar populations. For clusters $<$5 Gyr old, we expect that the analysis will be more challenging than for the old clusters because the stellar populations may change significantly over only a $<$1 Gyr age range.  This age sensitivity means that  it is crucial to test the IL method on younger clusters in order to assess how well the ages and abundances can be constrained, and to further develop the IL analysis strategy if needed.

\section{Observations and Data Reduction}
\label{sec:obs}

\begin{deluxetable}{rrrrrrrrrr}
\tablecolumns{10}
\tablewidth{0pc}
\tablecaption{Observation Log and Estimated S/N \label{tab:observations}}
\tablehead{
\colhead{Cluster} &\colhead{Exposure (s)}   &\multicolumn{3}{c}{S/N (pixel$^{-1}$)} \\&&\colhead{4380 \AA} &\colhead{6100 \AA} & \colhead{7550 \AA} }
\startdata

NGC 2019 &	20,218	&  30   & 70 & 70 \\
NGC 2005 &	22,010 	&  20 & 50 & 60\\
NGC 1916 &	22,785	&  40 & 80 & 90\\
NGC 1978 &	50,150 	& 20 & 50 & 60\\
NGC 1718 &	19,560	& 80 & 60 & 80 \\
NGC 1866 &	41,230	& 50 & 80 & 70 \\
NGC 1711 &	18,350	& 40 & 50 & 50 \\
NGC 2100 &	20,800	& 50 & 70 & 70 \\
NGC 2002 &	18,357	& 60 & 110 & 140 \\
\enddata
\end{deluxetable}

\begin{deluxetable}{rrrrr}

\tablecolumns{5}
\tablewidth{0pc}
\tablecaption{Cluster Structural Parameters And Fraction Observed  \label{tab:frac_obs}}
\tablehead{
\colhead{Cluster} &\colhead{c\tablenotemark{1}}&\colhead{R$_{King}$\tablenotemark{1}}   &\colhead{R$_{core}$\tablenotemark{2}} &\colhead{Fraction Observed} \\ &\colhead{arcsec}&\colhead{arcsec} &\colhead{arcsec} &  }

\startdata

NGC 2019 &      1.68  &  2.56   & 3.61 & 0.45 \\
NGC 2005 &      1.53  &  2.96 & 3.63 & 0.60\\
NGC 1916 &      1.47  &  3.02 & 3.35 & 0.62 \\
NGC 1978 &       \nodata & \nodata& \nodata&0.05-0.10 \\
NGC 1718 &      1.33  & 10.32 & 8.52 & 0.23 \\
NGC 1866 &      1.65  & 11.97  & 14.15 & 0.14 \\
NGC 1711 &      1.44  & 8.35 & 8.78 & 0.16  \\
NGC 2100 &      1.74  & 4.23 & 5.02 & 0.40 \\

\enddata

\tablecomments{1. From \cite{2005ApJS..161..304M}. The concentration parameter is defined as c $\equiv$log$_{10}$(R$_{tidal}$/R$_{King}$)., 2. Cluster core radii from \cite{2003MNRAS.338...85M}.   }

\end{deluxetable}

All of the data for the LMC training set clusters, with the exception of NGC 1718, were obtained using the echelle spectrograph on the 2.5 m du Pont telescope at Las Campanas during  dark time in 2000 December and 2001 January. The wavelength coverage of these spectra is approximately 3700--7800 \AA.  

As done for the MW training set clusters, the LMC cluster cores were scanned to obtain IL spectra \citepalias{2008ApJ...684..326M}. Because of the greater distance to the LMC clusters, we scanned a  $12\times12$ arcsec$^{2}$ or  $8\times8$ arcsec$^{2}$ region of  the cluster cores, instead of the $32\times32$ arcsec$^{2}$ region scanned in MW clusters.    Our spectra were reduced with standard IRAF\footnote{IRAF is distributed
  by the National Optical Astronomy Observatories, which are operated
  by the Association of Universities for Research in Astronomy, Inc.,
  under cooperative agreement with the National Science
  Foundation.} routines, combined with the scattered-light subtraction described in \citetalias{2008ApJ...684..326M}. The observational technique and data reduction are described in further detail in \citetalias{2008ApJ...684..326M} and  \citetalias{milkyway}.

NGC 1718 was observed with the MIKE double echelle spectrograph \citep{2003SPIE.4841.1694B} on the 6.5 m Magellan Clay telescope in 2006 November.   The integrated light spectrum was obtained by scanning the central $12\times12$ arcsec$^{2}$ region  of NGC 1718 using a modification to the telescope guiding program provided by D. Osip. 
The blue side wavelength coverage is 3350--5050 \AA, and the red side wavelength coverage is 4800--9000 \AA.  We used a slit size of  $1".0\times5".0$, and  3$\times$2 pixel on-chip binning.   The data were reduced using the MIKE Redux pipeline \citep{mikeredux}, an extended library of spectroscopic routines building on the SDSS pipeline and developed for a number of instruments by J. X. Prochaska.\footnote{ http://www.ucolick.org/~xavier/IDL/index.html}

Exposure times for each cluster and approximate S/N values at three regions in each spectrum are listed in Table~\ref{tab:observations}.  For all clusters we primarily use the higher S/N data $>$ 4400 \rAA for our abundance analysis. Example spectra for the clusters are shown in Figure~\ref{fig:lmcspectra}.

\begin{deluxetable*}{rrrrccccccccc}
\scriptsize
\tablecolumns{13}
\tablewidth{0pc}
\tablecaption{Line Parameters and Integrated Light Equivalent Widths for LMC GCs \label{tab:t4_stub}}
\tablehead{\colhead{Species} & \colhead{$\lambda$}   & \colhead{E.P.} & \colhead{log gf}   & \colhead{EW} & \colhead{EW} & \colhead{EW} & \colhead{EW} & \colhead{EW}& \colhead{EW}& \colhead{EW}& \colhead{EW}& \colhead{EW}\\ 
\colhead{} & \colhead{(\AA)} & \colhead{(eV)} &\colhead{} & \colhead{(m\AA)} &\colhead{(m\AA)} &\colhead{(m\AA)} &\colhead{(m\AA)} &\colhead{(m\AA)} &\colhead{(m\AA)} &\colhead{(m\AA)} &\colhead{(m\AA)} &\colhead{(m\AA)} 
\\ & & & & \colhead{1916}&\colhead{2005}&\colhead{2019}&\colhead{1978}&\colhead{1718}&\colhead{1866}&\colhead{1711}&\colhead{2100}&\colhead{2002}}

\startdata

 Fe  I &6265.141 &  2.180 & -2.532 & \nodata & \nodata & \nodata & \nodata & 86.3 & \nodata & \nodata & 87.5 & \nodata\\
 Fe  I &6311.504 &  2.830 & -3.153 & \nodata & \nodata & \nodata & \nodata & \nodata & 54.7 & \nodata & \nodata & \nodata\\
 Fe  I &6322.694 &  2.590 & -2.438 & \nodata & 36.3 & 47.7 &128.7 & \nodata & 58.9 & 47.4 & 63.4 & \nodata\\
 Fe  I &6330.852 &  4.730 & -1.640 & \nodata & \nodata & \nodata & \nodata & \nodata & 23.7 & \nodata & \nodata & 33.0\\
 Fe  I &6335.337 &  2.200 & -2.175 & 74.2 & 58.6 & 67.4 &135.6 & 74.1 & 67.3 & 53.5 & \nodata & \nodata\\
 Fe  I &6336.830 &  3.690 & -0.667 & 59.5 & \nodata & \nodata &134.6 & 65.8 & 68.7 & 40.4 & \nodata & \nodata\\
 Fe  I &6353.849 &  0.910 & -6.360 & \nodata & \nodata & \nodata & \nodata & \nodata & 86.5 & \nodata & \nodata & \nodata\\
 Fe  I &6355.035 &  2.840 & -2.328 & 60.7 & \nodata & 49.6 & \nodata & 68.0 & 46.5 & \nodata & 93.1 & \nodata\\
 Fe  I &6380.750 &  4.190 & -1.366 & \nodata & \nodata & \nodata & \nodata & \nodata & 42.7 & \nodata & \nodata & \nodata\\
 Fe  I &6392.538 &  2.280 & -3.957 & \nodata & \nodata & \nodata & \nodata & 30.2 & \nodata & \nodata & \nodata & \nodata\\

\enddata
\tablecomments{Table 4 is  presented in its entirety at the end of the text. ~ Lines listed twice correspond to those measured in adjacent orders with overlapping wavelength 
coverage.}
\normalsize
\end{deluxetable*}

We have calculated the total V-band flux contained in the scanned region for each cluster (with the exception of NGC 1978) using the surface brightness profiles uniformly measured by \citet{2005ApJS..161..304M}.  The cluster surface brightness profile parameters and core radii  \citep{2003MNRAS.338...85M} are listed with the calculated fraction of flux observed in Table~\ref{tab:frac_obs}.  A surface brightness profile was unavailable for NGC 1978.  To estimate the flux we observed we compare  the HST photometry (kindly provided by A. Mucciarelli) to the other clusters in our MW and LMC training set.   While the exact fraction for NGC 1978 is uncertain, it is clearly small and in the range of 5 to 10$\%$.  The S/N of our data for NGC 1978 is also lower than for most of the other clusters in the sample due to its smaller total luminosity and surface brightness.  The combination of lower data quality and undersampling makes analysis of NGC 1978 particularly challenging, which we discuss further in \textsection~\ref{sec:int}.

\section{Abundance Analysis}
\label{sec:analysis}

The method that  we have developed for measuring detailed abundances from the integrated light spectra of GCs is extensively described in \citetalias{2008ApJ...684..326M}, \citetalias{milkyway} and \citet{m31paper}. We briefly review that strategy   In \textsection~\ref{sec:lines}-\textsection~\ref{sec:oldcmd}.
  In \textsection~\ref{sec:old} we discuss the results for the older clusters of the LMC training set. We also develop an additional technique  to address incomplete sampling effects, which we also apply to the intermediate age and younger clusters in  \textsection~\ref{sec:int} and \textsection~\ref{sec:young}.

\subsection{EWs and Line Lists}        
\label{sec:lines}

As in all of our previous work on IL spectra, we use the semi-automated program GETJOB \citep{1995AJ....109.2736M} to measure absorption line equivalent widths (EWs)  for individual lines in the IL spectra.  Continuum regions for each spectral order are interactively fit with low order polynomials and line profiles are fit with single, double, or triple Gaussians.  We take special care in continuum placement and attention to line blending due to the nature of IL spectra.  Line lists were taken from \cite{2009A&A...497..611M}, \citetalias{milkyway},  \citet{m31paper}, and references therein.  The lines and EWs used in our final analysis for each cluster can be found in Table~\ref{tab:t4_stub}.

\subsection{Constructing CMDs and EW Synthesis}        
\label{sec:cmds}

Our abundance analysis technique allows us to calculate [Fe/H] solutions for any synthetic cluster population, without a priori knowledge of the true cluster population. The synthetic populations are used to synthesize IL EWs, which are compared to our observed IL EWs during our abundance analysis.  We construct synthetic populations using the single age, single metallicity theoretical isochrones from the Teramo\footnote{Teramo isochrones downloadable at http://albione.oa-teramo.inaf.it/}
   group \citep{2004ApJ...612..168P,2006ApJ...642..797P,2007AJ....133..468C}.  The isochrones we employ also contain an extended asymptotic giant branch (AGB),  $\alpha-$enhanced low$-$temperature opacities calculated according to \citet{ 2005ApJ...623..585F}, and 
 mass$-$loss parameter of $\eta$=0.2. 

 The Teramo isochrones are tabulated with two treatments of convection at the border of  convective stellar cores, which can be important because the choice of  treatment can affect the morphology of main sequence turnoffs and  blue loop populations of younger clusters \citep{2004ApJ...612..168P}.  
 The first, most simplistic treatment, is used in  the  ``canonical'' tracks.  In this case, convective instability is treated  using the Schwarzschild criterion, which results in  a convective core with a clean boundary that is determined by the input stellar physics.   
 The second treatment, used in the ``non-canonical'' tracks, parametrizes a certain amount of mechanical convective overshooting past the Schwarzschild boundary.  This overshooting can occur if  fluid elements maintain some velocity past the Schwarzschild boundary, and/or if there is some rotationally induced mixing beyond the boundary \citep{2000A&A...361..101M}.   Convective overshooting effectively increases the size of the stellar core, resulting in a brighter luminosity and shorter lifetime. 
Note that the presence or degree of convective overshooting only affects stars with masses larger than $\sim$1.1$\msol$, because only these stars are hot enough to develop a convective core during the H-burning phase.   This means that the treatment of convective overshooting only impacts the CMDs of  clusters younger than $\sim$3 Gyr.

 In \citetalias{milkyway} and  \citet{m31paper}  we analyzed clusters with ages $>$5 Gyr, and chose to use  the Teramo canonical evolutionary tracks.  This choice did not significantly affect our analysis because of the older ages and cooler temperatures of the stars in these clusters. 
  While the presence or degree of convective overshooting required to match observations is still under debate \citep{1999AJ....118.2839T,2002A&A...385..847B},
 \citet{2007AJ....133.2053M} found the non-canonical Teramo isochrones to be a better match  than the canonical  isochrones to the turnoff region of  their high precision HST photometry of NGC 1978. Therefore, in this work we have  chosen  to use the non-canonical isochrones for the younger clusters in order to be consistent with results from the best available photometry.   In \textsection \ref{sec:discussion} we briefly discuss uncertainties due to our choice of isochrones.

In our analysis, synthetic CMDs  are created by combining the model isochrones with an initial mass function (IMF) of the form in  \citet{2002Sci...295...82K}.   For the MW clusters in \citetalias{2008ApJ...684..326M} and \citetalias{milkyway},
we made an additional correction for mass segregation in the cluster cores due to the small fraction of the stellar populations we observed.
 For the LMC training set clusters we do not make a mass segregation correction.
This correction is not necessary in the older LMC clusters because we have observed  a larger fraction of the cluster population, and it is not necessary  for the younger clusters because we do not expect mass segregation due to dynamical evolution to be as significant an effect over their shorter lifetimes, as  the timescales for central and half-mass relaxation time are several hundred Myr and $\sim$3 Gyr, respectively \citep[e.g.][] {1992AJ....104.1086F}.  For simplicity and self-consistency, in this analysis we ignore the possibility of primordial mass segregation \citep{2009ApJ...698..615V}.

Synthetic CMDs are created for the available range of age and metallicity of the Teramo isochrones.  Each CMD is divided into $\sim$25 equal flux boxes containing stars of similar properties.   The properties of a flux-weighted  ``average'' star for each box are used in the IL EW synthesis, which we perform with ILABUNDS \citepalias{2008ApJ...684..326M}. ILABUNDs employs the 2010 version of  the spectral synthesis code MOOG \citep{1973ApJ...184..839S}.  We use the ODFNEW and AODFNEW  model stellar atmospheres of Kurucz
\footnote{The models are available  from R. L. Kurucz's Website at  http://kurucz.harvard.edu/grids.html}
  \citep[e.g.][]{2004astro.ph..5087C} for all abundance analysis.  
  
To begin the analysis of any cluster, we calculate a mean [Fe/H]  abundance from all available Fe I  lines for the  large  grid of synthetic CMDs.  We note that because we  measure far fewer Fe II lines  than Fe I lines,  we do not use Fe II lines to constrain the best-fitting CMD.  The abundances we measure from Fe II lines are therefore presented in \citetalias{paper4}.    We  next use the quality of the Fe I abundance solution to constrain the best-fitting age and abundance for each cluster.   As in our previous work, all abundances are calculated under the assumption of local thermodynamic equilibrium (LTE).
We only include Fe I lines with EWs$<$150 m\rAA   \citepalias[see][]{milkyway} in order to avoid line saturation problems.

\subsection{Determining the  Best-Fitting CMD}
\label{sec:oldcmd}

The first step in this procedure is to determine one synthetic CMD for each age that has the best-fitting metallicity.  Our analysis of the MW training set \citepalias{milkyway} has shown that the best strategy for consistently identifying the best-fitting CMD for each age is to require that the abundance used in constructing the isochrones be consistent with the abundance recovered by our analysis \citep[see also][]{m31paper}.   This is essentially taking advantage of the fact that the dependence of the RGB on metallicity is relatively well understood \citep[e.g.][]{2005ARA&A..43..387G}, and that  RGB stars have an important influence on Fe I line strengths in IL spectra.
Hereafter, we refer to the solutions where [Fe/H]$_{\rm{iso}}$=[Fe/H]$_{\rm{cluster}}$ as ``self-consistent'' Fe abundance solutions.   Often, it is clear that the self-consistent solution lies at a metallicity in between two synthetic CMDs in our grid.  When this occurs, we calculate an isochrone with the appropriate metallicity according to the interpolation scheme recommended by \cite{2006ApJ...642..797P}.

After constraining the best-fitting synthetic CMD properties from an initial grid of $\sim$70 possibilities  to a single  [Fe/H] for each age, we are able to isolate the most appropriate age(s) out of these 8 solutions using Fe line diagnostics. These diagnostics, which are commonly used in standard stellar abundance analyses,  relate to the quality of the [Fe/H] solutions.  In particular, a stable [Fe/H] solution should not depend on the parameters of individual lines (excitation potentials, wavelengths, or reduced EWs\footnote{Reduced EW 
$\equiv$ log(EW / $\lambda$) }), and the  standard deviation of the [Fe/H] solution  should be small.  Correlations of [Fe/H] with excitation potential (EP), wavelength, and reduced EW are indicative of improper distributions of stellar temperatures or gravities in the synthetic CMDs as compared to the actual clusters.  In general, these effects are difficult if not impossible to disentangle without additional constraints, although we will discuss them in the context of incomplete sampling in \textsection~\ref{sec:2019}.  For the purposes of constraining the best-fitting CMD, we consider synthetic CMDs that minimize these correlations to be most representative of the cluster stellar population.  To that end, for each synthetic CMD solution, we perform a linear least-squares fit to [Fe/H] versus EP, wavelength, reduced EW.  We then use the magnitude of the slope in these fits for our Fe line diagnostics, along with the standard deviation of the mean [Fe/H] of the Fe I  lines.  We find that these 4 diagnostics are usually strongly correlated with each other, which provides a means to identify the most appropriate synthetic CMD using Fe lines alone.

\subsection{Results: Old Clusters}
\label{sec:old}
For old clusters ($>$10 Gyrs), the analysis method developed in Papers I and II  and described above can unambiguously identify an accurate CMD for the analysis of the integrated light spectra.  Several of the clusters in the LMC sample discussed here are in that age range.    
The LMC sample presents some challenges to this basic analysis in that even old clusters in this sample are in some cases sparsely sampled due to low surface brightness or low total mass.  For that reason, our analysis results can be improved even for these old clusters by  allowing for a statistically incomplete sampling of the CMD.    We present an additional strategy for allowing for incomplete sampling, and address the  issues for each cluster below. 

\subsubsection{NGC 2019}
\label{sec:2019}

NGC 2019 is an old, moderately metal-poor cluster with a  total luminosity of  \mv$\sim-8$ \citep{1998MNRAS.300..665O,2008MNRAS.385.1535P}.  However, it is also a good test case for demonstrating the need for accounting for sampling uncertainties in an integrated light analysis.  We calculate that we have observed $\sim$45\% of the total flux of NGC 2019, which means that this cluster has the most sampling incompleteness   of the old clusters in the LMC training set because it has the fewest number of stars in each synthetic CMD.

\begin{figure}
\centering
\includegraphics[angle=90,scale=0.3]{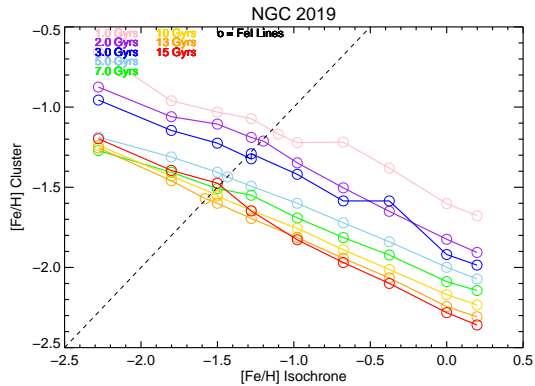}
\caption[Abundance solutions for Fe I lines for NGC 2019]{Abundance solutions for Fe I lines for NGC 2019.  Colors correspond to CMDs of different ages, as labeled.  Self-consistent solutions are found where the output [Fe/H]$_{\rm{cluster}}$ solution equals the input [Fe/H]$_{\rm{iso}}$ abundance of the isochrone. These solutions fall on the dashed black 1:1  line. }
\label{fig:2019 Fe plot} 
\end{figure}

\begin{figure}
\centering
\includegraphics[scale=0.5]{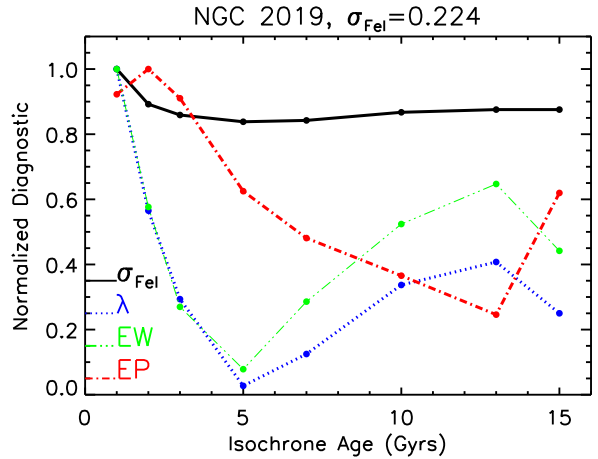}
\caption[Fe abundance diagnostics  for self-consistent solutions  for NGC 2019]{Fe abundance diagnostics for self-consistent solutions for NGC 2019. Each diagnostic has been normalized to its maximum, or worst, value in order to show all diagnostics on the same scale. Solid black line shows $\sigma_{\rm{Fe}}$, dotted blue line is the slope in [Fe/H] vs. $\lambda$, red dash-dot line is the slope in [Fe/H] vs. EP, and the  green dash-triple dot line is the slope in [Fe/H] vs. reduced EW (log(EW/wavelength)).  The best solutions are found for ages $>$ 5 Gyr. }
\label{fig:2019 diag} 
\end{figure}

We begin our analysis by determining the best-fitting CMD following our standard method.  In Figure~\ref{fig:2019 Fe plot}, we show the [Fe/H] solutions for the grid of synthetic CMDs described in \textsection~\ref{sec:cmds}, which span a range in age of 1-15 Gyr, and  range in metallicity of  $-2.6 \leq$[Fe/H]$ \leq+0.2$.  The CMDs with self-consistent [Fe/H] solutions (i.e. [Fe/H]$_{\rm{iso}}$=[Fe/H]$_{\rm{cluster}}$), lie on the dashed line in Figure~\ref{fig:2019 Fe plot}.  By requiring self-consistency in the [Fe/H] solution, we have narrowed down the grid of synthetic CMDs to one acceptable CMD for each age.    To determine which of the 8 self-consistent CMDs is the best match to the stellar population of NGC 2019,  we next compare the Fe line diagnostics.  These diagnostics are shown in Figure~\ref{fig:2019 diag}.  Each diagnostic has been normalized to its worst value in order to show all 4 diagnostics on the same scale.  The diagnostics in  Figure~\ref{fig:2019 diag} show similarities to the diagnostics found for other old clusters in the MW training set in \citetalias{milkyway} and for  old clusters in M31 \citep{m31paper}.  The best diagnostics for NGC 2019 are found for ages $\geq$5 Gyr.   While 5 to 15 Gyr is a much broader range in age than we found for MW clusters,  it can be seen in Figure~\ref{fig:2019 Fe plot}  that the difference in [Fe/H] between the 5 and 15 Gyr synthetic CMDs is only $\sim$0.15 dex, which means that the metallicity of NGC 2019 is already very well constrained.

We note that, as in our previous work,  we find that the best solutions are for ages $>$5 Gyr, even though the  behavior  of the diagnostics fluctuates somewhat, or in some cases the diagnostics don't completely agree with each other.   This is expected, as there are some properties of the theoretical stellar populations that don't exactly match the true stellar populations in the cluster, and the different diagnostics are sensitive to different stellar types.  More uncertain aspects of the isochrones can include  AGB stars, which we discuss below for NGC 2019, or blue horizontal branch stars and  blue stragglers, which we discuss in more detail for NGC 2005 in \textsection \ref{sec:2005}.

We next investigate the effect of incomplete sampling on the age and abundance solutions for NGC 2019.  We note that for old, low luminosity clusters, the most important sampling uncertainties  will be   fluctuations in the number and type of evolved stars,  particularly luminous AGB stars and RGB stars  \citep{2000A&AS..146...91B}.   Here we will explore tests to improve  our solution  by accounting for statistical fluctuations in luminous evolved stars.  Although these tests are probably unnecessary for the analysis of more distant, and thus better sampled, extragalactic clusters, they can be applied with no a priori knowledge of the CMD, and therefore can be easily generalized for the analysis of any cluster.

\begin{deluxetable}{r|rrr|rrr}
\tablecolumns{7}
\tablewidth{0pc}
\tablecaption{Synthetic CMD Comparison  \label{tab:cmdcomparison}}
\tablehead{
&\multicolumn{3}{c}{Averaged}& \multicolumn{3}{c}{Monte Carlo}\\
\colhead{Box}&\colhead{$M_{V}$} &\colhead{$B-V$}  &\colhead{N$_{stars}$} &\colhead{$M_{V}$} &\colhead{$B-V$} &\colhead{N$_{stars}$}  }
\startdata

1&7.178 &0.865 &19033.70  &  7.188 &0.868 &18867\\
2&6.211 &0.640 &7343.58  &  6.215 &0.641 &7439\\
3&5.640 &0.520 &4460.14   & 5.651 &0.521 &4325\\
4&5.208 &0.455 &2913.89   & 5.231 &0.458 &3025\\
5&4.836 &0.414 &2185.18   & 4.863 &0.417 &2169\\
6&4.488 &0.384 &1596.35   & 4.520 &0.386 &1596\\
7&4.165 &0.360 &1234.24   & 4.200 &0.363 &1212\\
8&3.875 &0.346 &853.19   & 3.901 &0.347 &867\\
9&3.592 &0.349 &661.95   & 3.616 &0.348 &665\\
10&3.290 &0.404 &492.02   & 3.333 &0.389 &508\\
11&2.876 &0.593 &340.62   & 2.977 &0.567 &370\\
12&2.056 &0.674 &160.46   & 2.231 &0.665 &187\\
13&1.310 &0.719 &81.82   & 1.478 &0.708 &99\\
14&0.653 &0.771 &44.14   & 0.888 &0.750 &55\\
15&0.101 &0.830 &26.12   & 0.319 &0.805 &33\\
16&$-$0.235 &0.871 &19.25  & $-$0.104 &0.854 &22\\
17&$-$0.743 &0.948 &12.11  & $-$0.453 &0.902 &16\\
18&$-$1.204 &1.033 &7.90  & $-$0.897 &0.975 &11\\
19&$-$1.623 &1.124 &5.35  & $-$1.289 &1.051 &8\\
20&$-$2.009 &1.225 &3.82  & $-$1.815 &1.173 &5\\
21&$-$2.344 &1.339 &2.80  & $-$2.455 &1.387 &3\\
22&$-$2.606 &1.448 &1.88  & 0.442 &0.569 &36\\
23&0.185 &0.629 &31.98   & 0.430 &0.532 &35\\
24&0.430 &0.530 &35.19   & 0.400 &0.531 &34\\
25&0.404 &0.528 &34.44   & $-$0.231 &0.732 &23\\
26&0.071 &0.653 &26.34   & $-$1.552 &1.048 &6\\
27&$-$1.258 &0.981 &7.93 &  $-$2.676 &1.514 &3\\
28&$-$2.583 &1.520 &2.28&\nodata&\nodata&\nodata \\ 
\enddata

\end{deluxetable}

In order to allow the synthetic CMD stellar populations to vary in a meaningful way, we use a Monte Carlo technique to statistically populate the cluster IMF with discrete numbers of stars.  The total number of stars in each synthetic CMD is normalized so that the total flux in the CMD is consistent with the observed  \mv  of the cluster, modulated by the observed flux fraction.   When reducing the total flux of the cluster to be consistent with the observed region, we have assumed that stars of different masses are evenly distributed throughout the cluster.
An important difference between generating the synthetic CMDs using a Monte Carlo method versus the technique used for the original analysis method developed in  \citetalias{2008ApJ...684..326M} and \citetalias{milkyway}, is that the Monte Carlo method creates  CMD boxes  with integer numbers of stars.   The original method creates average CMD boxes with non-integer numbers of stars, with the constraint that the number of stars in each box must be $\geq$ 1.0.  To illustrate this point, in Table~\ref{tab:cmdcomparison} we compare the synthetic CMD created using the original method for   a 10 Gyr, [Fe/H]=$-1.5$ isochrone, to a synthetic CMD created using the Monte Carlo technique.  As expected, the significant differences between the two CMDs occur for the luminous ($M_V \leq 0$) boxes representing the upper RGB and AGB stars.  Most of these boxes were created by averaging the properties of 10 stars or less, and thus are particularly sensitive to small number statistics.  In the example comparison in Table~\ref{tab:cmdcomparison}, the CMD created using the Monte-Carlo technique has one less box than the CMD created using the original technique, due to the redistribution of the luminous stars in boxes 17 through 28.

To approximately assess the impact of sampling uncertainties, we have decided to create 100 synthetic CMDs for each isochrone of interest using the Monte-Carlo technique.  We will refer to these CMDs created using the Monte-Carlo technique as ``CMD realizations.''  For clarity, we will refer to synthetic CMDs created using the original technique as ``average CMDs.''

\begin{figure}
\centering
\includegraphics[angle=90,scale=0.35]{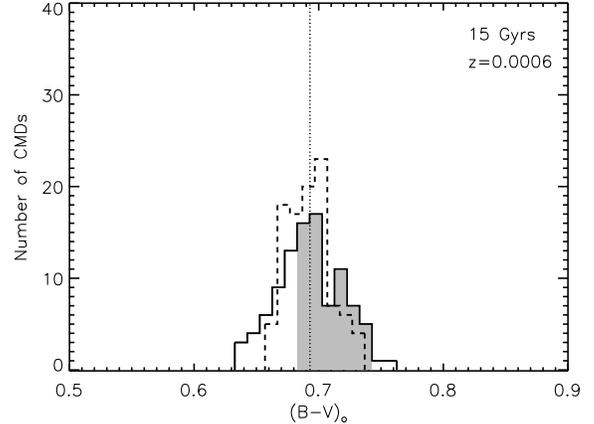}
\caption[Histogram of integrated \bvo~ color for 100 CMD realizations of a 15 Gyr, z=0.0006, (Fe/H=$-1.5$)  isochrone]{Histogram of integrated \bvo~ color for 100 CMD realizations of a 15 Gyr, z=0.0006, ([Fe/H]=$-1.5$) isochrone.  Solid black line shows the histogram for a population where the total flux in stars has been normalized to 45\% of a \mv$=-8$  cluster, which is appropriate for our integrated light spectrum of NGC 2019.  Dashed black line shows the histogram for a population normalized to 100\% of a \mv$=-8$ cluster.  CMDs with \bvo~ color consistent with the observed, reddening-corrected $B-V$ of NGC 2019 are shaded in gray. The \bvo~ of the average CMD is marked by the dotted line, and is consistent with the peak in both histograms for the CMD realizations, as it represents an average population.  }
\label{fig:2019 histo} 
\end{figure}

One way to evaluate the effect of under-sampling the population is to compare the level of statistical fluctuations in the intrinsic, integrated  color, which is defined here as  ``\bvo.''   
For example, \cite{1999A&AS..136...65B,2000A&AS..146...91B} evaluated the statistical fluctuations in the  integrated colors of cluster CMDs that were created using a Monte Carlo technique similar to the procedure used here.  They found that the integrated colors showed large variations for  CMDs created with less than 30,000 stars.
We show the variation in \bvo~ for 100 CMD realizations of one of the best-fitting CMDs for NGC 2019 in    Figure ~\ref{fig:2019 histo}. 
   The CMD realizations, shown by the solid line,  were created using the best-fitting CMD with an age of  15 Gyr, and [Fe/H]=$-1.5$, and show a spread in  \bvo~ of $\sim$0.13 mag.  
   Note that the number of stars in each of the CMDs included in the solid line histogram were normalized to correspond to 45\% of the flux of a \mv$\sim-8$ cluster, which corresponds to the observed flux of NGC 2019.   To evaluate the spread in \bvo~ that results from our incomplete sampling,  we also create CMD realizations normalized to 100\% of a  \mv$\sim-8$ cluster.  These more luminous CMD realizations are shown by the dashed line histogram in Figure ~\ref{fig:2019 histo}.  The spread in \bvo~ of these more luminous CMD realizations  is $\sim$0.08 mag, which, as expected, is a smaller range.   As a sanity check, we also mark the \bvo~ of the average CMD created using the original technique with a dotted line in Figure~\ref{fig:2019 histo}.  We note that the \bvo$=0.69$ of the average  CMD is consistent with the \bvo~ of the peak in both CMD realization histograms,  which we would expect because this CMD represents an average population for the isochrone.

\begin{figure}
\centering
\includegraphics[angle=90,scale=0.3]{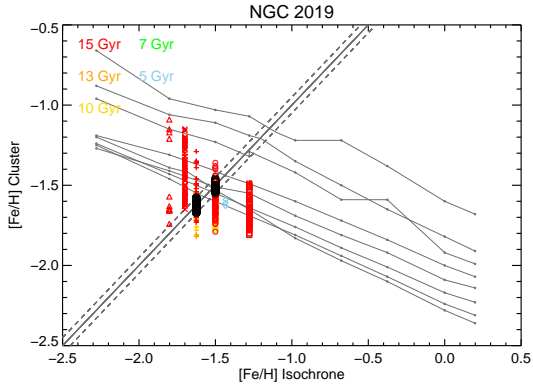}
\caption[Same as Figure~\ref{fig:2019 Fe plot}, with the addition of the Fe I abundance results for the CMD realizations of NGC 2019]{Same as Figure~\ref{fig:2019 Fe plot}, with the addition of the Fe I abundance results for the CMD realizations of NGC 2019.  Red points correspond to 15 Gyr CMD realizations.  Only CMD realizations consistent with the observed $B-V$ color of NGC 2019 are shown.   Black circles on the solid black 1:1 line denote CMD realizations that satisfy the self-consistent criterion [Fe/H]$_{\rm{\rm{iso}}}$=[Fe/H]$_{\rm{cluster}}$, and are therefore possible solutions for the population of NGC 2019. }
\label{fig:2019samples} 
\end{figure}

After creating CMD realizations for a given isochrone, an appropriate subset of CMD solutions can be identified by applying other observable constraints.  As a first constraint, we impose the self-consistency requirement used in our original method.  This constraint must be reapplied to the [Fe/H] abundance solutions for the CMD realizations because the redistribution of luminous, cool giants in the CMD realizations results in a spread in the derived output [Fe/H].  The results for the CMD realizations of NGC 2019 are shown  in Figure~\ref{fig:2019samples}, with 
 the  average CMD solutions  from Figure~\ref{fig:2019 Fe plot} for reference.   We have marked the CMD realizations that meet the self-consistency requirement as black circles in Figure~\ref{fig:2019samples}.

\begin{figure*}
\centering
\includegraphics[scale=0.7]{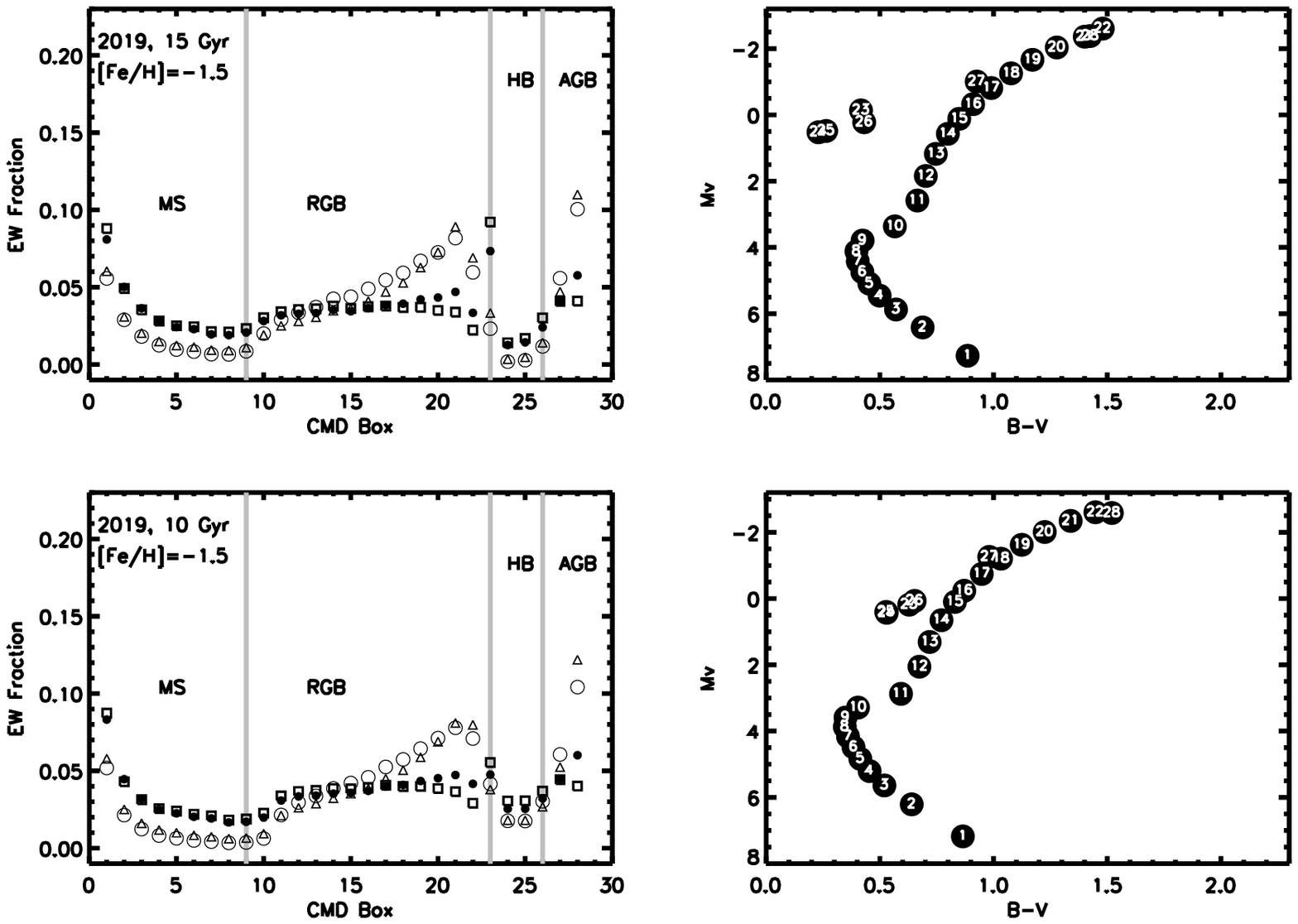}
\includegraphics[scale=0.7]{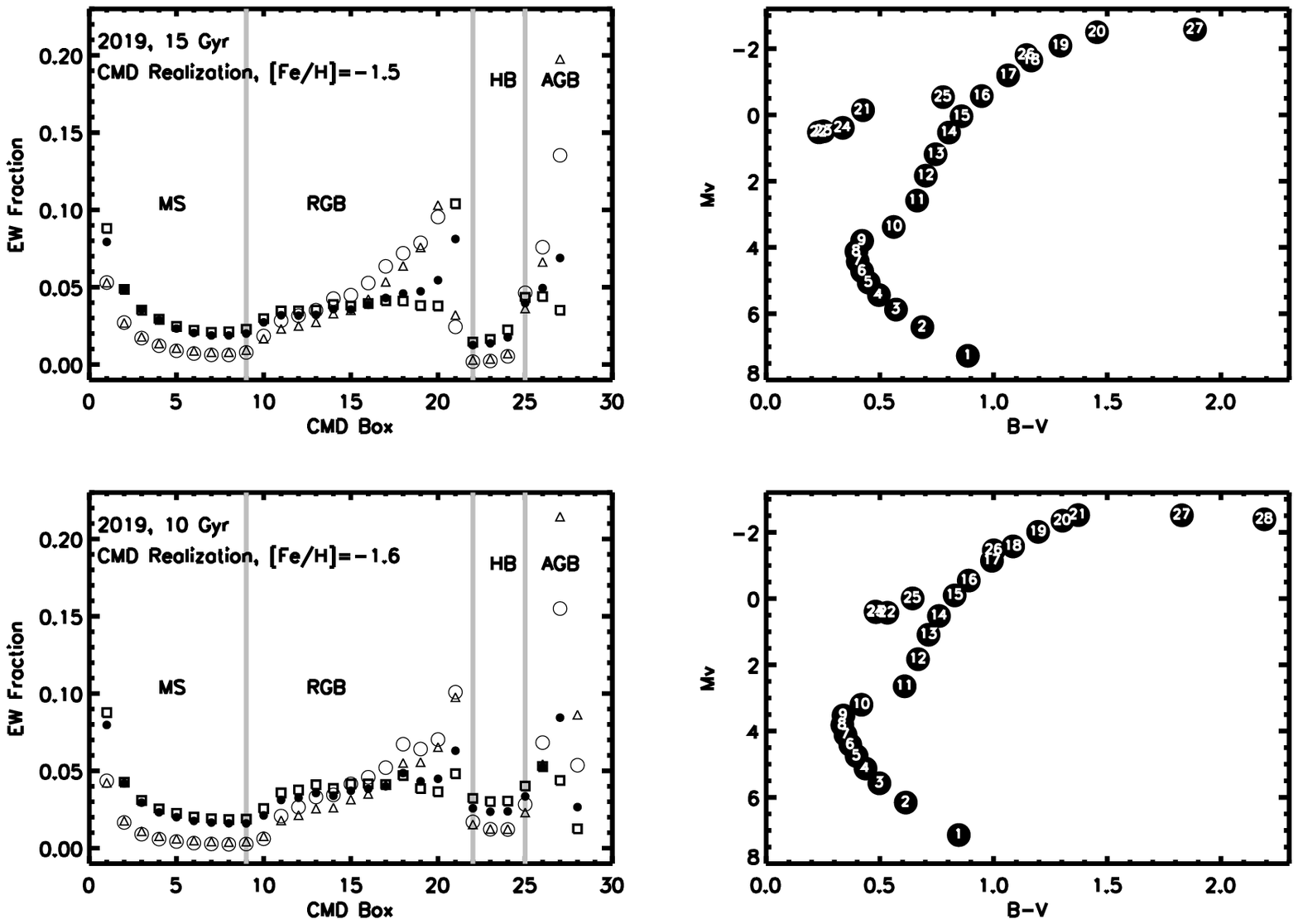}

\caption{\footnotesize Fe I EW strength for synthetic CMD boxes.  Left panels show the contribution of each synthetic CMD box to the total EW fraction of an Fe I line, and right panels show the associated synthetic CMDs, with the box numbers labeled.   In the left panels,  for an EP comparison, the 5307 \rAA Fe I line (EP=1.60 eV), and 5383 \rAA Fe I line (EP=4.3eV), are shown by open and filled circles respectively.   For a wavelength comparison, the  4494 \rAA Fe I line (EP=2.20 eV), and 6677 \rAA Fe I line (EP=2.70) are shown by  open squares and  open triangles, respectively.  The RGB and AGB CMD boxes dominate the contribution to the Fe I line EWs. Top two rows correspond to the average CMDs with ages of 15 and 10 Gyr, respectively, and [Fe/H]=$-1.5$.  Third row corresponds to the best-fitting CMD realization for NGC 2019 with an age of 15 Gyr and [Fe/H]=$-1.5$, and the bottom row corresponds to the best-fitting CMD realization with an age of 10 Gyr and [FeH]=$-1.6$. }
\label{fig:2019fracew} 
\end{figure*}

The spread in derived [Fe/H] is not unexpected,  because cool giants have a strong influence on the flux-weighted Fe I EWs, which was discussed in \citetalias{2008ApJ...684..326M}.     To  clarify this point in the context of sampling uncertainties,    in  Figure~\ref{fig:2019fracew} we  show the EW fraction contributed from different regions of the CMD for different Fe lines.  The lines  have been chosen to illustrate the importance of different types of stars on different lines. Four lines are plotted, which have very different combinations of wavelength and EP.  This  emphasizes how redistributions of luminous, cool giants can affect the [Fe/H] vs. EP and [Fe/H] vs. wavelength diagnostics.
  In general,  Figure~\ref{fig:2019fracew} demonstrates that the RGB and AGB CMD boxes dominate ($\sim$80\%) the Fe I EWs.  More subtle is the fact that high EP and bluer wavelength Fe I lines are slightly  less sensitive to AGB and RGB boxes than low EP and redder wavelength lines.   The high EP and bluer wavelength lines are more sensitive to the main sequence (MS) and horizontal branch (HB) boxes than the low EP and redder wavelength lines, because they have stars at hotter temperatures.  The sensitivity of the Fe I line abundances to small redistributions of flux between the AGB, RGB and HB means that different CMD realizations result in a spread in mean [Fe/H], but also that the Fe line diagnostics are in principle  sensitive to whether a statistically rare CMD realization results in an improvement in the overall stability of the [Fe/H] solution.

From this exercise it is apparent  that there may be self-consistent  CMD realizations that exist   at different  [Fe/H]  than our original solution.  It is therefore necessary to follow the procedure outlined above for a range of isochrone metallicity (and ages if warranted).  In the case of NGC 2019, we have done this for isochrones with  [Fe/H]=$-1.8$ to $-1.3$ and for ages between 5 to 15 Gyr, which can be seen in Figure~\ref{fig:2019samples}.

\begin{figure}
\centering
\includegraphics[angle=90,scale=0.35]{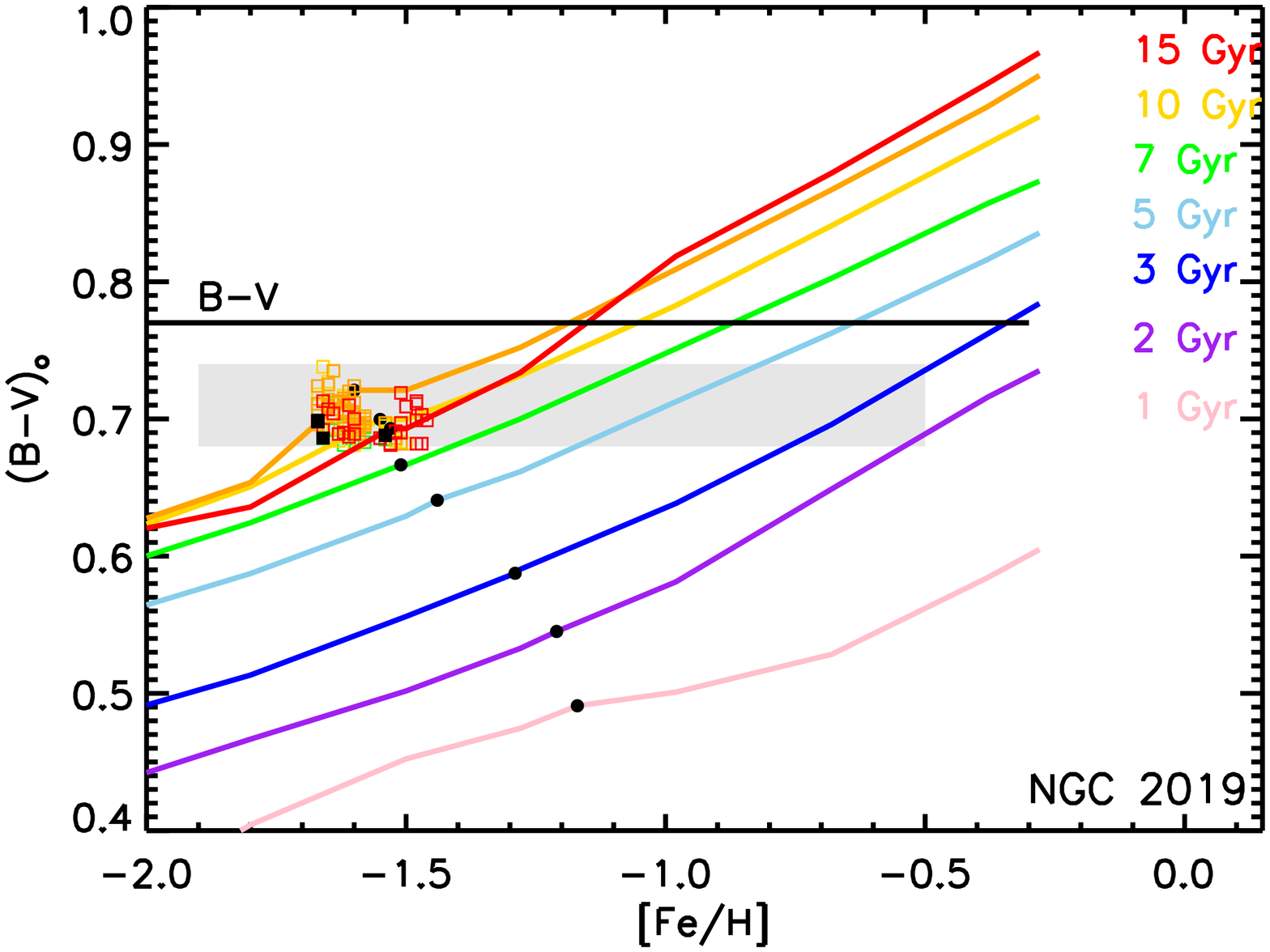}
\caption[Integrated \bvo~ colors calculated from the grid of synthetic CMDs for NGC 2019]{Integrated \bvo~ colors calculated from the grid of synthetic CMDs for NGC 2019, shown as a function of [Fe/H]$_{\rm{iso}}$ for each age. Colors denote CMDs with the  same ages as in Figure~\ref{fig:2019 Fe plot}.  Black circles show the  [Fe/H]$_{\rm{cluster}}$ and age solutions determined for the average CMD analysis.  Solid horizontal line corresponds to the observed $B-V$ of NGC 2019 from Table~\ref{tab:lmctable}. Gray shaded region corresponds to the region in \bvo~ that is consistent with the reddening corrected $B-V$, with photometric uncertainties.  Colored square points correspond to the \bvo~ and [Fe/H]$_{\rm{cluster}}$ of self-consistent CMD realizations for NGC 2019, while black solid squares correspond to the best-fitting CMD realization for each age.    }
\label{fig:2019colors} 
\end{figure}

After we have eliminated CMD realizations based on our [Fe/H] self-consistency requirement, we can eliminate additional CMDs based on their consistency with the cluster's observed, reddening-corrected integrated color.
CMD realizations for NGC 2019  that are consistent with the observed $B-V$ from the catalog of 
\citet{2008MNRAS.385.1535P}, and the $E(B-V)$ determined by \citet{1998MNRAS.300..665O} are shaded in gray in the  histogram in  Figure ~\ref{fig:2019 histo}. In this case, we note that the \bvo~ of the average CMD is also consistent with the observed $B-V$ of NGC 2019,  which is a hint that the average CMD may already be a reasonable match. 
The original average CMD solutions at ages of  5 and 7 Gyr have integrated \bvo~ colors that are inconsistent with the observed $B-V$.  This is apparent in Figure~\ref{fig:2019colors}, in which the average 5 and 7 Gyr solutions, which are marked as filled black circles, are outside the shaded gray region corresponding to the observed $B-V$.   While  there are some CMD realizations with an age of 7 Gyr that do satisfy the \bvo~ requirement,  in general there are many fewer acceptable 7 Gyr CMD realizations than there are for ages $>$10 Gyr.   In other words, while NGC 2019 may have an age of 5 Gyr, this possibility is statistically unlikely according to our analysis.

\begin{figure}
\centering
\includegraphics[scale=0.5]{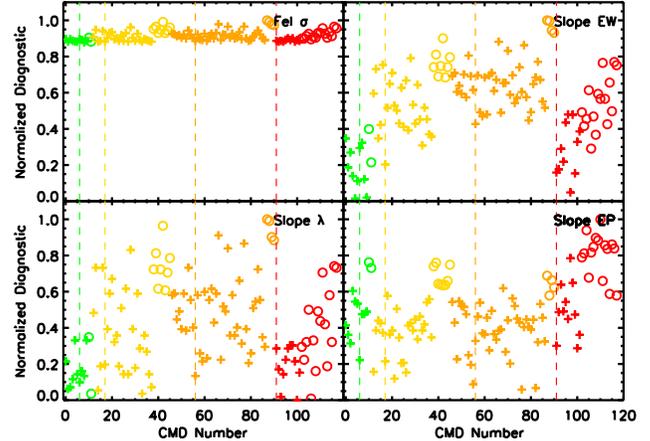}
\caption[Normalized diagnostics for NGC 2019 CMD realizations that satisfy color and Fe I self-consistency criteria. ]{Normalized diagnostics for NGC 2019 CMD realizations that satisfy color and Fe I self-consistency criteria.  Symbols and colors are the same as in Figure ~\ref{fig:2019samples}. CMDs are arranged by increasing age, and then by increasing [Fe/H]$_{\rm{iso}}$ for each age.   The best solution for each age is shown by a dashed line, and corresponds to the solution that best minimizes all four diagnostics at once. For NGC 2019, the best solution overall is for an age of 10 Gyr, and [Fe/H]$=-1.6$. }
\label{fig:2019newdiag} 
\end{figure}

We find 120 CMD realizations for isochrones with ages of 7$-$15 Gyr  and 
[Fe/H]$_{\rm{iso}}=-1.6$ to $-1.5$ that are consistent with the  observed \bvo~ of NGC 2019 and also result in a self-consistent [Fe/H] solution.   We next identify the best-fitting CMD(s) for each age using the Fe diagnostics introduced in \textsection~\ref{sec:oldcmd}. In Figure ~\ref{fig:2019newdiag} we show  the  normalized Fe I diagnostics for the 120 CMD realizations, which we have designated CMDs 0 through 119.  As was the case for the original 8 average CMD solutions, the most stable CMD realizations are those that minimize all of the diagnostics simultaneously.  We highlight the best-fitting,  most stable solution for each age with dashed lines in Figure~\ref{fig:2019newdiag}.   In the case of NGC 2019 we find that for ages of 7, 10 and 15 Gyrs the best-fitting CMD realization has an input [Fe/H]$_{\rm{iso}}=-1.6$, which is 0.1 dex lower than the  [Fe/H]$_{\rm{iso}}$ of the best-fitting average CMD solutions.  For an age of 13 Gyrs the best-fitting solution has an input [Fe/H]$_{\rm{iso}}=-1.5$, which is the same  as for the original average CMD solution.  The statistical line-to-line scatter in Fe abundance, or $\sigma_{\rm{Fe}}$, is 0.20 dex.

\begin{figure}
\centering
\includegraphics[scale=0.5]{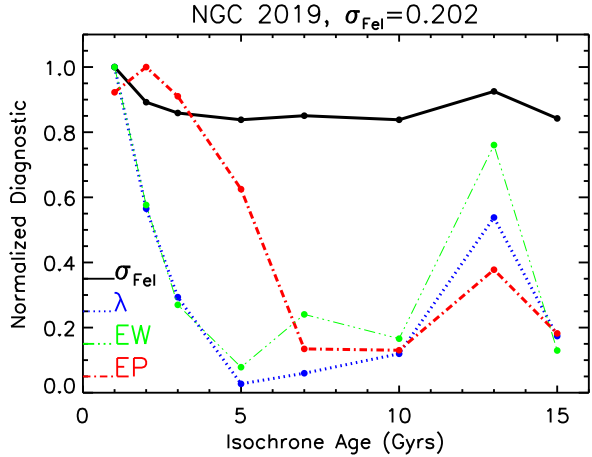}
\caption[Same as Figure~\ref{fig:2019 diag}, except the solutions shown for ages of 7 to 15 Gyr correspond to the solutions for the  best-fitting CMD realizations.]{Same as Figure~\ref{fig:2019 diag}, except the solutions shown for ages of 7 to 15 Gyr correspond to the solutions for the  best-fitting CMD realizations.  The diagnostics for these ages are more correlated with each other than for the average CMD solutions, and the best solutions are found for 7$-$15 Gyrs.  }
\label{fig:new2019 diag} 
\end{figure}

After constraining the 120 self-consistent CMD realizations to a single best-fitting CMD realization for each age between 7 to 15 Gyr, we can compare the normalized diagnostics in order to identify the best-fitting age(s) for NGC 2019, similar to what we did for the average CMD solutions at the beginning of this section.   Figure~\ref{fig:new2019 diag} shows the diagnostics for the original 1 to 5 Gyr solutions compared to the diagnostics for the best-fitting CMD realizations at ages of 7 to 15 Gyr.   The main differences between  Figure~\ref{fig:2019 diag} and Figure~\ref{fig:new2019 diag} are that the diagnostics for ages of 7 to 15 Gyrs are clearly much more strongly correlated with each other, and that the CMDs with ages of 7, 10 and 15 Gyrs result in more stable [Fe/H] solutions than the original average CMD solutions.  Our final solution is constrained to two possible synthetic CMDs with ages of 7 to 10 Gyrs.

\begin{figure*}
\centering
\includegraphics[angle=90,scale=0.2]{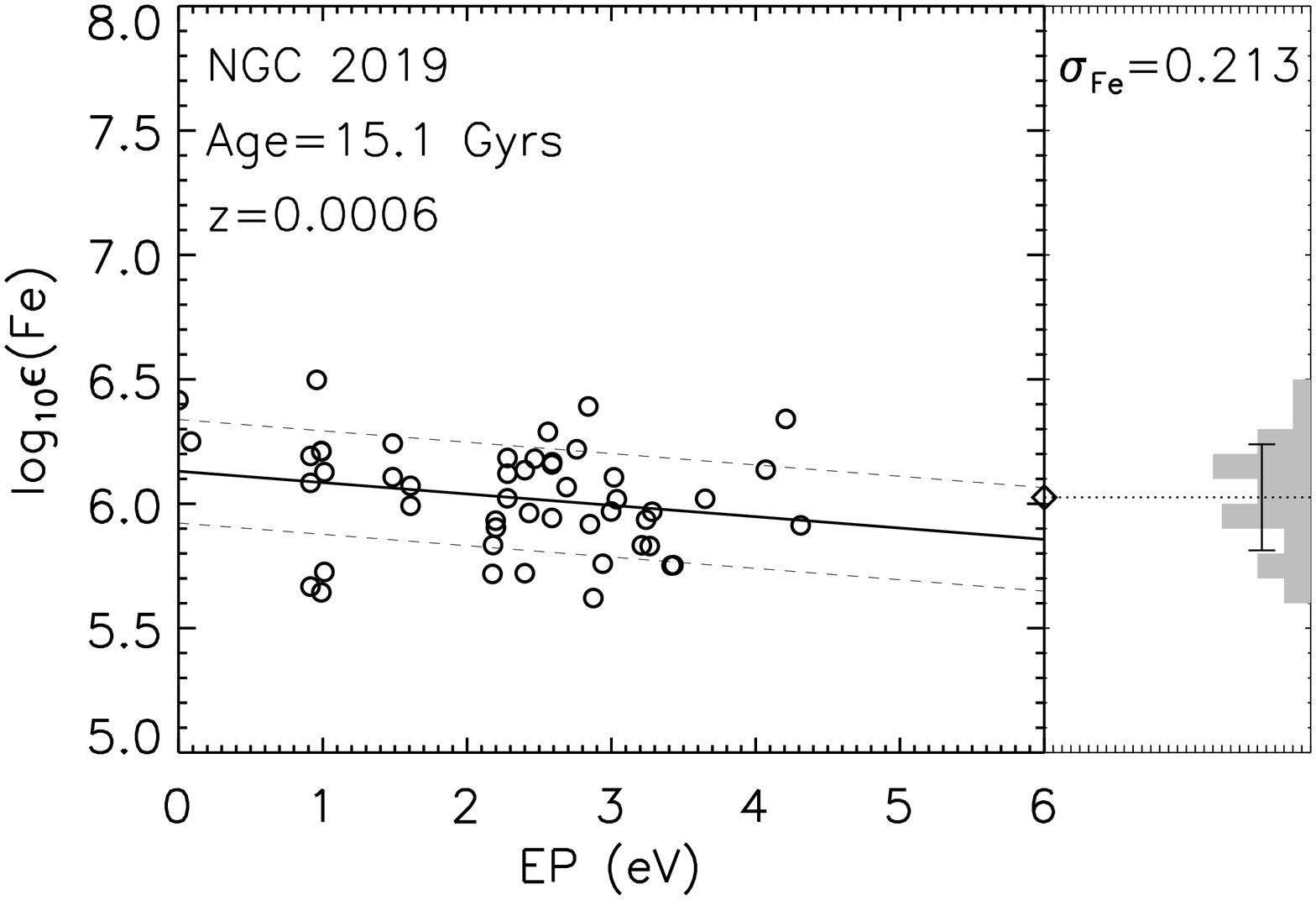}
\includegraphics[angle=90,scale=0.2]{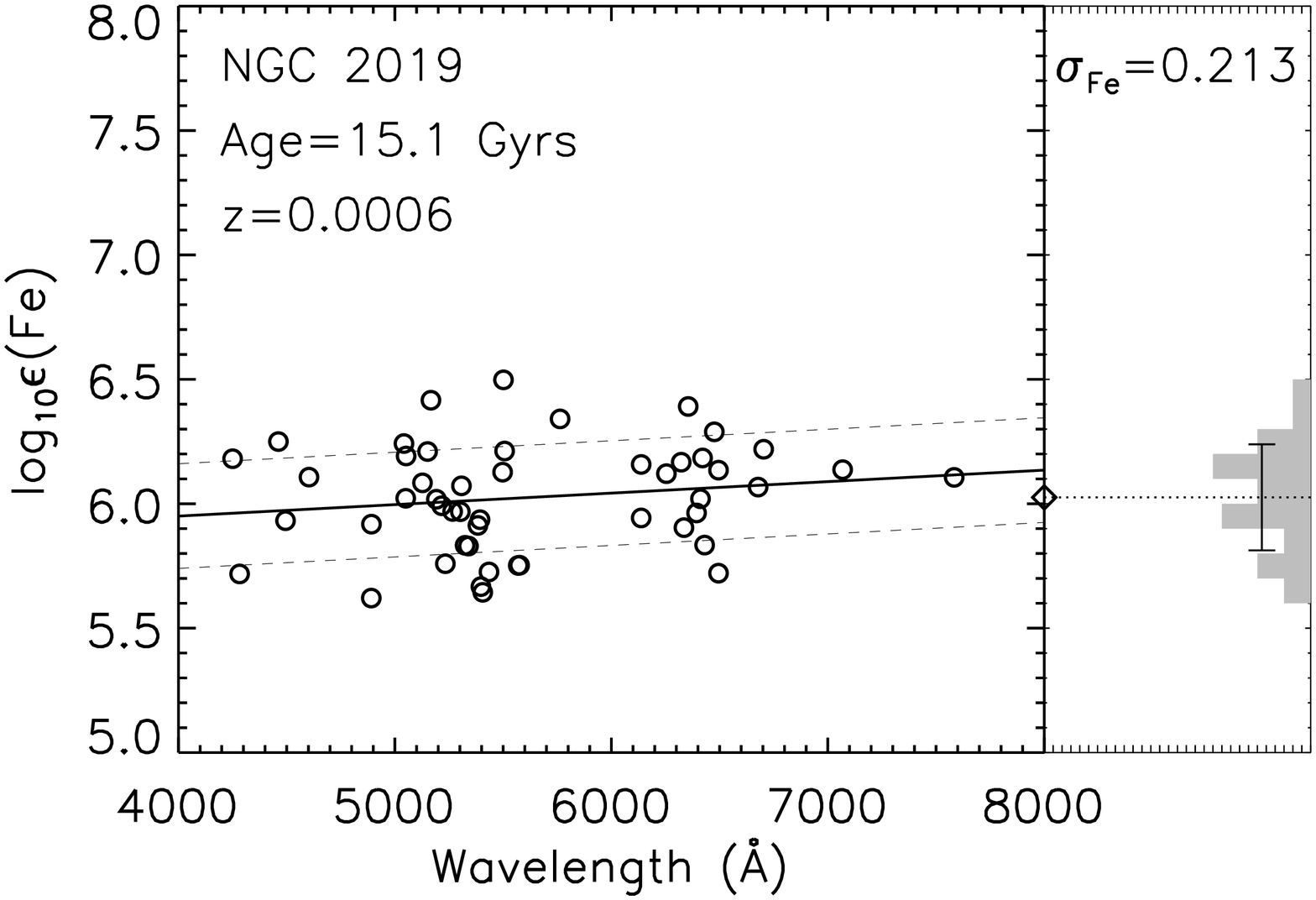}
\includegraphics[angle=90,scale=0.2]{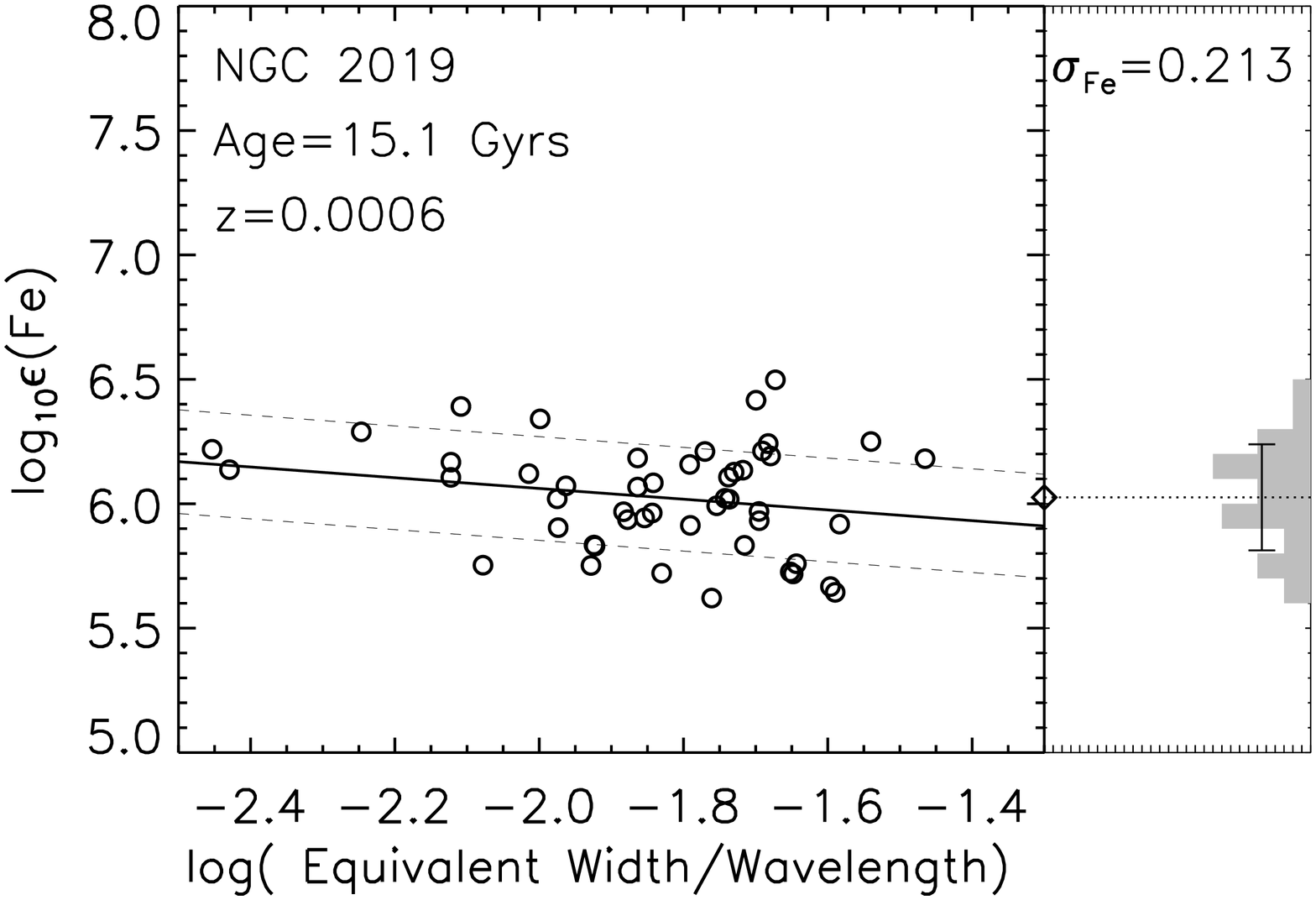}
\includegraphics[angle=90,scale=0.2]{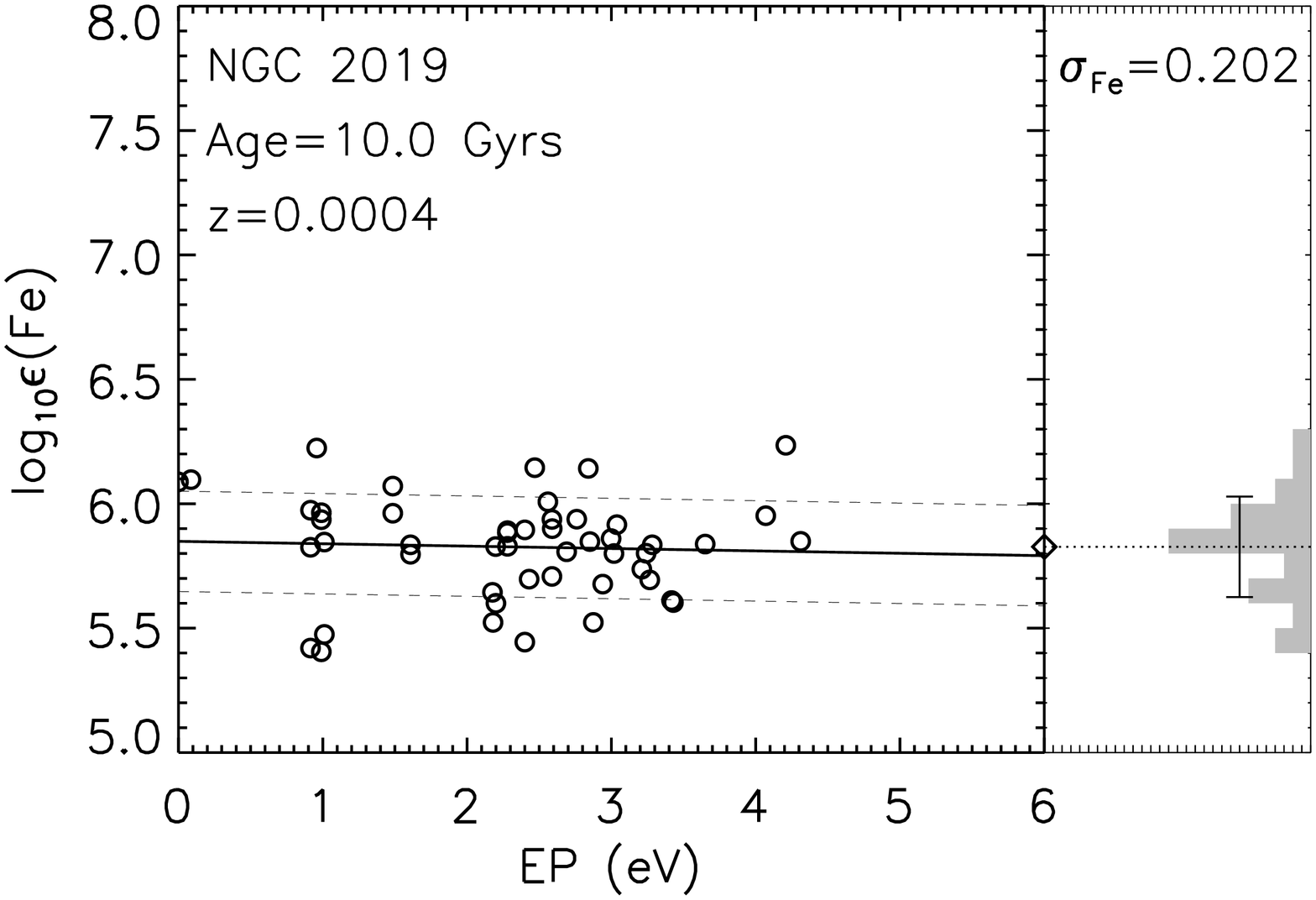}
\includegraphics[angle=90,scale=0.2]{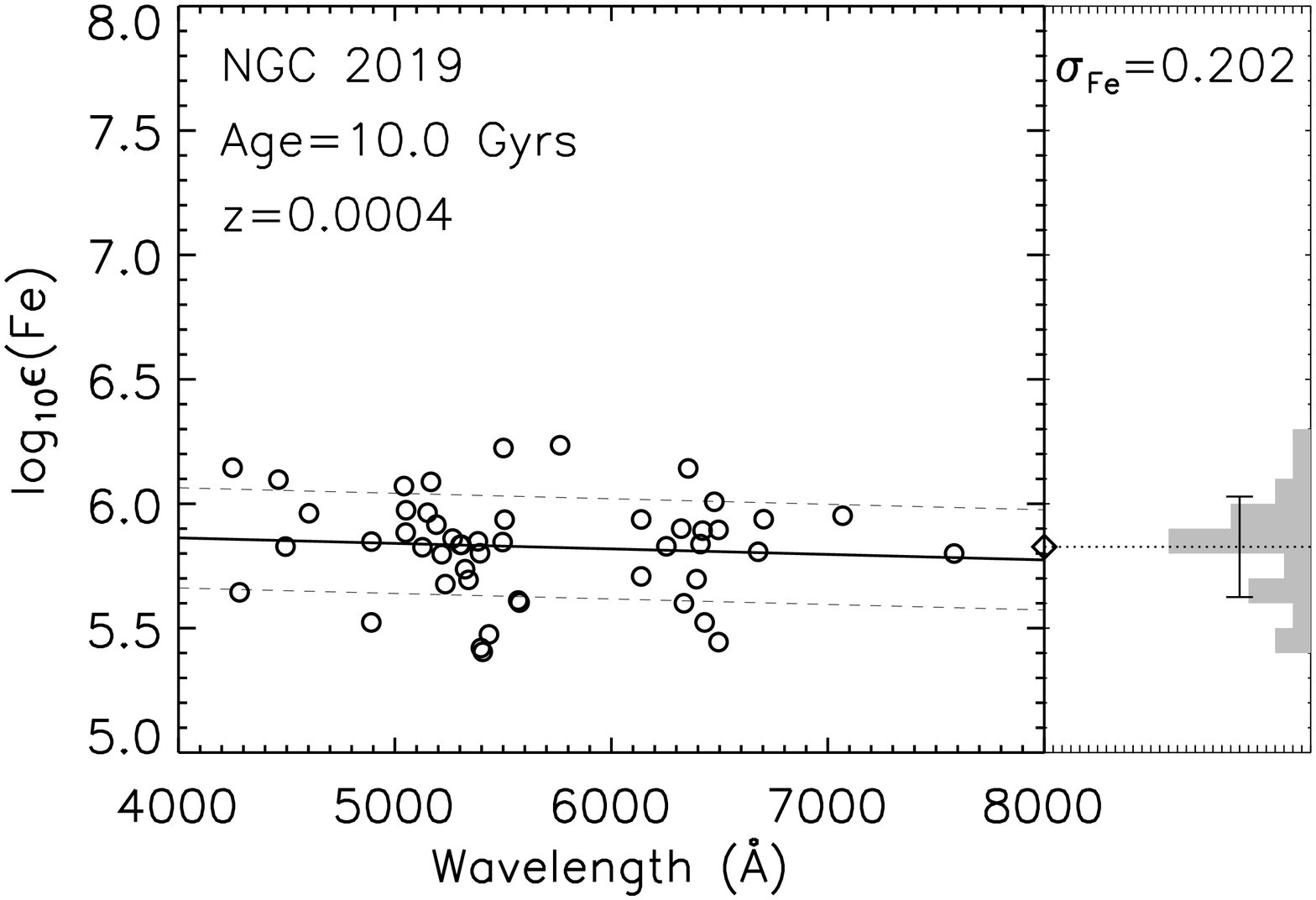}
\includegraphics[angle=90,scale=0.2]{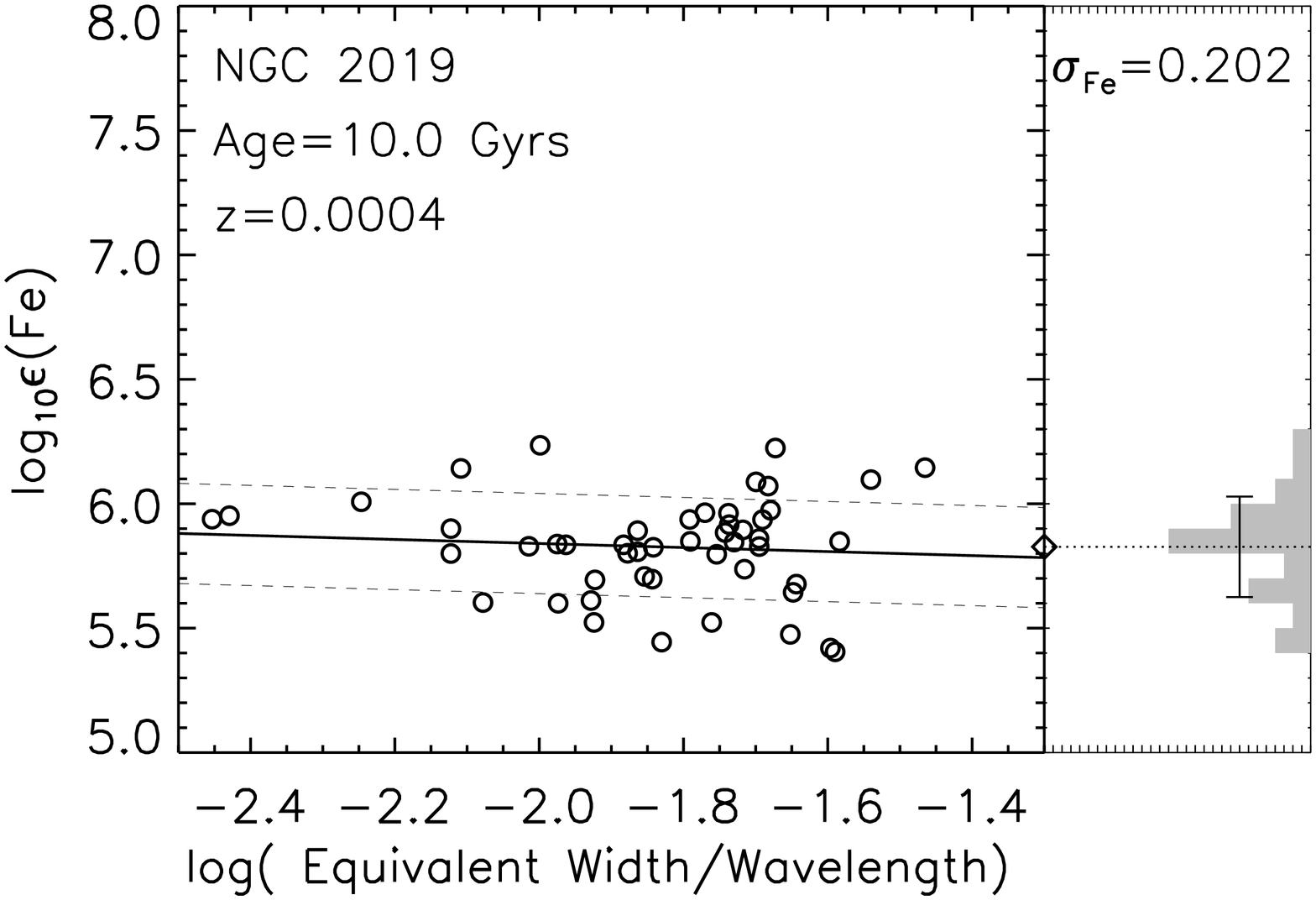}
\caption{Individual diagnostic plots for NGC 2019, where Fe I  lines are marked by dark circles.    The solid line shows the linear fit to the Fe I lines and dashed lines show the 1 $\sigma$ deviation of points around the fit.  Diamonds mark the average Fe I abundance.  The original average CMD solution using a 15 Gyr, [Fe/H]$-1.5$ isochrone is shown in the top panels.  The final solution for a best-fitting CMD realization using a 10 Gyr, [Fe/H]=$-1.6$ CMD isochrone is shown in the bottom panels.  The final solution (bottom row) has a smaller $\sigma_{\rm{Fe}}$ and  smaller dependence on EP, wavelength, or reduced EW than the original solution. }
\label{fig:2019 diagnostics} 
\end{figure*}

 For reference, in Figure~\ref{fig:2019 diagnostics} we show the [Fe/H] solutions as a function of EP, wavelength and reduced EW for the original average 15 Gyr solution, as well as the best-fitting 10 Gyr CMD realization.   It is visually apparent in Figure~\ref{fig:2019 diagnostics} that for the 10 Gyr CMD realization the $\sigma_{\rm{Fe}}$ is reduced and the dependence of [Fe/H] on EP, wavelength and reduced EW is significantly smaller. Therefore,  we choose to use the 10 Gyr CMD realization as our final best-fitting CMD, which will be used for calculating the abundances of all other elements in \citetalias{paper4}.

We are able to find a much more stable solution for NGC 2019 when allowing for sampling uncertainties because of the different contributions  of  RGB and AGB stars to the flux in different   Fe lines.  
In principle, we can not only identify the most stable solution, but also evaluate whether  statistically rare, but cool AGB stars may be present in a cluster. 
To understand why,  it is helpful to revisit the solutions for NGC 2019 in Figure~\ref{fig:2019fracew}. The right panels of Figure~\ref{fig:2019fracew} show the synthetic CMDs for the original 15 and 10 Gyr solutions, as well as a 15 Gyr CMD realization at the same [Fe/H] of the original solutions, and the best-fitting 10 Gyr CMD realization with a lower [Fe/H].
The original 10 and 15 Gyr solutions are very similar except for the  position of the HB boxes, and therefore there are only subtle differences in the Fe I EW fractions in the left panels.   However,   for both  the 15 Gyr  and 10 Gyr CMD realizations there are very red, very luminous, and very cool average star boxes on the AGB that are not present in either of the original CMDs.   AGB stars at these types of  luminosities and temperatures are relatively rare, so  for less massive clusters, boxes at these positions in the CMD do not appear  for the average CMDs that we construct using our original technique.   It is evident from Figure~\ref{fig:2019fracew} that not only do  the AGB stars in the very luminous and very cool boxes contribute significantly to the Fe I EWs, but they contribute especially to the low EP and redder wavelength  Fe I line EWs. 
 In the case of NGC 2019, for the original average CMD solutions we found higher [Fe/H] at redder wavelengths, as well as higher [Fe/H] at lower EP.  The   addition of  significant line strength in very red and very cool stars in the synthesized EWs  means that lower Fe abundances are required to match the observed EWs for the redder wavelength and  lower EP lines.  Therefore the result of adding significant cool AGB flux is a decrease in the dependence of  [Fe/H] on  EP and wavelength. 
We emphasize that  although the predictions of the properties of very cool AGB stars (or M stars) in the isochrones are still uncertain to some degree, our final result does not depend on determining the temperatures and luminosities of these stars precisely.  Instead we have isolated a subset of solutions that have $4-10\%$ of the  flux  averaged at  approximately  the same temperature and luminosity.

In conclusion, we have calculated age and abundance solutions for synthetic CMDs with a range of age and [Fe/H] for NGC 2019, and found that a stellar population that is 5 to 15 Gyr in age, and has [Fe/H]$=-1.5$ provides the most self-consistent solution using our original technique.  In this section we have described a supplementary  technique to evaluate the level of sampling incompleteness for our observations, and performed tests to assess whether the [Fe/H] solutions can be improved by allowing for statistical fluctuations in the stellar population.   In the case of NGC 2019, we find that these tests allow us to  improve our constraint on the age to 7 to 10 Gyr, as well as determine that a CMD realization with  [Fe/H]=$-1.6$ that includes rare and cool  AGB stars is a more appropriate match to the stellar population.  As shown in Figure \ref{fig:2019 diagnostics}, the line-to-line scatter is $\sigma_{\rm{Fe}}=0.20$ dex.  For solutions between 7 and 10 Gyr, we find a maximum difference in [Fe/H] of 0.08 dex, and therefore we quote an uncertainty due to the age of the cluster, $\sigma_{\rm{age}}$, of $\pm$0.04 dex.  In \textsection \ref{sec:discussion} we  discuss the final error analysis for all clusters, including statistical uncertainties  and systematic errors.

\subsubsection{NGC 2005}
\label{sec:2005}

The next LMC training set cluster we analyze is NGC 2005.  This and the remaining clusters are all analyzed as outlined for NGC 2019, above.  Here we only summarize the final result for the best-fitting CMD solutions.

\begin{figure}
\centering
\includegraphics[scale=0.5]{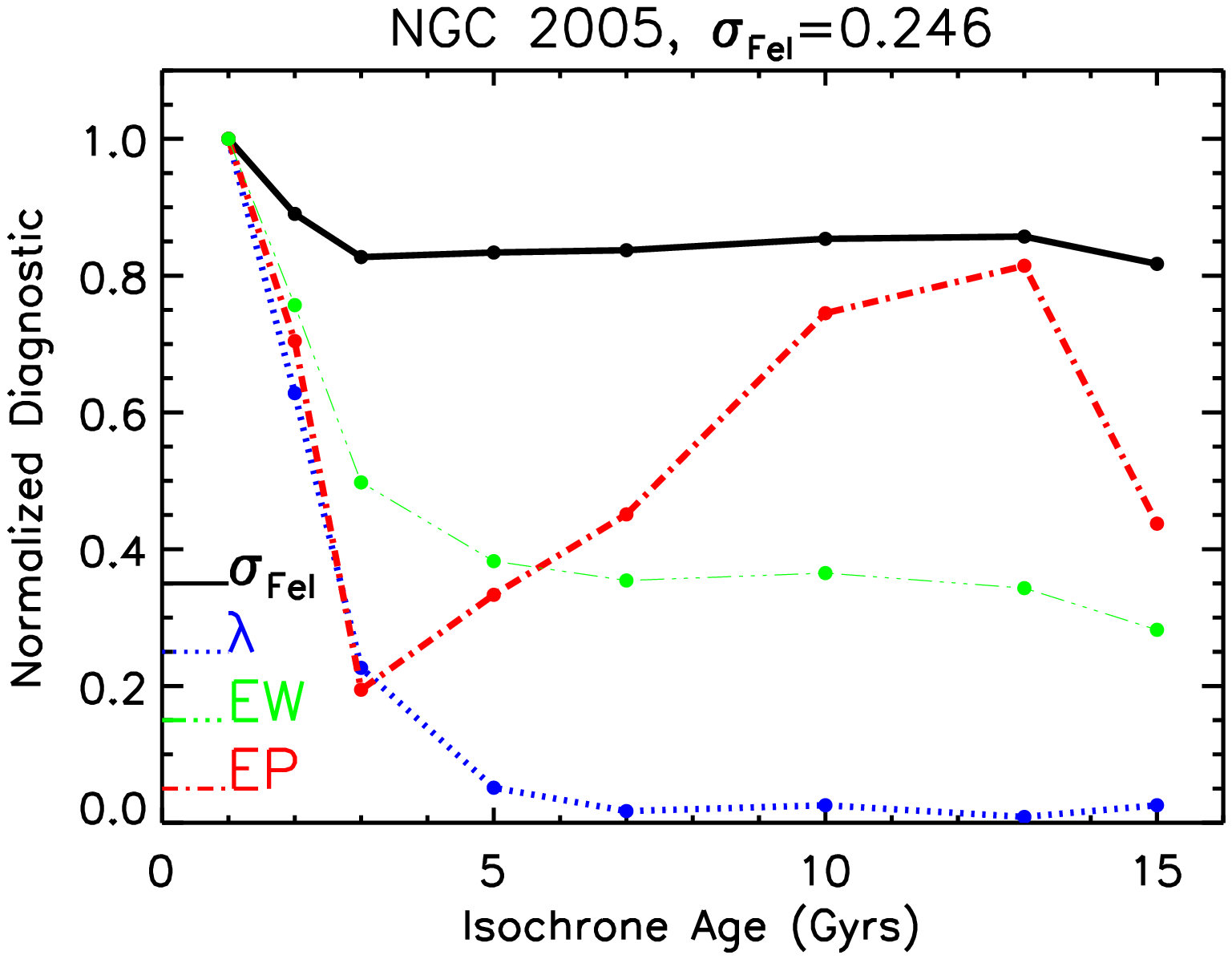}
\caption[Same as Figure~\ref{fig:2019 diag} for NGC 2005.]{Same as Figure~\ref{fig:2019 diag} for NGC 2005. Best solutions are found for ages $>$5 Gyr.  }
\label{fig:2005 diag} 
\end{figure}

\begin{figure}
\centering
\includegraphics[angle=90,scale=0.35]{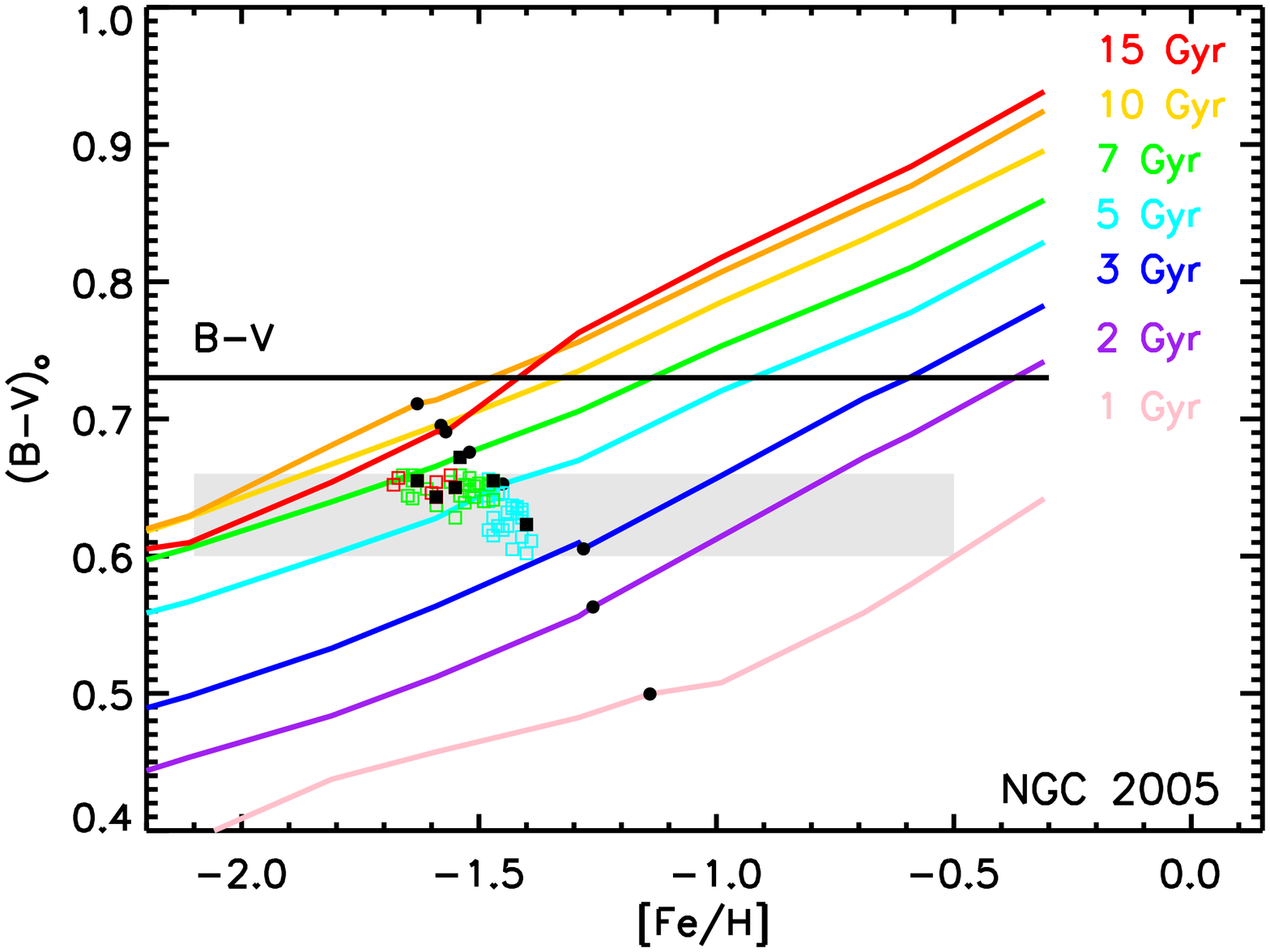}
\caption[ Same as Figure~\ref{fig:2019colors}  for NGC 2005.]{ Same as Figure~\ref{fig:2019colors}  for NGC 2005.}
\label{fig:2005colors} 
\end{figure}

The diagnostics for the 8, self-consistent, average CMD solutions for NGC 2005 are shown in  Figure~\ref{fig:2005 diag}.  Like NGC 2019,  the diagnostics for the 1 to 3 Gyr CMDs for NGC 2005 show that these are poor solutions. However, there is very little difference in the  quality of the solutions between 5 and 15 Gyr, because the RGB morphology is very similar in this age range.    The 10 to 15 Gyr average CMD solutions result in [Fe/H]$\sim-1.6$, and  the individual diagnostics for the 15 Gyr average CMD solution show a  small dependence of [Fe/H]  with wavelength, as well as a more significant dependence  of [Fe/H] with EP and reduced EW.

The line-to-line scatter of the best abundance solutions is $\sigma_{\rm{Fe}}\sim$0.25 dex, which is higher than what we typically obtained for clusters in the MW training set.   As discussed in \textsection~\ref{sec:trainingSet}, this is probably a consequence of the lower S/N of the LMC training set, and makes NGC 2005 an interesting test case for analysis of clusters with low S/N data.  In addition, NGC 2005 is an interesting test case for our new technique to address sampling uncertainties because it has a very blue horizontal branch (HB).  In our previous work, we have found that the presence of blue HB stars in a cluster does not significantly affect the [Fe/H] solution because blue HB stars are so hot that they contribute a very small fraction to the IL Fe I EWs.  However, in \cite{m31paper}, we discussed how blue HB stars \textit{can} have a significant effect on the $B-V$ of a cluster, which can be important here because we take the $B-V$ into consideration during our tests for sampling uncertainties.  
 We note that, along with very blue HB stars, blue stragglers are not included in the isochrones used in our analysis.  While blue stragglers can also affect the integrated $B-V$ to some degree, it must be kept in mind that they are much less luminous ($\sim$2 magnitudes in V) than blue HB stars, even though they also have very high temperatures.  They are also less numerous than blue HB stars \citep{2004ApJ...604L.109P}.  We estimate that blue HB stars can account for at most 5-15 \% of the total flux, and blue stragglers can only account for  at most 2 \% of the flux. 
  This means that blue stragglers do not have as significant an effect on the integrated color as blue HB stars.  It also means that any conclusions we reach regarding blue HB stars can be extended to blue stragglers as well, to a lesser degree.  Finally, it is important to keep in mind that the hot temperatures of blue stragglers also mean that they contribute very little to the Fe I EWs.  Thus our analysis is not very sensitive to omitting blue stragglers,  in contrast to low resolution line index strategies that utilize Balmer line indexes. 
  In the rest of this section we discuss how the blue HB in particular can affect our sampling uncertainty analysis and keep in mind that our conclusions are  generalizable to  less luminous blue straggler stars as well.

NGC 2005 is less luminous than NGC 2019 by $\sim$0.5 V mag, and  we calculate that we have observed 60\% of the total flux.    Although NGC 2005 is better sampled than NGC 2019,  the 15 Gyr  synthetic CMDs have a comparable number of stars  ($\sim$40,000) to those of NGC 2019 because it is less luminous overall.  Thus, it is  possible that incomplete sampling could be affecting our analysis.

First we calculate abundances for CMD realizations created from isochrones with  [Fe/H]$=-2.1$ to $-1.5$ and ages of 5 to 15 Gyr.     The range in [Fe/H] of the CMD realizations  is small, and the self-consistent solutions have  [Fe/H]$\sim-1.6$. 
We next isolate a subset of CMD realizations with a  \bvo~ consistent with the observed  values of $B-V$=0.73 and $E(B-V)$=0.1 of \citet{2008MNRAS.385.1535P}, and  \citet{1998MNRAS.300..665O}, respectively.  The resulting \bvo~ range allowed for the CMD realizations is shown by the shaded gray area  in Figure~\ref{fig:2005colors}. Inspection of Figure~\ref{fig:2005colors} and Figure~\ref{fig:2019colors} reveals that although we derive comparable [Fe/H] for both NGC 2005 and NGC 2019,  the reddening corrected $B-V$ for NGC 2005 is bluer than that of NGC 2019, which means that the  10 to 15 Gyr  average CMD solutions are  inconsistent with the observed, reddening corrected $B-V$.   There are three possible explanations for this color mismatch: the $E(B-V)$ is not accurate, NGC 2005 has an age of  5 to 7 Gyr, or blue HB stars are affecting the observed color. As mentioned above, we know from the HST CMD of  \citet{1998MNRAS.300..665O} that NGC 2005 has a blue HB and that blue HB stars could change the \bvo~ of the synthetic CMDs by $\sim0.1$ dex \citep{m31paper}, so this seems like it may be the right explanation.
To confirm this hypothesis,  we can compare the diagnostics for the self-consistent solutions over the entire age and color range.  We find that the best solutions with ages of 5 and 7 Gyr offer  only a marginal improvement in the diagnostics over the average CMD solutions at these ages, which implies that a younger age for NGC 2005 is not preferred.  For the older ages, we find that the best-fitting CMD realizations have the same \bvo$\sim0.67$ as the average CMD solutions, and  result in only small improvements in the diagnostics.  This confirms that the solution for NGC 2005 was well determined using our original technique, despite the presence of blue HB stars,  and that the age range and \bvo~ are a good match.   Moreover, by allowing the \bvo~ of the CMD realizations to include redder colors, we have also demonstrated that significant flux in cool AGB stars is not needed to improve the solution for NGC 2005.

\begin{figure*}
\centering
\includegraphics[angle=90,scale=0.2]{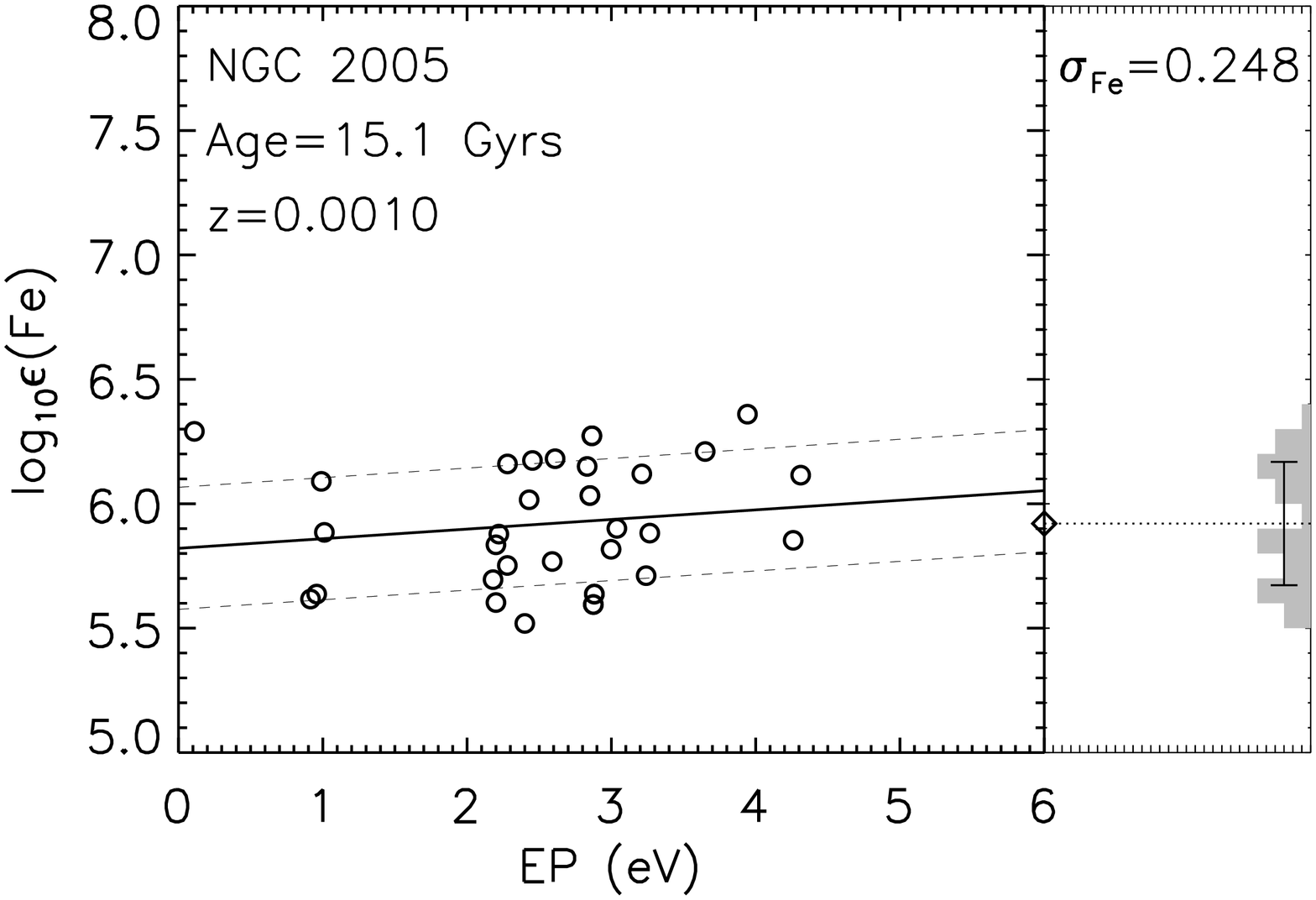}
\includegraphics[angle=90,scale=0.2]{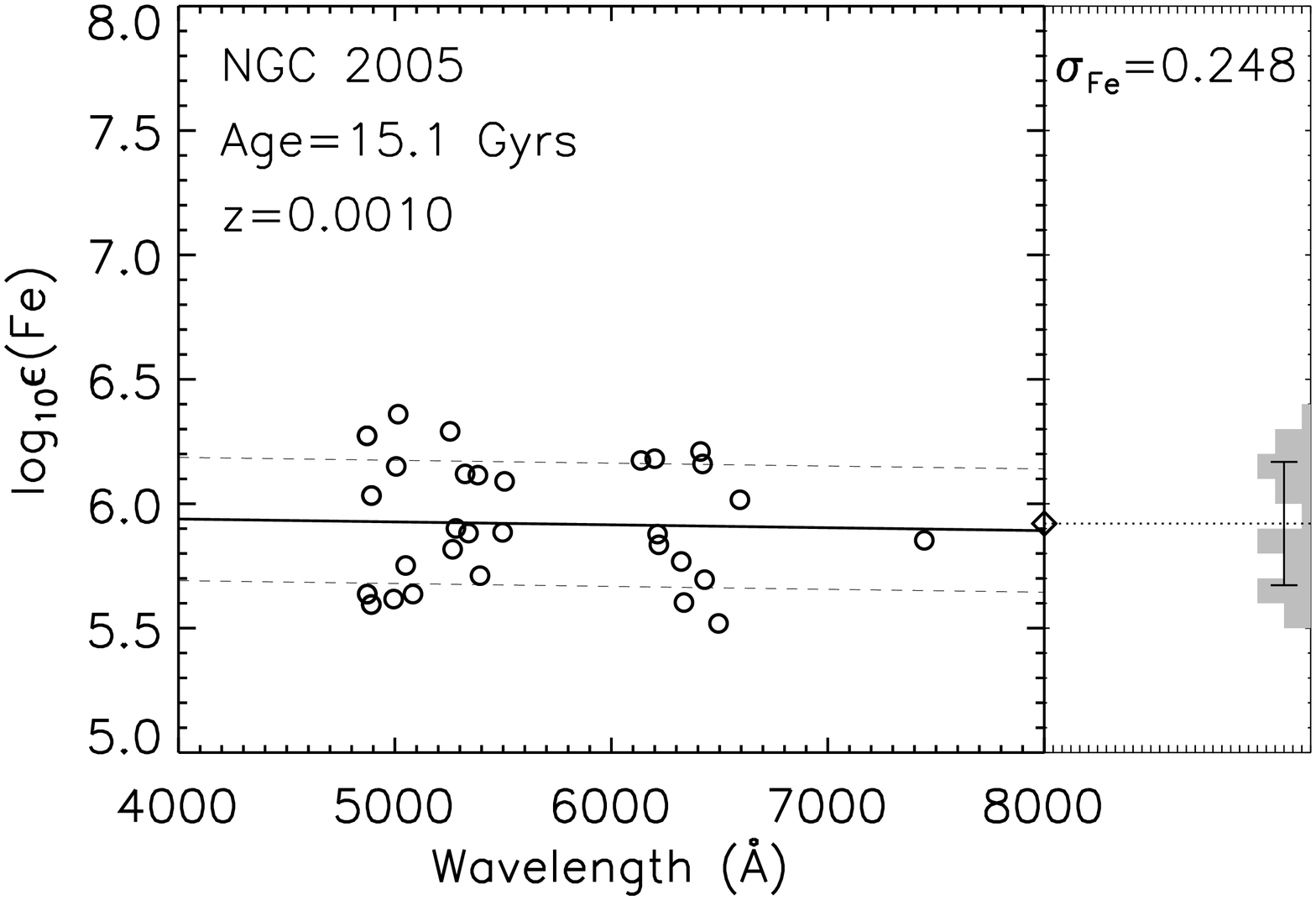}
\includegraphics[angle=90,scale=0.2]{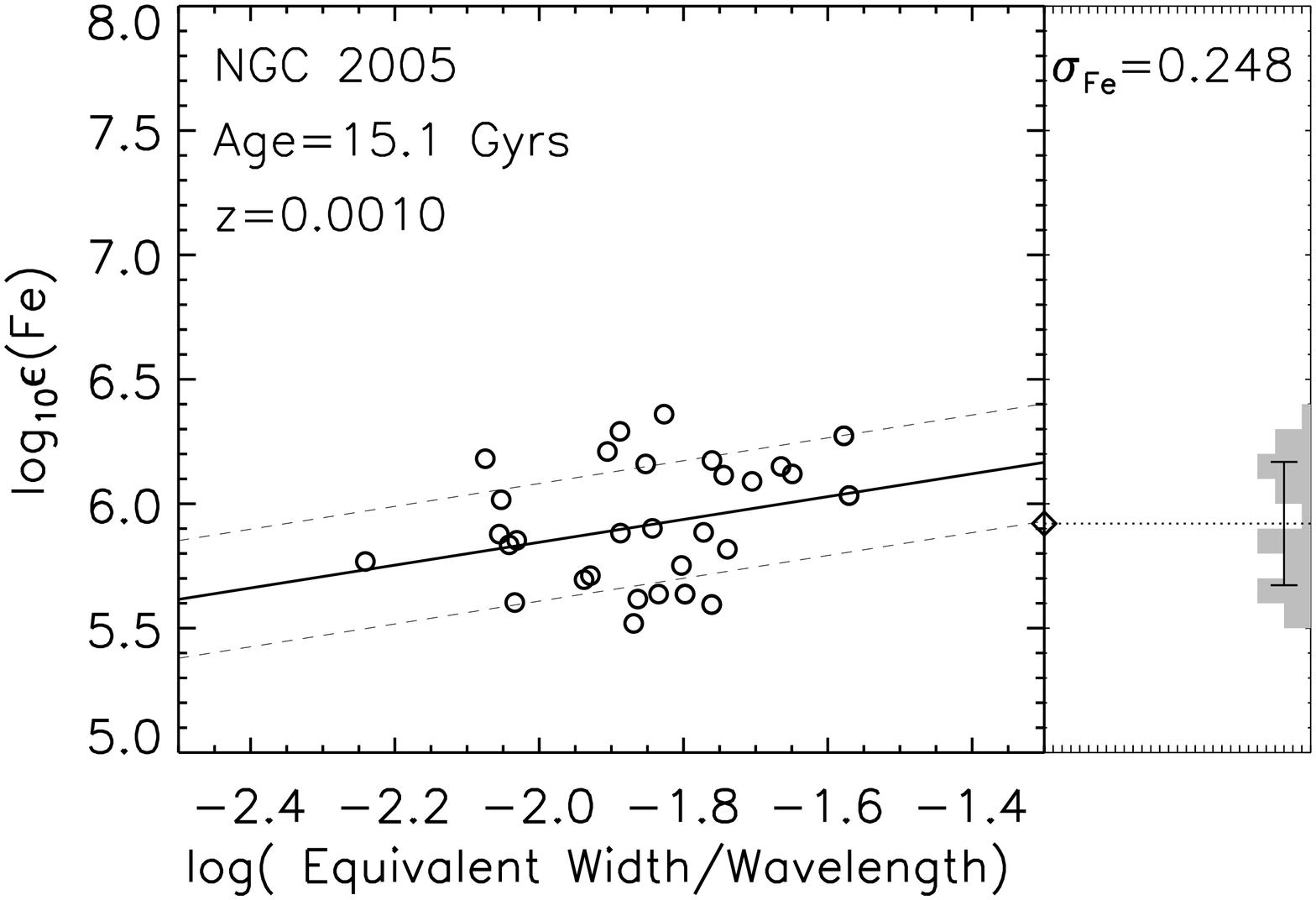}
\includegraphics[angle=90,scale=0.2]{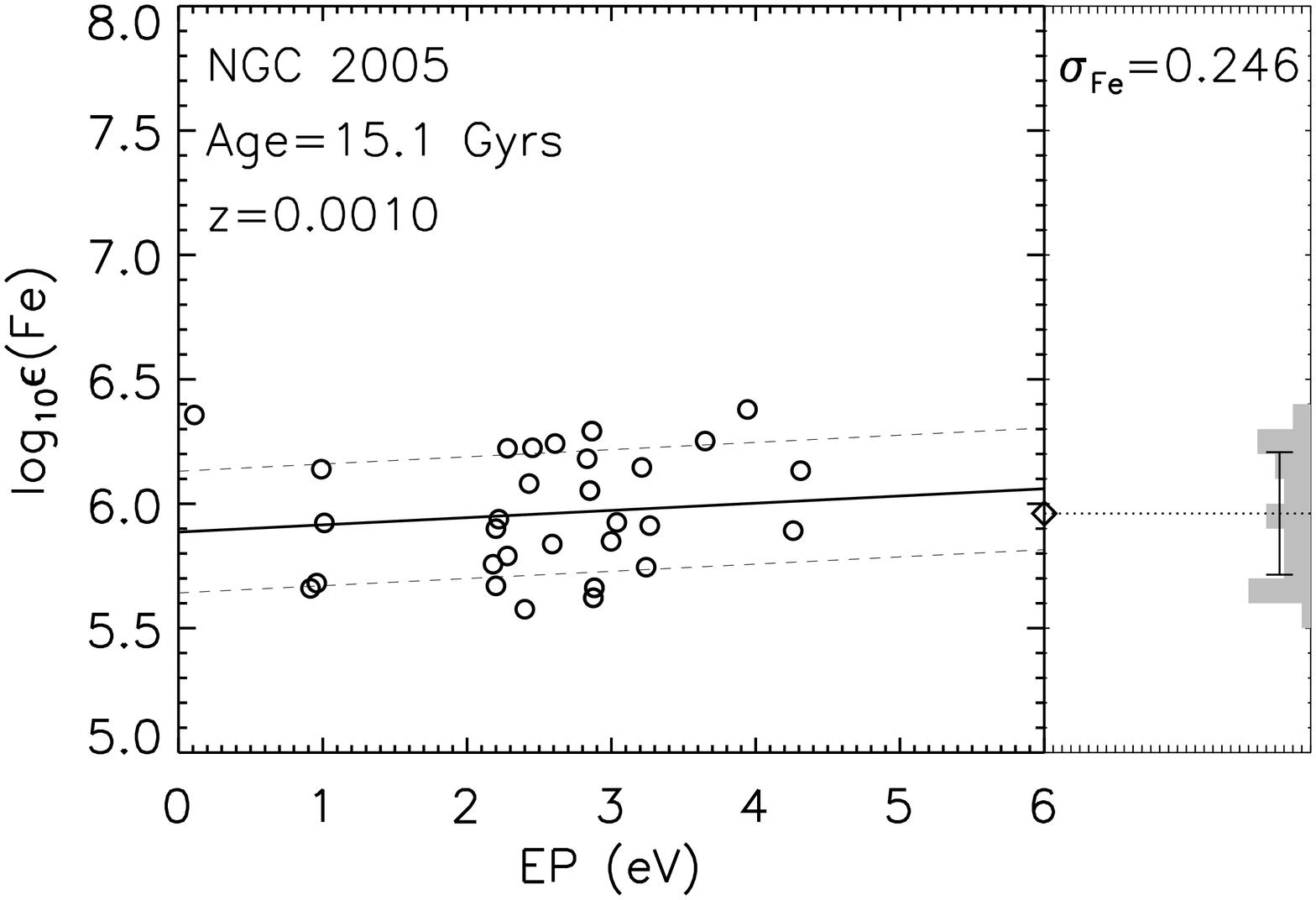}
\includegraphics[angle=90,scale=0.2]{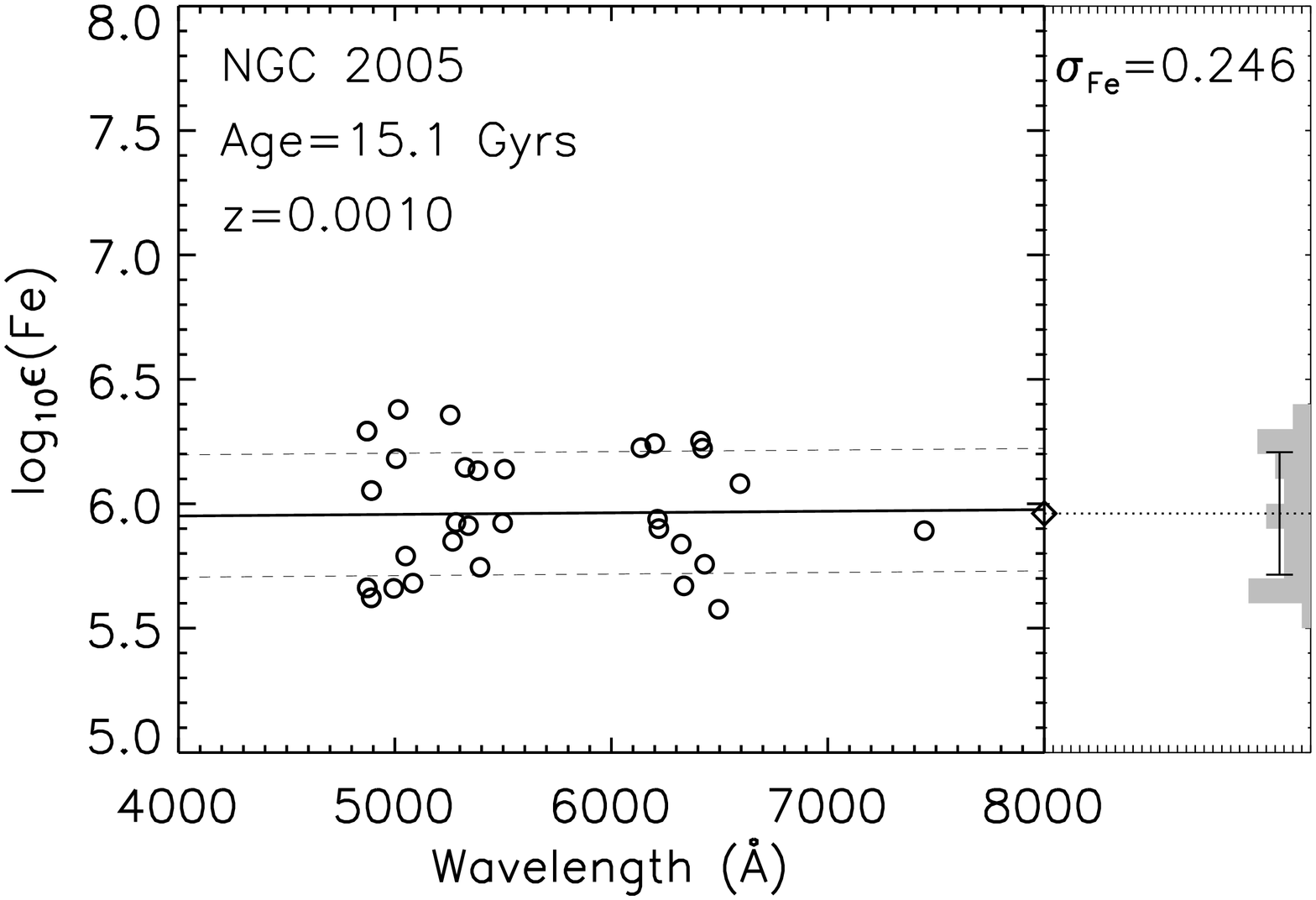}
\includegraphics[angle=90,scale=0.2]{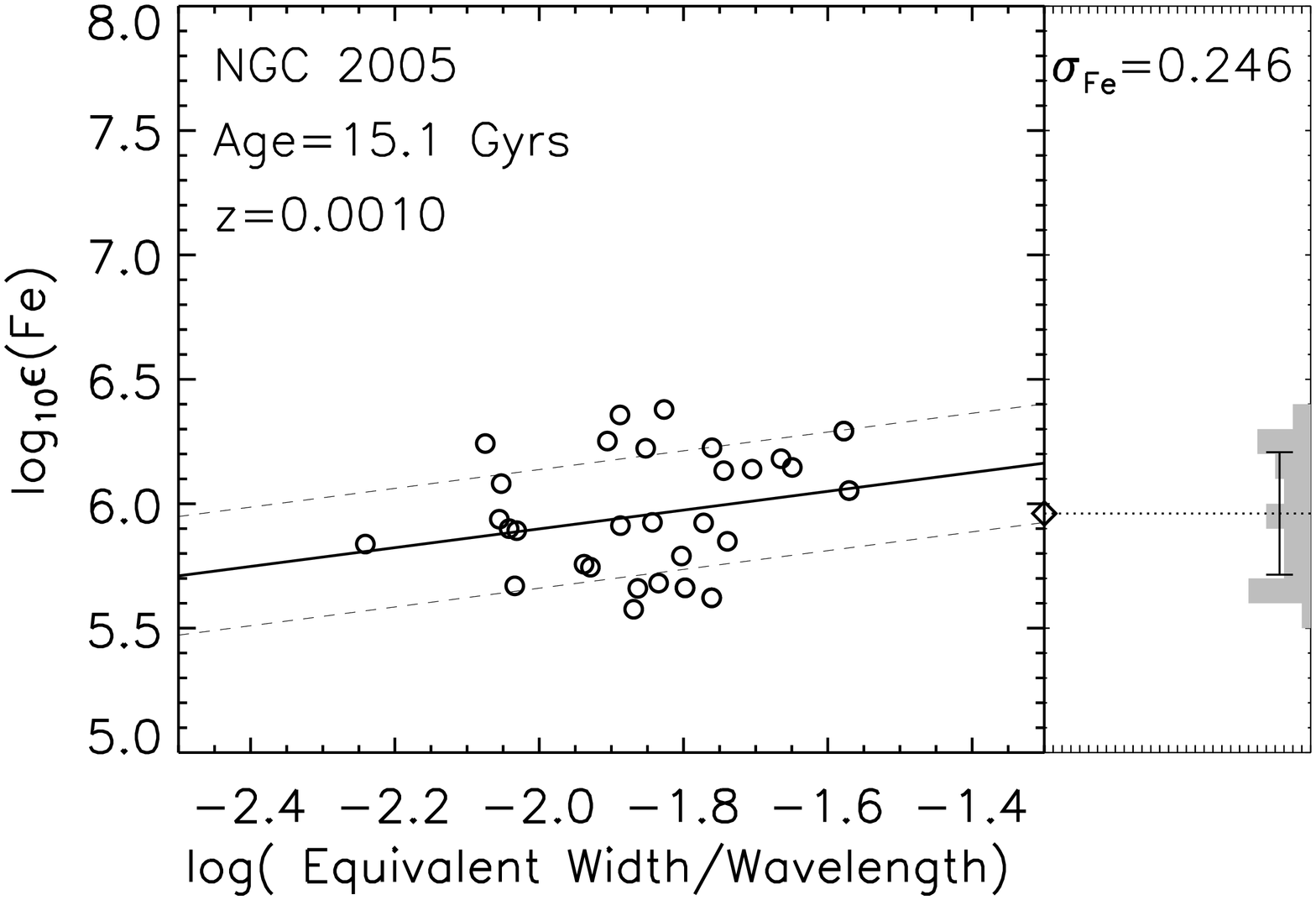}
\caption[Same as Figure~\ref{fig:2019 diagnostics} for NGC 2005]{Same as Figure~\ref{fig:2019 diagnostics} for NGC 2005.  The top panels correspond to the solution for the self-consistent average CMD solution for a 15 Gyr, [Fe/H]$=-1.5$ isochrone, and the bottom panels correspond to the solution for the  best-fitting CMD realization from a 15 Gyr, [Fe/H]=$-1.5$ isochrone.  The solution shown in the bottom panels has a smaller $\sigma_{\rm{Fe}}$, and reduced  [Fe/H] dependence on wavelength, EP,  and reduced EW than the original solution. }
\label{fig:2005 diagnostics} 
\end{figure*}

In conclusion, the final best-fitting CMD we determine  for NGC 2005 has an age of  15 Gyr, and [Fe/H]$=-1.6$, although in general we find that we can only constrain the age of NGC 2005 to a 5 to 15 Gyr range.  We show the individual Fe line diagnostics for the best-fitting 15 Gyr CMD realization in the bottom panels of Figure~\ref{fig:2005 diagnostics}, so that they can be visually compared to the 15 Gyr average CMD solutions in the top panels.  When compared to the average CMD solution, we find that the best-fitting CMD realization has a slightly reduced $\sigma_{\rm{Fe}}$,  and the dependence of [Fe/H] on wavelength has disappeared.  The dependence of [Fe/H] on EP and reduced EW has decreased, but still remains.  It is impossible to tell whether the remaining dependency of [Fe/H] on EP and reduced EW is  due to incorrectly modeling the HB, blue stragglers,  a more insidious sampling incompleteness, interloping stars, or simply a result of the lower S/N of the data.

It is apparent from the analysis of NGC 2005 that it is important to be aware of the impact of blue HB stars on the integrated $B-V$ of a cluster when comparing it to the \bvo~of the synthetic CMDs. However, while blue HB stars may be important for matching the $B-V$, cooler and redder stars continue to be more important for deriving self-consistent, stable [Fe/H] solutions.  Moreover, our tests for sampling uncertainties are adequate to determine whether significant flux from cool AGB stars is necessary for a stable [Fe/H] solution.   Despite the issues discussed for NGC 2005, we find that the [Fe/H] is well determined and is within $\sim0.1$ dex of [Fe/H]=$-1.6$ over the 5$-$15 age range found for the best-fitting CMD solutions.

\subsubsection{NGC 1916}
\label{sec:1916}

\begin{figure}
\centering
\includegraphics[scale=0.5]{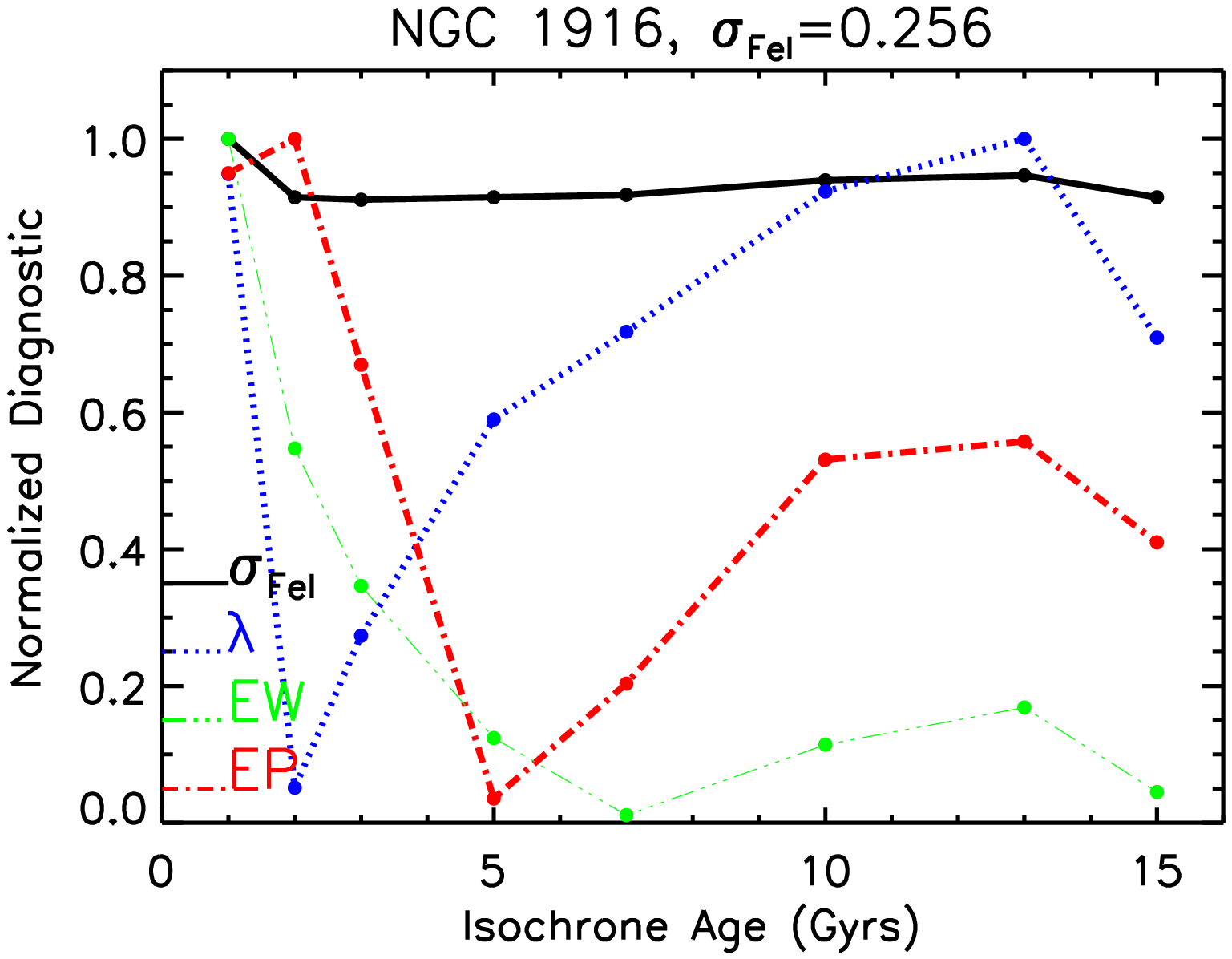}
\caption[Same as Figure~\ref{fig:2019 diag} for NGC 1916.]{Same as Figure~\ref{fig:2019 diag} for NGC 1916. Best solutions are found for ages $>$5 Gyrs. }
\label{fig:new1916 diag} 
\end{figure}

\begin{figure}
\centering
 \includegraphics[angle=90,scale=0.35]{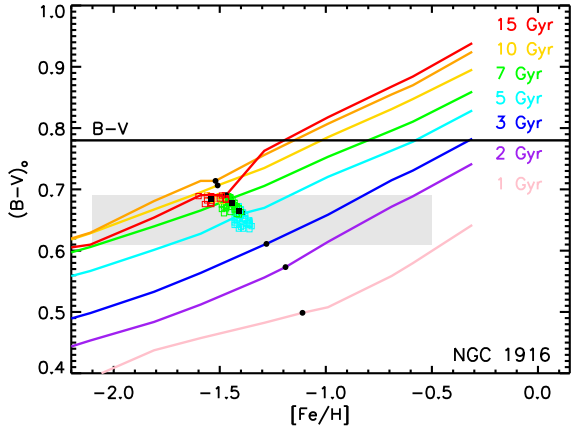}
\caption[ Same as Figure~\ref{fig:2019colors} for NGC 1916.]{ Same as Figure~\ref{fig:2019colors} for NGC 1916. }
\label{fig:1916colors} 
\end{figure}

NGC 1916 is the most massive old cluster in the LMC training set (\mv$=-8.96$), and we have observed a significant fraction ($\sim$60\%) of the total flux. For that reason, we do not expect that sampling uncertainties will have a large impact on  our solution.  Nevertheless, we follow the same analysis procedure as for NGC 2019 and NGC 2005 for clarity and to demonstrate the impact of sampling uncertainties on an old, well-sampled cluster. 

In Figure~\ref{fig:new1916 diag} we show the normalized Fe line diagnostics for the self-consistent average CMD solutions.  Like  NGC 2019 and NGC 2005, we find that the diagnostics for NGC 1916 are generally best for ages between 5 and 15 Gyrs.  The Fe I lines result in solutions of [Fe/H]$\sim-1.5$, with a best $\sigma_{\rm{Fe}}$=0.259, which is  comparable to the result for NGC 2005, but higher than that for NGC 2019.  The 15 Gyr solution shows a fairly significant dependence of [Fe/H] with wavelength, a small dependence of [Fe/H] with reduced EW, and negligible dependence of [Fe/H] with EP.

\begin{figure*}
\centering
\includegraphics[angle=90,scale=0.2]{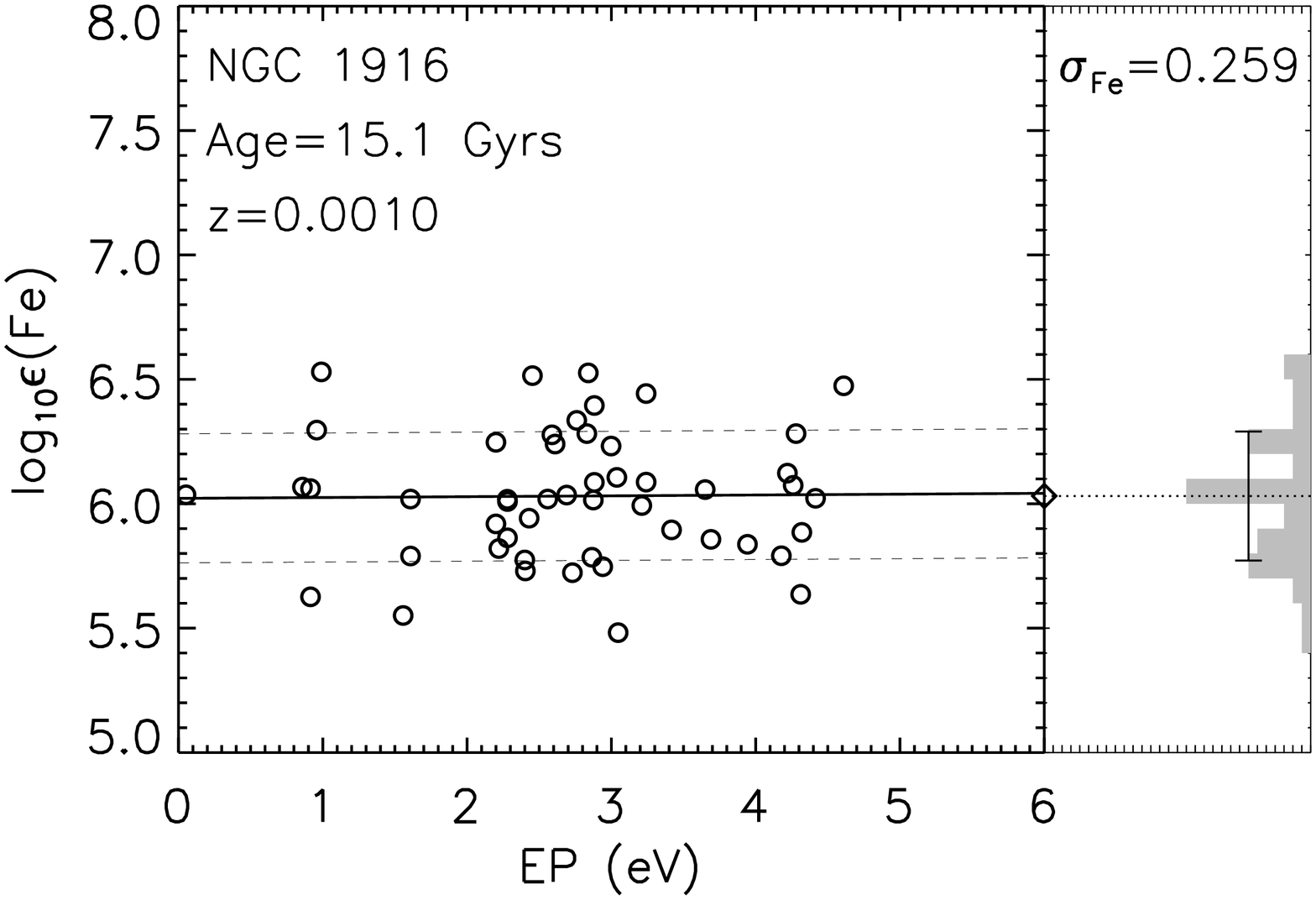}
\includegraphics[angle=90,scale=0.2]{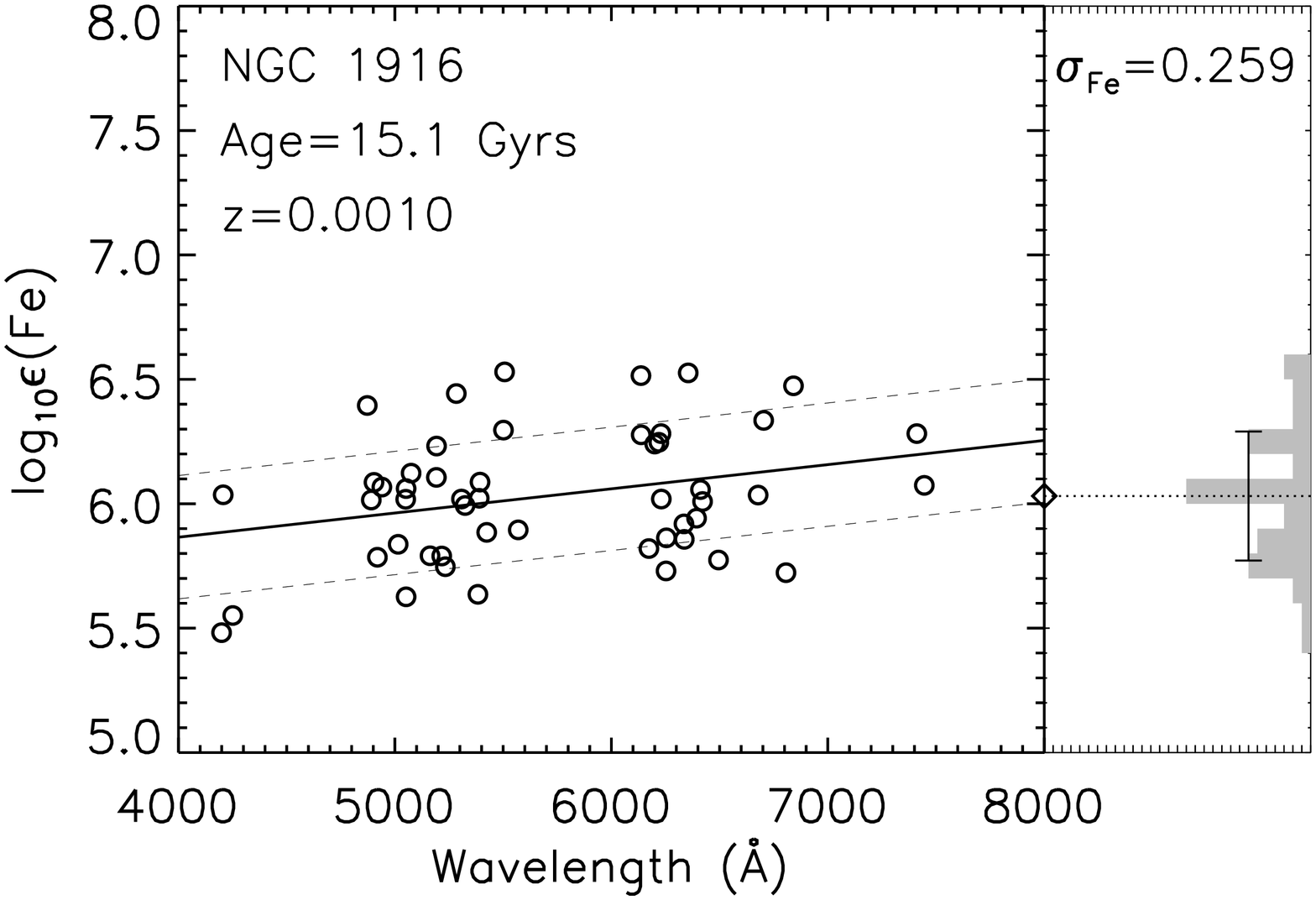}
\includegraphics[angle=90,scale=0.2]{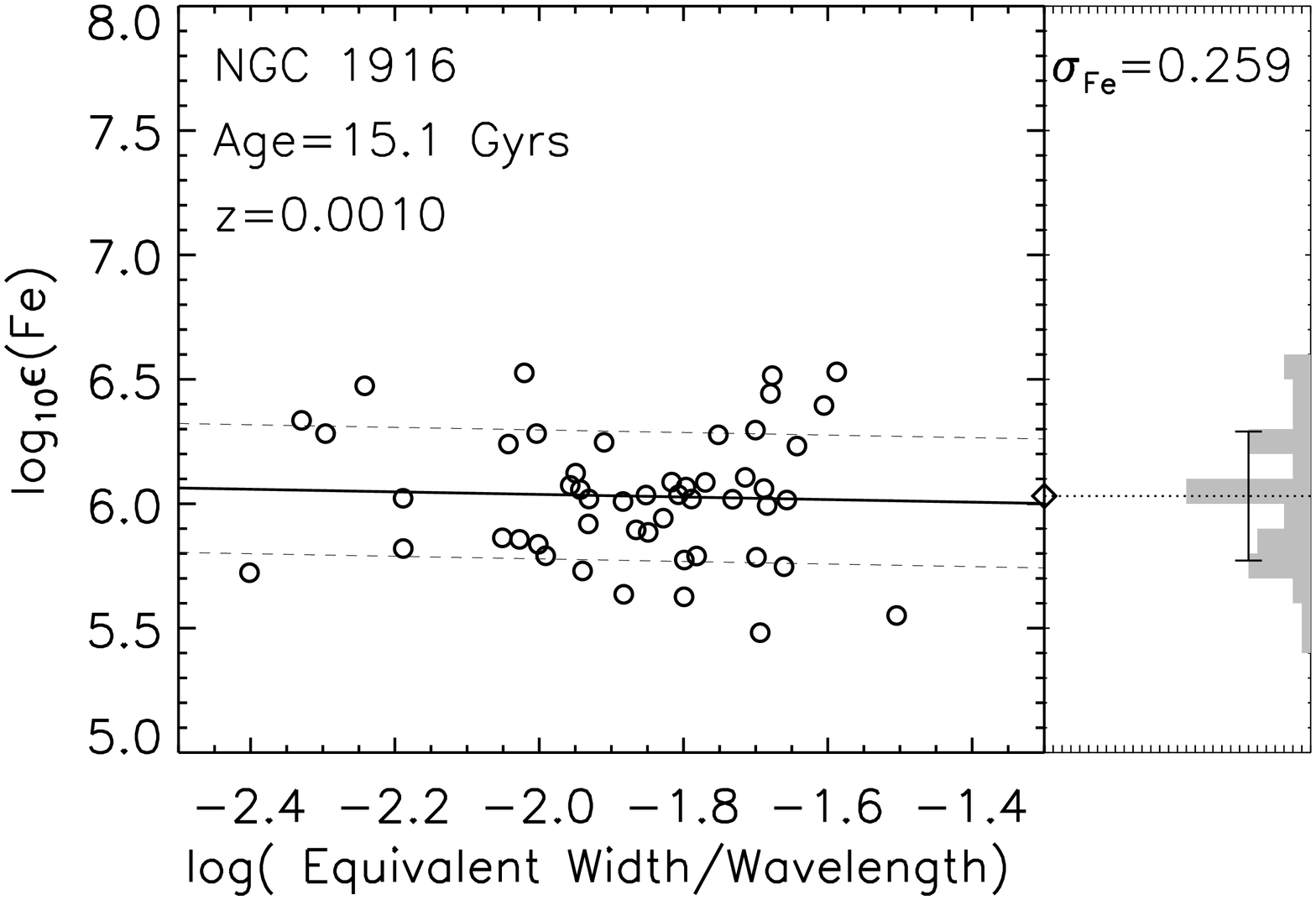}
\includegraphics[angle=90,scale=0.2]{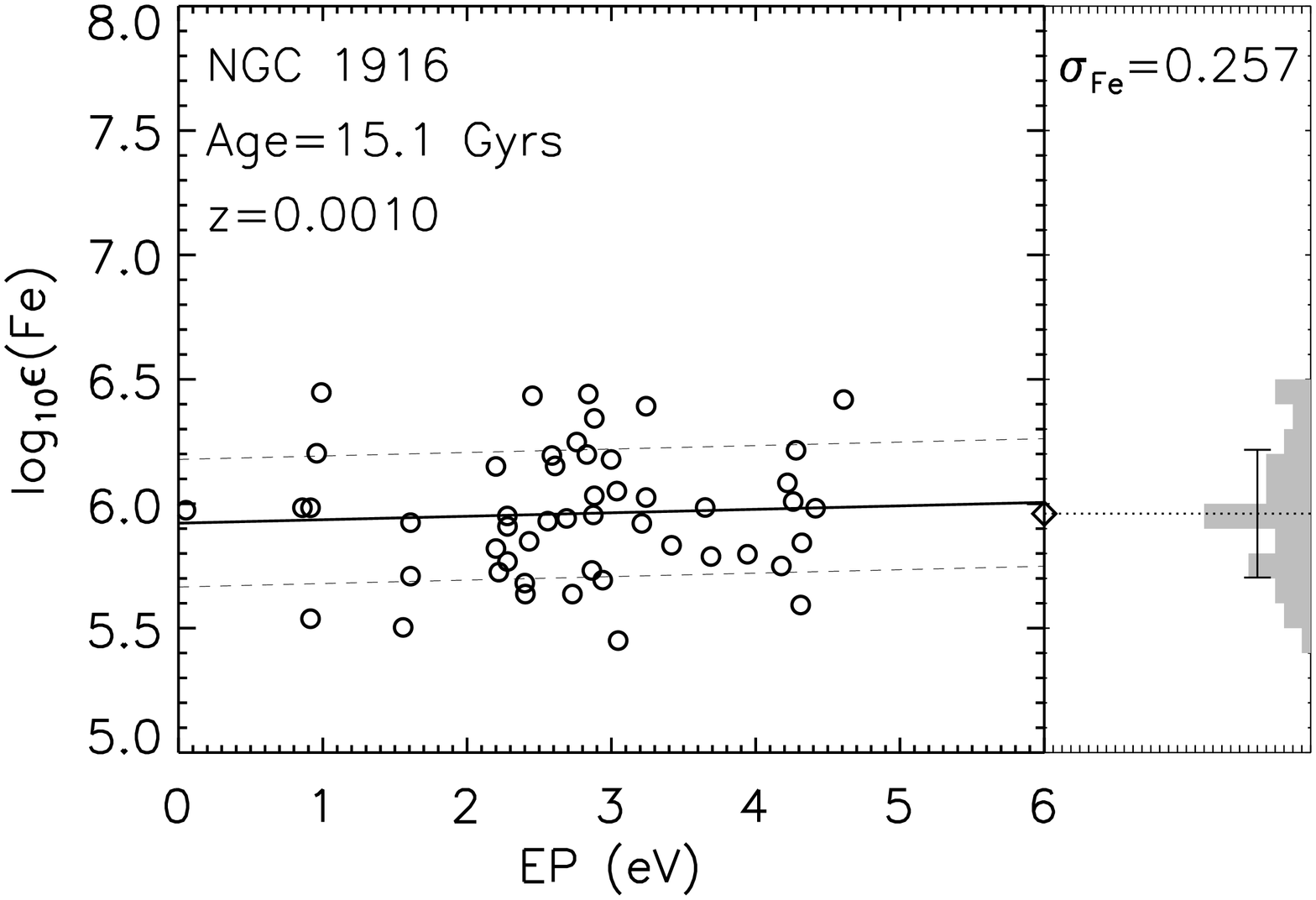}
\includegraphics[angle=90,scale=0.2]{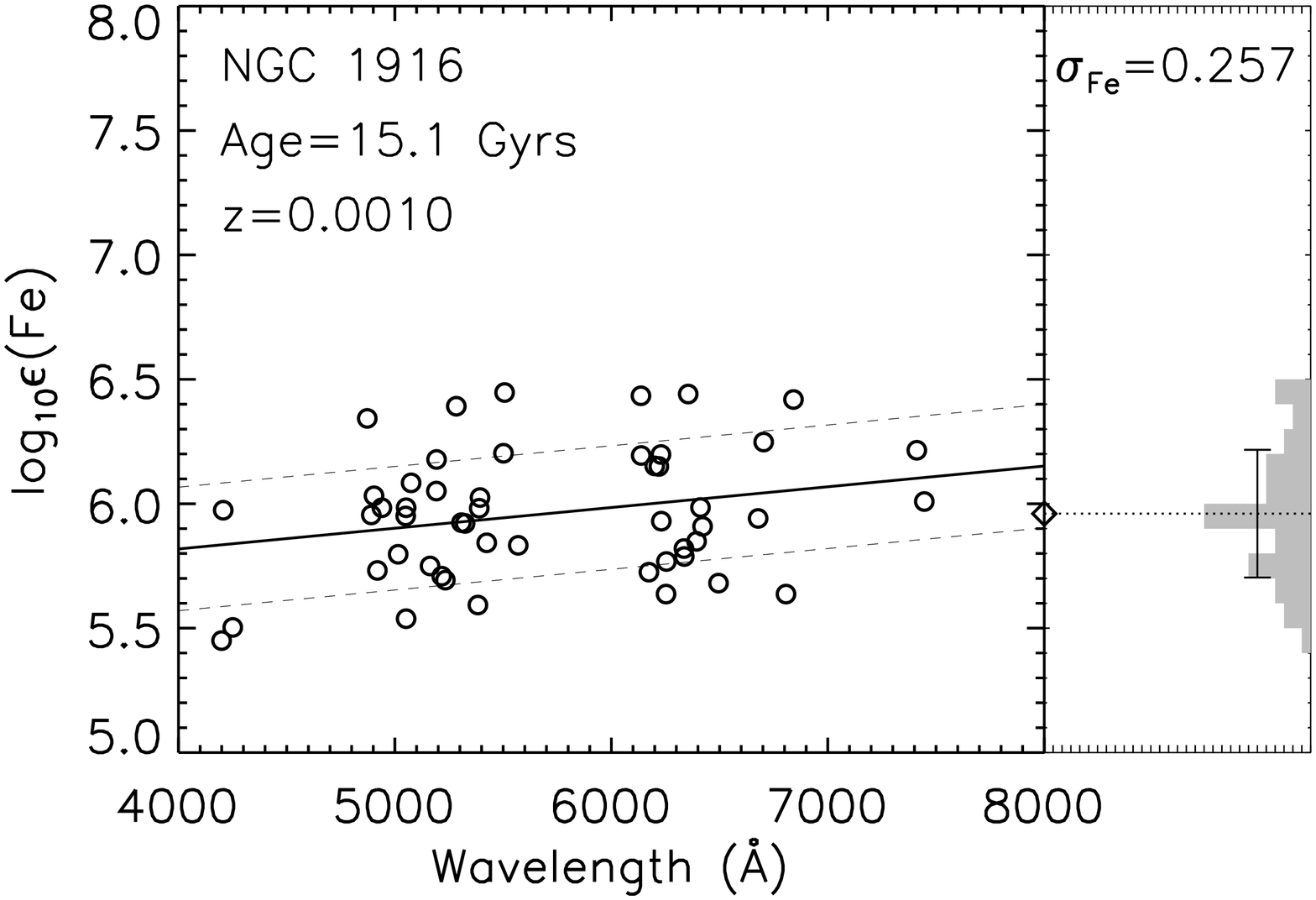}
\includegraphics[angle=90,scale=0.2]{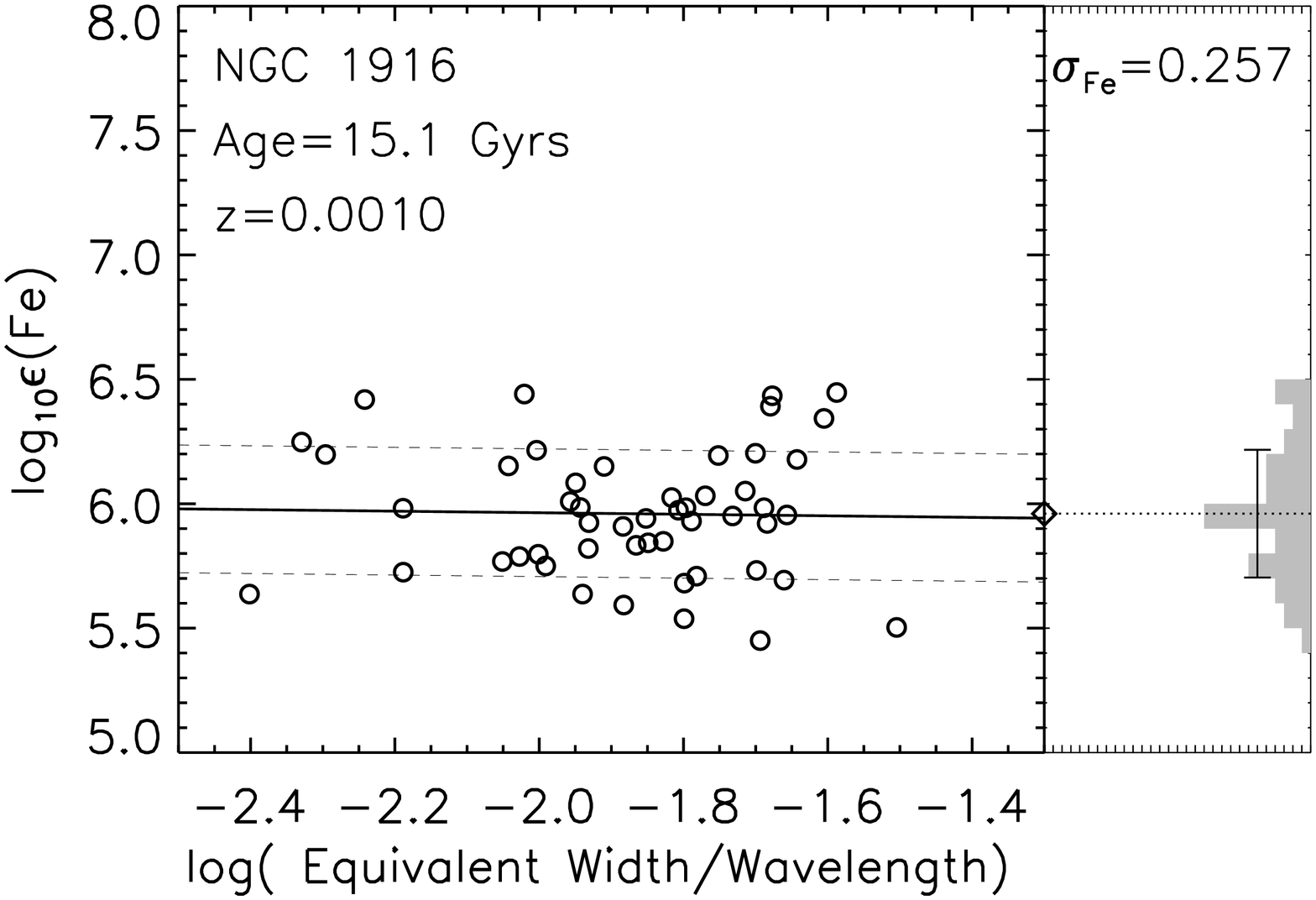}

\caption[Same as Figure~\ref{fig:2019 diagnostics} for NGC 1916]{Same as Figure~\ref{fig:2019 diagnostics} for NGC 1916.  The top panels correspond to the solution for the self-consistent average CMD solution for a 15 Gyr, [Fe/H]$=-1.5$ isochrone, and the bottom panels correspond to the solution for the  best-fitting CMD realization from a 15 Gyr, [Fe/H]=$-1.5$ isochrone. The solution in the bottom panels has a smaller $\sigma_{\rm{Fe}}$ than the original solution.}
\label{fig:1916 diagnostics} 
\end{figure*}

As for NGC 2005, \citet{1998MNRAS.300..665O} observed   a significant number of blue HB stars in NGC 1916,
although their analysis of the HST CMD was hampered by differential reddening, so a precise age, [Fe/H], and $E(B-V)$ were not determined.    We are in a unique position to  determine an accurate [Fe/H] and age for NGC 1916 because we create synthetic CMDs to represent the cluster populations and are therefore not sensitive to differential reddening effects.  
We  find that the  reddening corrected $B-V$ color  \citep{2008MNRAS.385.1535P} is inconsistent with the color of the 10 and 13 Gyr average CMD solutions, as shown in 
Figure~\ref{fig:1916colors}.  
Because of the uncertainty in the $E(B-V)$ and the effect of blue HB stars, we test CMD realizations with a  wide range in age and \bvo~ as we did for NGC 2005.  
 CMD realizations with self-consistent [Fe/H]  and marginal improvement  in the diagnostics are found  for  ages of 5, 7 and 15 Gyrs.   The most stable solutions have the same [Fe/H] as the average CMD solutions, and nearly identical  \bvo.  This means that the stellar population for NGC 1916 was already well constrained using our original technique.
 In Figure \ref{fig:1916 diagnostics} we show the original 15 Gyr average CMD solution, as well as the best-fitting solution from the 15 Gyr CMD realizations.  The latter shows a slightly smaller $\sigma_{\rm{Fe}}$, and slightly smaller dependence of [Fe/H] with wavelength and reduced EW than the original solution.

In summary, we find the best-fitting solutions for NGC 1916 have ages of  5 to 15 Gyrs, and a well-constrained  [Fe/H]$\sim-1.5$.  The fact that we cannot substantially improve the diagnostics by allowing for statistical fluctuations and a large range in \bvo~ suggests that any mismatch between the observed $B-V$  and  the synthetic CMD \bvo~  could simply be  due to a deficit of blue HB stars in the isochrones.   In any case,  both the preferred CMD population and the metallicity of NGC 1916 are well determined from our analysis. This  is an important demonstration of the utility of our method for determining the properties of GCs affected by differential reddening, which are otherwise difficult to study using other means.

\subsection{Results: Intermediate Age Clusters}
\label{sec:int}
We have tested our analysis method on two LMC clusters with ages of $\sim$1-2 Gyr.  Even without a priori knowledge of the cluster ages, it is immediately apparent using our original technique that these clusters are better matched by CMDs with ages $<$5 Gyrs.  As soon as this is apparent, we use a finer grid of ages for our synthetic CMDs 
between 0.5 Gyr and 5 Gyr because significant changes in the CMD stellar populations can occur on much shorter timescales (0.5-1 Gyr) than they do  for old, $>$5 Gyr CMDs.  For the initial synthetic CMD grid, we  use isochrones with ages of  0.5, 0.7, 1.0, 1.25, 1.5, 1.75, 2.0, 2.5, 3.0, 3.5, 5.0, 7.0, and 10 Gyr.
We discuss the determination of the best-fit CMDs below.

It is especially important to evaluate the effect  that sampling uncertainties can have on the age and abundance solutions for the intermediate age clusters.   This is because  they are both $\sim$2 Gyr in age, and therefore rapidly evolving,  and  because they are  the least luminous,  least massive,  clusters in both the LMC and MW training sets, and therefore the  most likely to suffer from statistical fluctuations in a small number of luminous stars.  At a fixed total luminosity a 1 Gyr cluster has fewer, but more massive, more luminous, stars than a 10 Gyr cluster \citep[e.g., see][]{1999A&AS..136...65B}.

\subsubsection{NGC 1718}
\label{sec:1718}

\begin{figure}
\centering
\includegraphics[angle=90,scale=0.35]{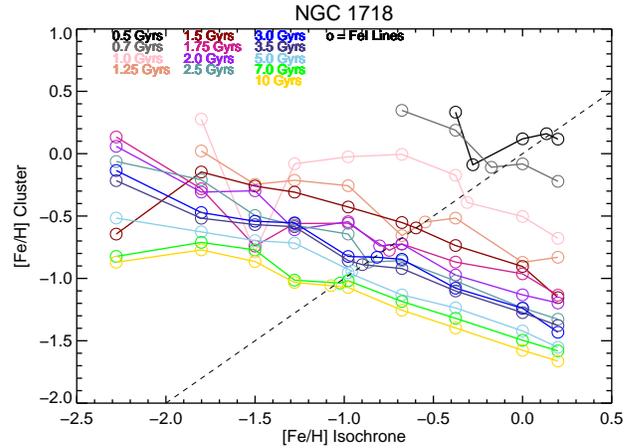}
\caption{Same as Figure~\ref{fig:2019 Fe plot} for NGC 1718. Additional ages for the grid of synthetic CMDs for intermediate age clusters are shown by the labeled colors.  }
\label{fig:Fe plots1718} 
\end{figure}

\begin{figure}
\centering
\includegraphics[scale=0.5]{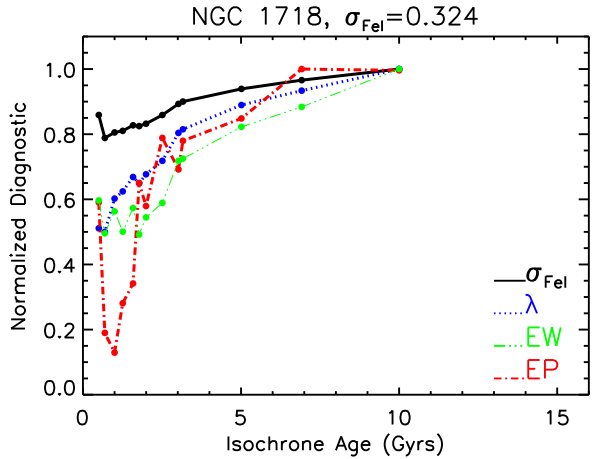}
\caption[Same as Figure~\ref{fig:2019 diag} for NGC 1718]{Same as Figure~\ref{fig:2019 diag} for NGC 1718.  The best solutions are found for ages of 0.7$-$1.25 Gyrs. }
\label{fig:CMD trends1718} 
\end{figure}

NGC 1718 is the least massive cluster in the training set (\mv$\sim-6.5$), and is estimated to have  an age of $\sim$2 Gyr \citep{2007A&A...462..139K}.   The mean [Fe/H] solutions we derive for the grid of  synthetic CMDs  are shown in Figure~\ref{fig:Fe plots1718}, and the diagnostics for the  self-consistent CMDs for each age are shown in Figure~\ref{fig:CMD trends1718}.  The best $\sigma_{\rm{Fe}}$=0.324, which is significantly larger than the $\sigma_{\rm{Fe}}$ we obtain for any of the old clusters in the sample.  However, even though the $\sigma_{\rm{Fe}}$ is large, it can clearly be seen in Figure~\ref{fig:CMD trends1718} that all of the diagnostics are consistently better for the younger ages than they are for the older ages.  This is the clearest demonstration to date  that  our abundance analysis method and  the Fe I lines alone can be used to constrain cluster age.

 We next investigate the limits we can place on the age of the cluster using this method, and whether we can successfully identify an age that is consistent with the age determined from resolved star photometry.    Figure~\ref{fig:CMD trends1718} shows that the minimum in $\sigma_{\rm{Fe}}$ occurs at 0.7 to 1.25 Gyrs, and that there is a corresponding minimum in the Fe I slope with EP and some indication of a minimum in  Fe I slope with EW at these ages.  However,  while there is some indication of a preferred solution in the appropriate age range around 1 Gyr,
the full range from 0.5 to 1.25 Gyr gives acceptable solutions.
The  1 Gyr  solution, with   [Fe/H]$\sim-0.4$,  has a small dependence of [Fe/H] with EP, but a  significant dependence of [Fe/H] with both wavelength and reduced EW. 
 This suggests that this  CMD is a poor  match to the true CMD of the cluster.  Again, we next explore how incomplete sampling may play a role.

\begin{figure}
\centering
\includegraphics[angle=90,scale=0.35]{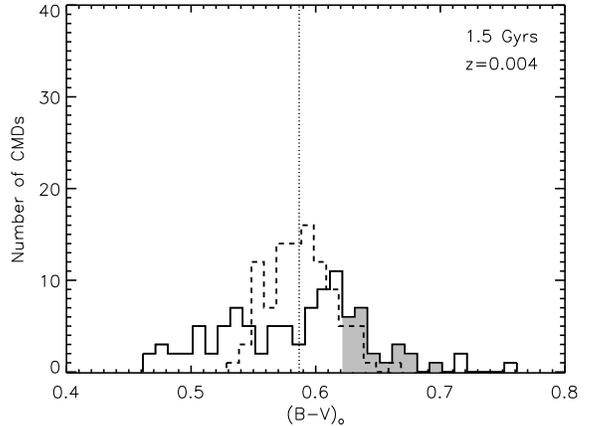}
\caption[Histogram of integrated \bvo~ color for 100 CMD realizations of a 1.5 Gyr, z=0.004  (Fe/H$=-0.66$) isochrone]{Histogram of integrated \bvo~ color for 100 CMD realizations of a 1.5 Gyr, z=0.004 ([Fe/H]=$-0.66$) isochrone.  Solid black line shows the histogram for a population where the total flux in stars has been normalized to 23\% of a \mv$=-6.5$  cluster, which is appropriate for our integrated light spectrum of NGC 1718.  Dashed black line shows the histogram for a population normalized to 100\% of a \mv$=-6.5$ cluster.  CMDs with \bvo~ color consistent with the observed, reddening-corrected $B-V$ of NGC 1718 are shaded in gray. The large spread of \bvo~ in the solid line histogram is an indication that our observations of NGC 1718 are significantly affected by sampling incompleteness.}
\label{fig:histo1718} 
\end{figure}

  In Figure~\ref{fig:histo1718} we show the histogram of \bvo~ for 100 CMD realizations of a \mv=$-6.5$ cluster, which corresponds to the total flux of NGC 1718 \citep{2007A&A...462..139K}.   Also shown is  the histogram for CMD realizations for 23\% of a \mv=$-6.5$ cluster, which corresponds to the fraction of NGC 1718 we have observed. The larger spread in \bvo~ of the CMD realizations in Figure~\ref{fig:histo1718} when compared to that for NGC 2019 in Figure~\ref{fig:2019 histo} demonstrates that statistical stellar population fluctuations  are a greater issue for NGC 1718.  It is also clear from the comparison of the two histograms for NGC 1718 in  Figure~\ref{fig:histo1718} that the sampling uncertainties are significantly exacerbated by the incomplete sampling of our observations. It is particularly worrisome that the peak is hard to distinguish for the less luminous CMD realizations.

For  NGC 1718, the large spread in \bvo~  is due to a combination of its  significantly smaller  luminosity,  its young age, and  the incomplete sampling of the observations.  
We are especially interested in evaluating the effect of its young age, as this will be most relevant to future work on unresolved clusters.
It will be impossible to disentangle the uncertainties due to the age of NGC 1718 from the uncertainties due to our observations, but this means 
this is a worst case scenario for future work on unresolved clusters.

\begin{figure}
\centering
\includegraphics[angle=90,scale=0.35]{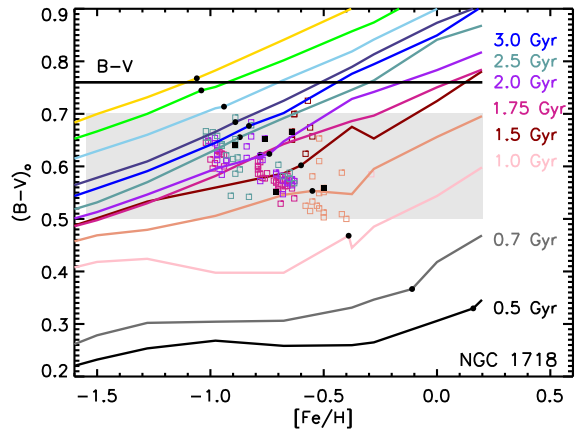}
\caption[Same as Figure~\ref{fig:2019colors} for NGC 1718]{Same as Figure~\ref{fig:2019colors} for NGC 1718.  Colors are the same as in Figure~\ref{fig:Fe plots1718}.  Shaded gray region corresponds to the range in \bvo~ allowed for the subset of CMD realizations used for abundance analysis.}
\label{fig:1718colors} 
\end{figure}

 We create CMD realizations for NGC 1718 with ages and [Fe/H] in the range 1 to 2.5 Gyr and $-0.98$ to $-0.26$, and include isochrones at the interpolated [Fe/H] solutions from the average CMDs. 
The CMDs result in a wide range of [Fe/H]  solutions, ranging from [Fe/H]$\sim-2$ to [Fe/H]$\sim0$.  This is a result of fairly drastic fluctuations in  luminous giant stars, which both cause the large scatter in \bvo~  seen in Figure~\ref{fig:histo1718} and have a large impact on the Fe I EWs.  Fortunately, this means that it is easier to use the Fe lines to identify CMD realizations that are viable solutions.
 For NGC 1718 we are able to narrow down the possible CMD realizations from $>$1000 to $\sim$120 by applying our usual requirement that   the solutions be self-consistent, and that they  have colors consistent with the cluster's observed properties.
  We include CMD realizations with 0.5$<$ \bvo $<$ 0.7 for abundance analysis, which is highlighted by the shaded gray region in Figure~\ref{fig:1718colors}.  The upper limit of this range  corresponds with the observed, reddening corrected $B-V$ of NGC 1718.  We extend the lower limit of the \bvo~ range to include more CMD realizations with ages of 1 to 1.5 Gyr, which have bluer colors.
We search a wider parameter space because of the significant trends in the original [Fe/H] solutions, and because  Figure~\ref{fig:CMD trends1718} suggests that the diagnostics at these ages are generally better.  We note that even with a lower limit of  \bvo=0.5, it is pretty clear from  Figure~\ref{fig:1718colors} that  CMDs with ages $<$1.0 Gyr can be ruled out  based on the observed $B-V$.

\begin{figure}
\centering
\includegraphics[scale=0.5]{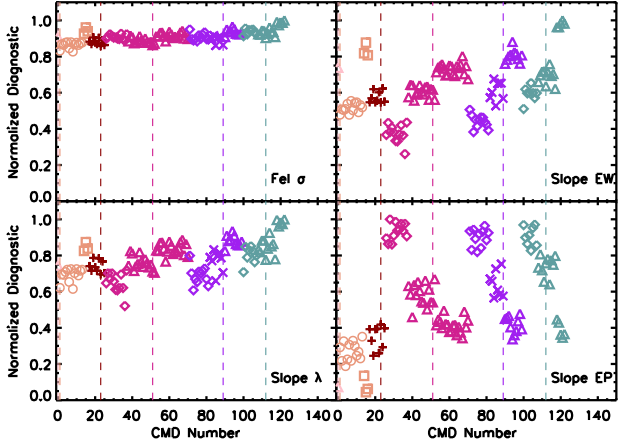}
\caption[Same as Figure~\ref{fig:2019newdiag} for NGC 1718]{Same as Figure~\ref{fig:2019newdiag} for NGC 1718.  Colors are the same as in Figure~\ref{fig:Fe plots1718}. Younger ages are overall better solutions. }
\label{fig:1718newdiag} 
\end{figure}

 In Figure~\ref{fig:1718newdiag} we show the Fe I diagnostics for the CMD realizations that have self-consistent [Fe/H] solutions, ordered first by increasing age and then by increasing [Fe/H] for each age. This figure shows that the $\sigma_{\rm{Fe}}$ tends to get larger with increasing age, as was found for the average CMDs in Figure~\ref{fig:CMD trends1718}, which implies that the younger ages are generally better solutions.  
   We pick the solution that best minimizes all of the diagnostics for each age, while no one solution ideally minimizes all diagnostics at once.  We find that the  best-fitting CMD realization at each age has the same [Fe/H] as the original average CMD solution, which implies that in this case incomplete sampling is not adding significant uncertainty to the best-fitting [Fe/H], despite the sampling concerns discussed above.

\begin{figure}
\centering
\includegraphics[scale=0.5]{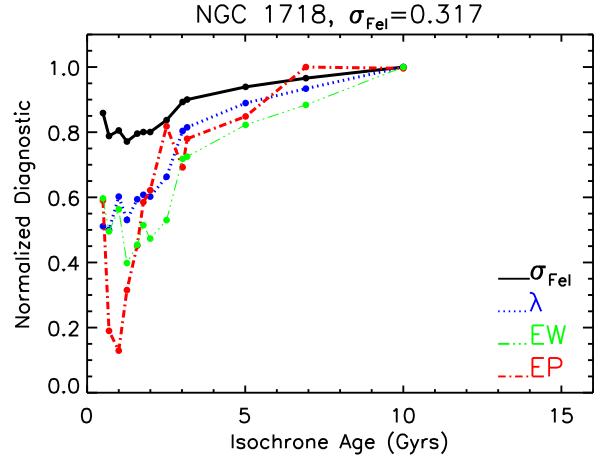}
\caption[Same as Figure~\ref{fig:CMD trends1718}, except the 1.25-2.5 Gyr solutions have been replaced by the solutions for the best-fitting CMD realization at these ages.]{Same as Figure~\ref{fig:CMD trends1718}, except the 1.25-2.5 Gyr solutions have been replaced by the solutions for the best-fitting CMD realization at these ages.  The best solutions are found for ages of 1$-$2.5 Gyr. }
\label{fig:CMD trends1718new} 
\end{figure}

After narrowing down the possible solutions from $\sim$120 self-consistent CMD realizations to  5 solutions, one for each age,    we compare these 5 solutions  to the 7 average CMD solutions for the other ages in the grid.  The diagnostics for these final 12 solutions are shown  in  Figure~\ref{fig:CMD trends1718new}.   We find that the case for an age of $\sim$1 Gyr is a little stronger than it was from Figure~\ref{fig:CMD trends1718}, due to an improvement in the $\sigma_{\rm{Fe}}$, and that the other diagnostics are more closely correlated with this improvement. However, while the 1 to 1.25 Gyr solutions are the most stable overall, the differences between  the solutions with ages between  1 to 2.5 Gyr are still small.  Therefore, our final age constraint for NGC 1718 is 1 to 2.5 Gyr.  This is a narrower range in age than we are generally able to constrain for old clusters, but because NGC 1718 is young, the stellar population changes considerably over this 2.5 Gyr timescale. 
Accordingly, the [Fe/H] solution is less constrained than for older clusters in the training set, although the solutions for the older clusters spanned a larger range in age.
Explicitly,  the 1.0 Gyr solution has [Fe/H]=$-0.39$, and the 2.5 Gyr solution has [Fe/H]=$-0.89$.  To convey this range of possible solutions in age and corresponding abundance, we average the two [Fe/H] results and quote an uncertainty due to the age solution, $\sigma_{\rm{age}}$,  of $\pm0.25$ dex.

\begin{figure*}
\centering
\includegraphics[angle=90,scale=0.2]{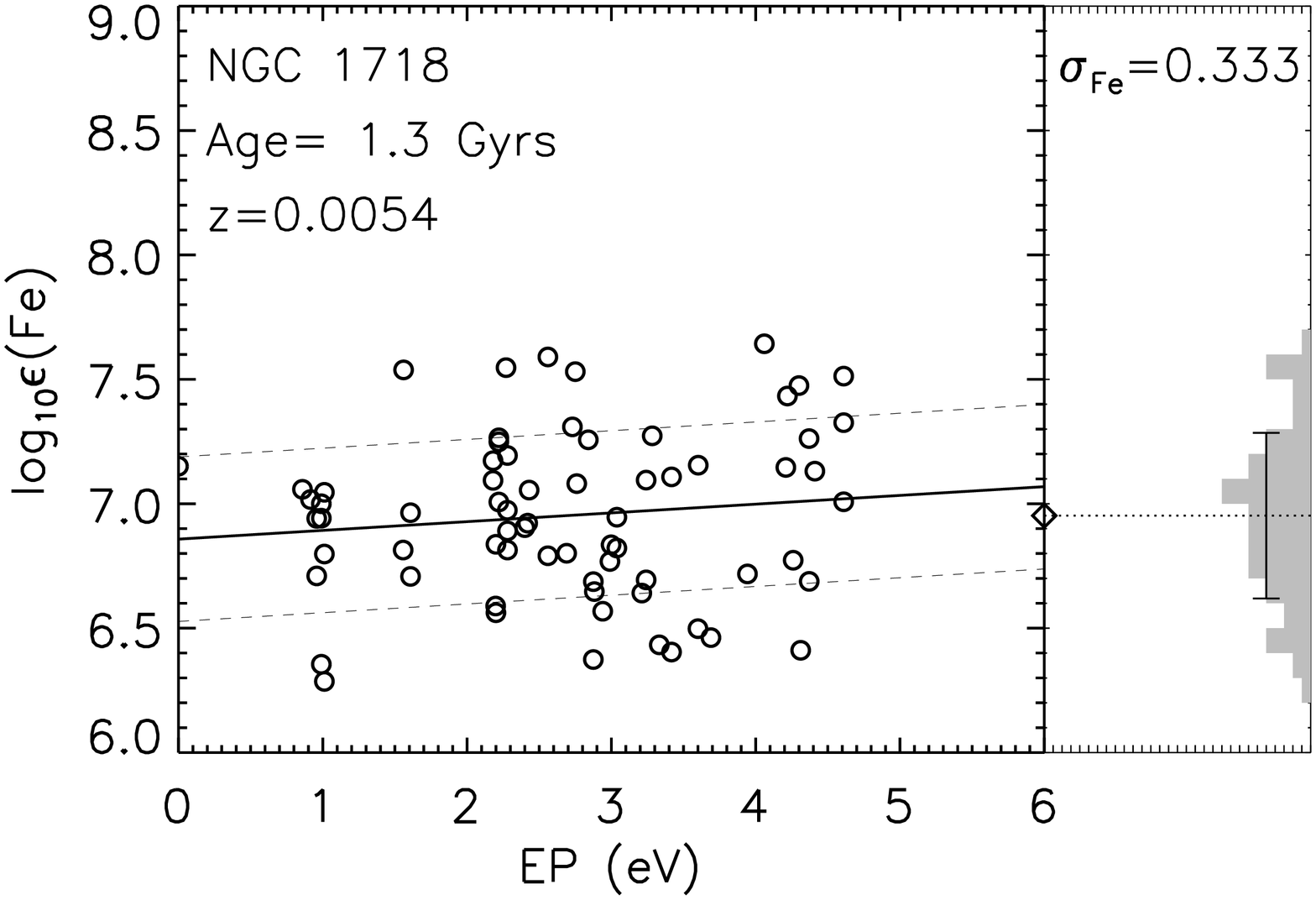}
\includegraphics[angle=90,scale=0.2]{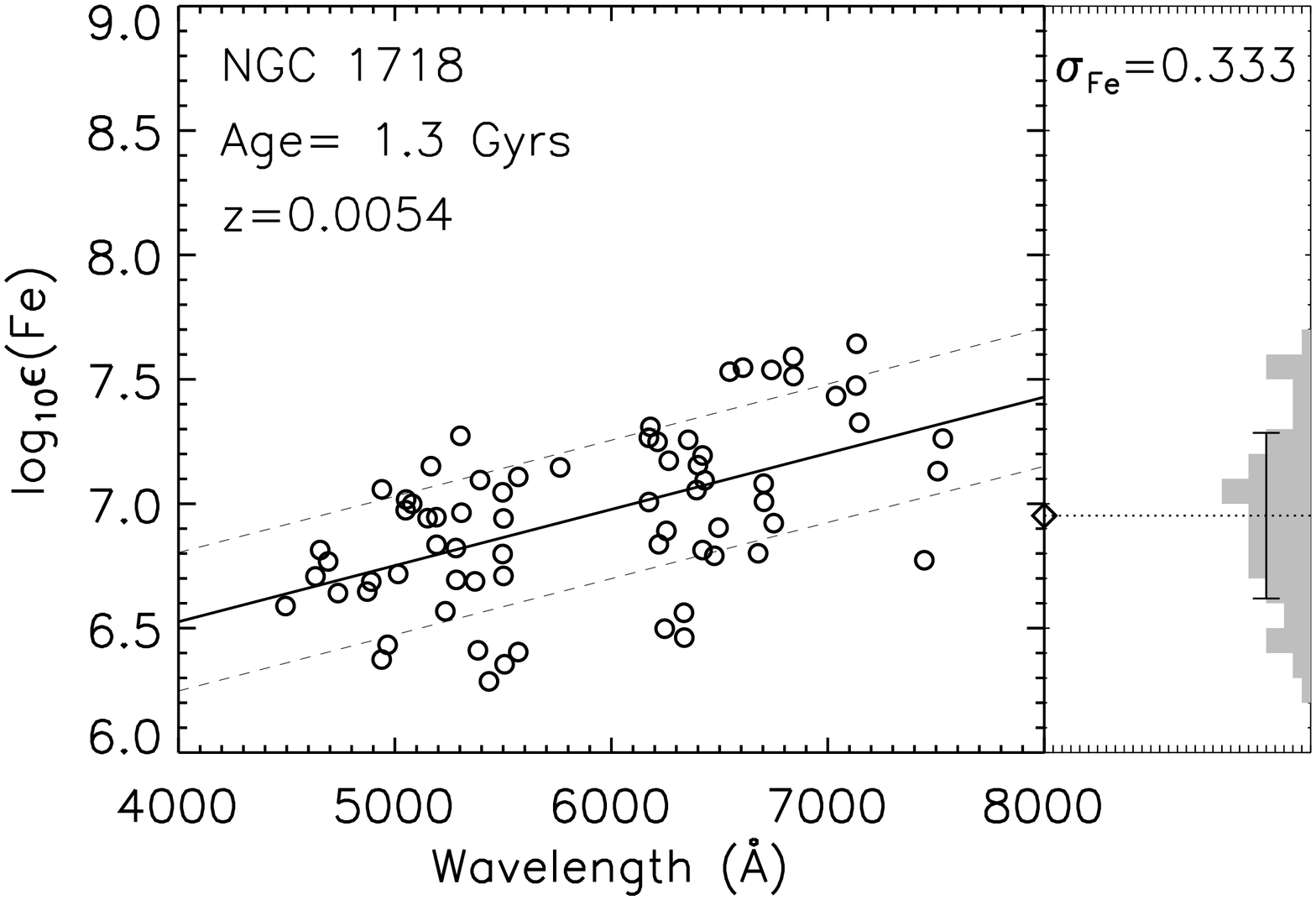}
\includegraphics[angle=90,scale=0.2]{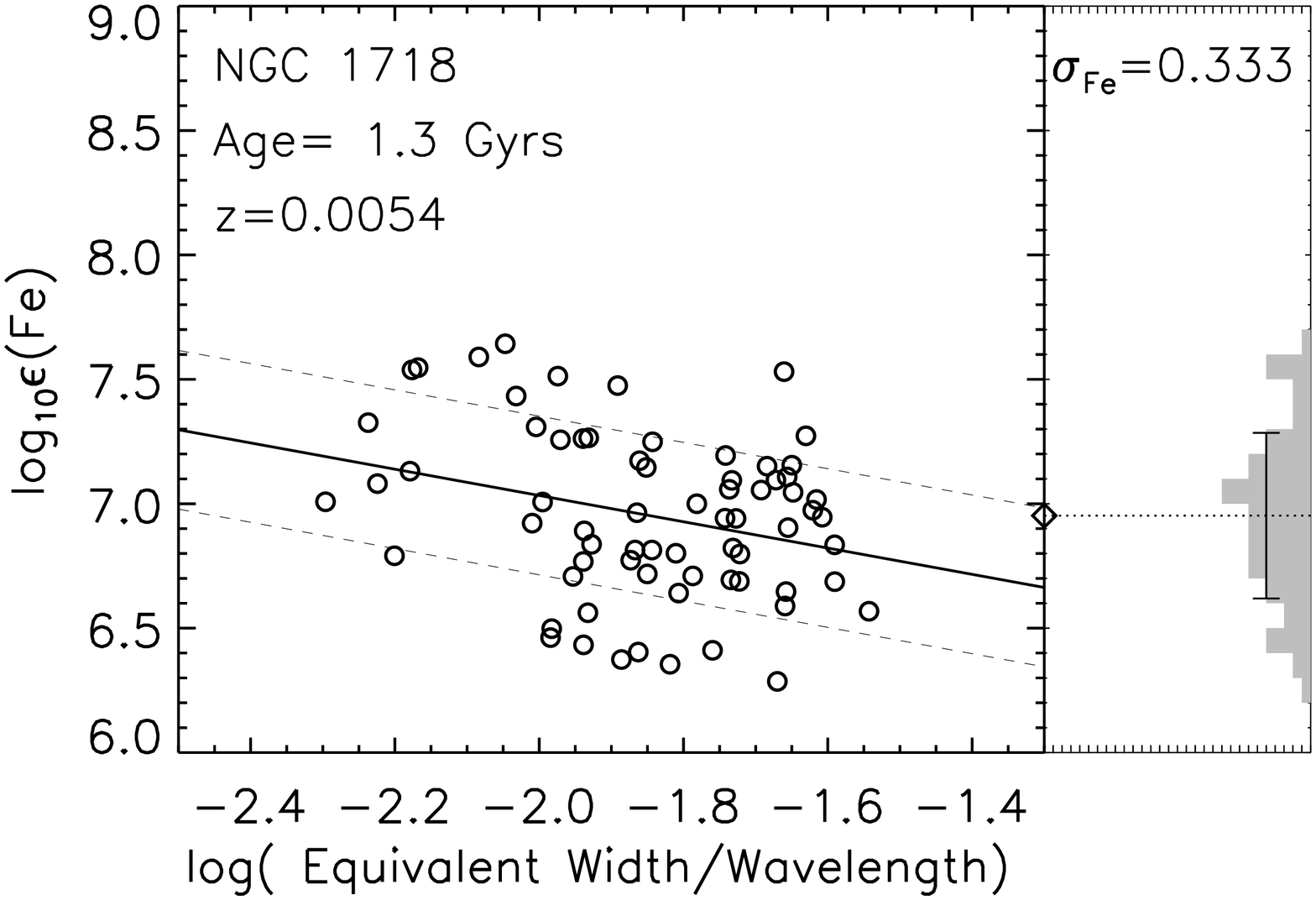}
\includegraphics[angle=90,scale=0.2]{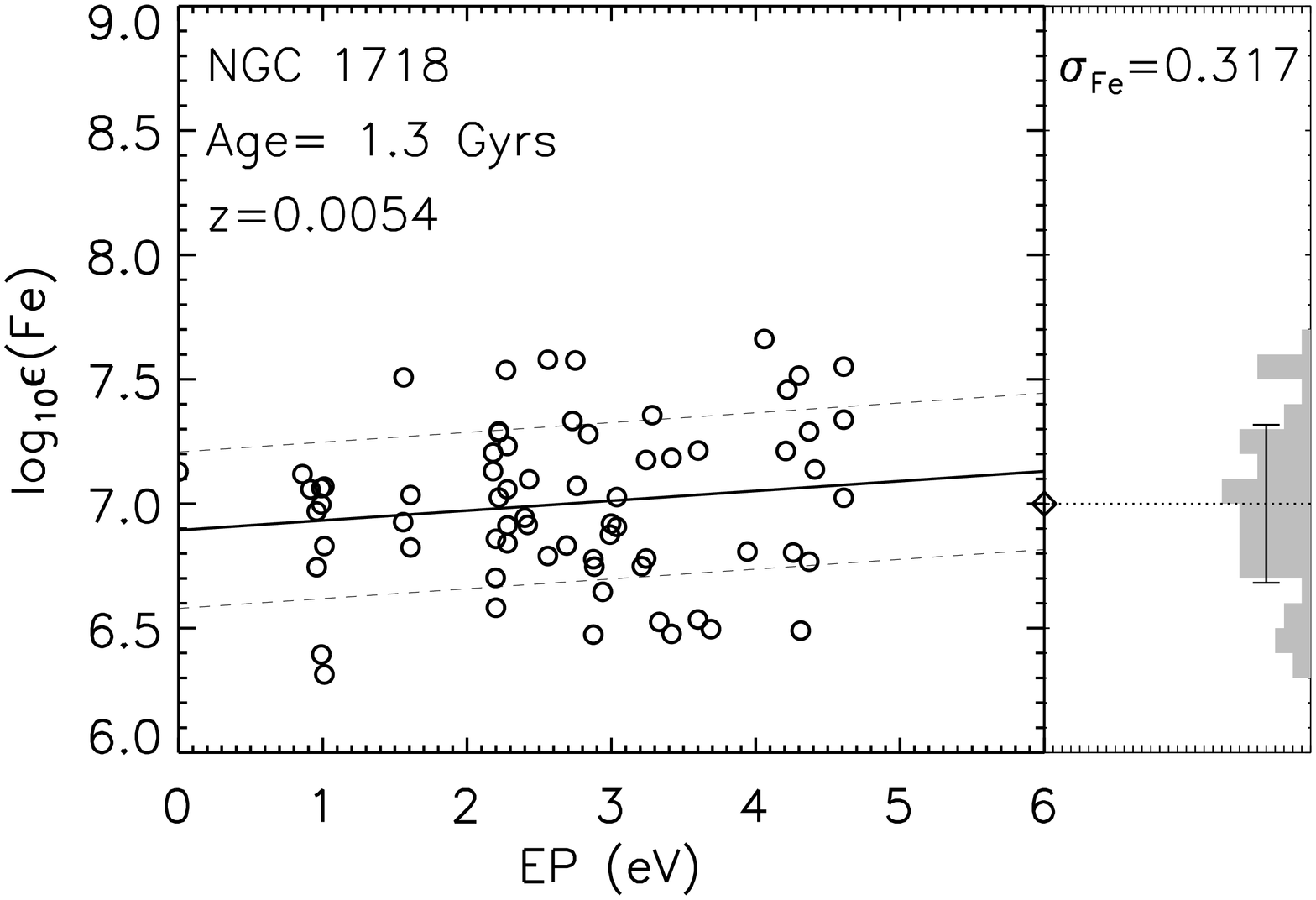}
\includegraphics[angle=90,scale=0.2]{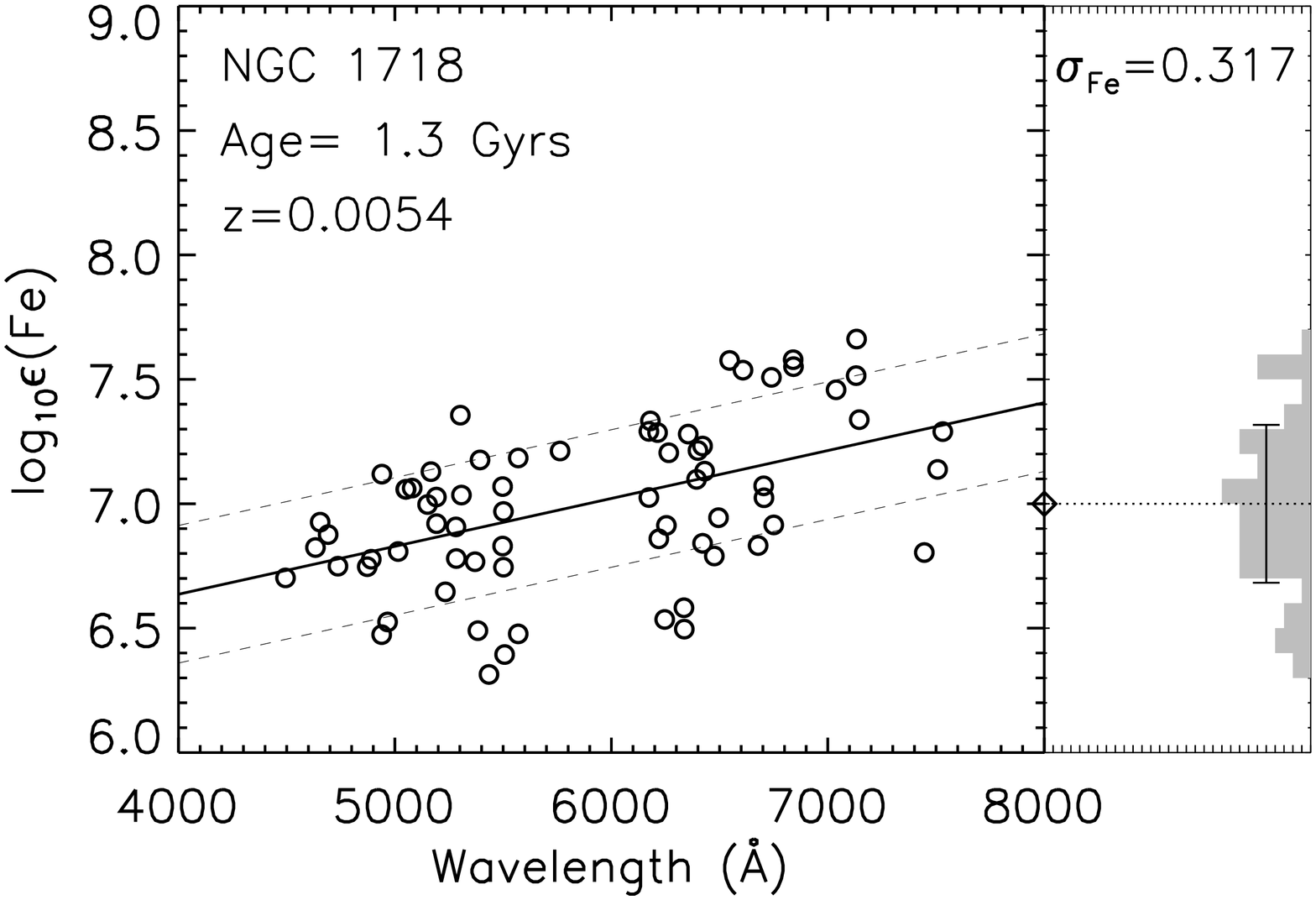}
\includegraphics[angle=90,scale=0.2]{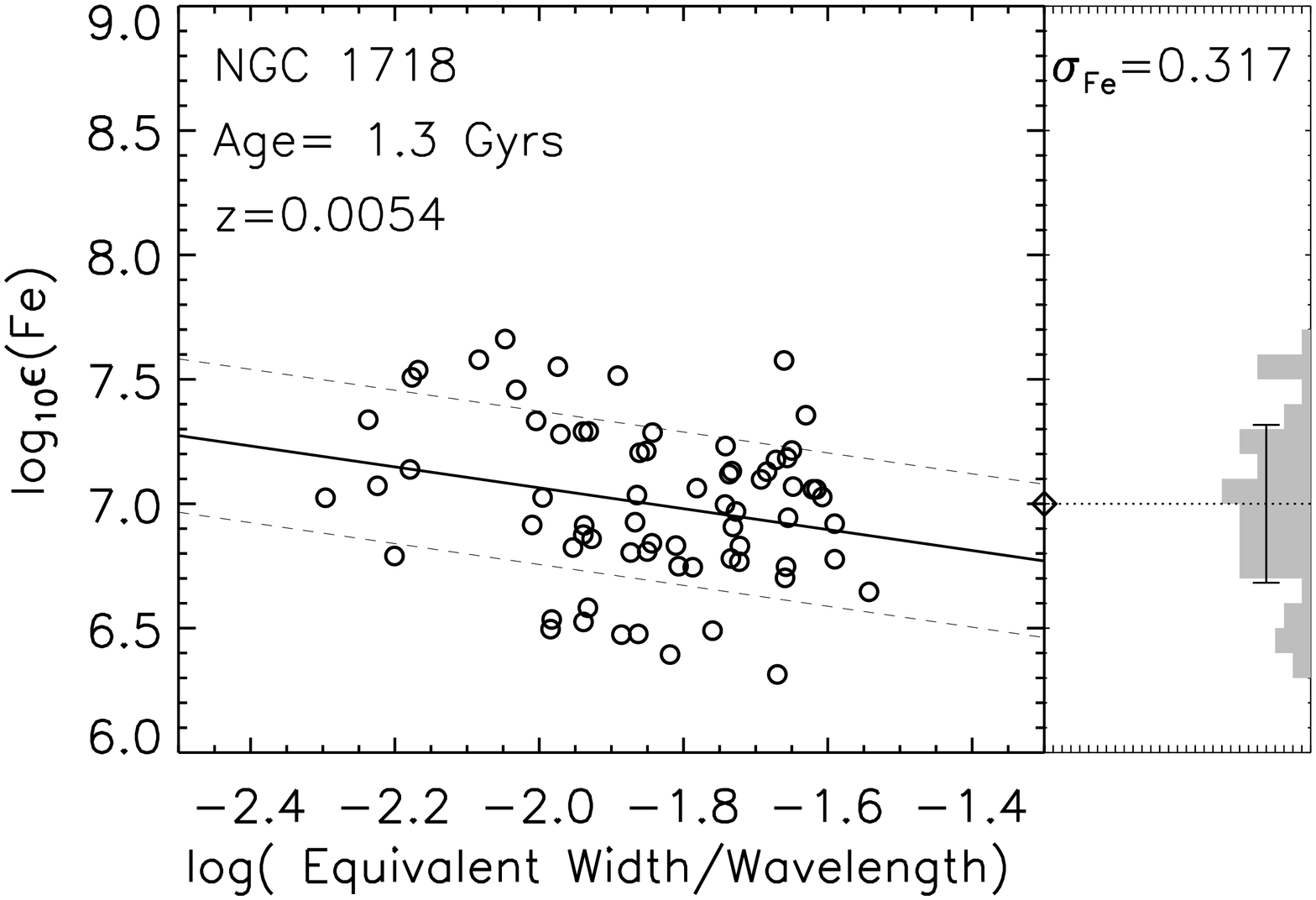}

\caption[Same as Figure~\ref{fig:2019 diagnostics} for NGC 1718]{Same as Figure~\ref{fig:2019 diagnostics} for NGC 1718.  The top panels correspond to the average CMD solution using a 1.25 Gyr, [Fe/H]=$-0.55$ isochrone, and the bottom panels correspond to the best-fitting CMD realization using a 1.25 Gyr, [Fe/H]=$-0.55$ isochrone. The solution in the bottom panels has a smaller $\sigma_{\rm{Fe}}$ and slightly smaller [Fe/H] dependence on EP, wavelength, and reduced EW.}
\label{fig:1718 diagnostics} 
\end{figure*}

To give a more qualitative impression of the difference between the original solution and the final solutions given above, in Figure~\ref{fig:1718 diagnostics} we compare  the individual diagnostics for the average CMD solution with an age of 1.25 Gyr, compared to the best-fitting CMD realization with an age of 1.25 Gyr.  The most significant improvement in the bottom panels when compared to the top panels is a reduced $\sigma_{\rm{Fe}}$.   Because there is still a significant dependence of [Fe/H] with EP, wavelength, and reduced EW, it is  clear  that our best-fitting stellar population is still not an ideal match to the true population of the cluster.

In summary, the analysis of NGC 1718 demonstrates that we are able to distinguish between clusters that are $\sim$1 Gyr in age from clusters that are $>$5 Gyr in age using our abundance analysis method.  We are able to confidently constrain the age of NGC 1718 to a range of 1$-$2.5 Gyr using the stability of the [Fe/H] solution and the observed integrated color, albeit with a larger [Fe/H] uncertainty than for older clusters. 
Our final abundance constraint is 
  [Fe/H]=$-0.64$, with an uncertainty due to the age of $\sigma_{\rm{age}}= \pm 0.25$ dex.  The final error analysis is summarized in \textsection \ref{sec:discussion}.   As discussed above, the larger uncertainty in [Fe/H] could be refined with higher quality data and more complete sampling.  While sampling uncertainties can cause a large spread in \bvo, we are able to constrain a small subset of solutions by requiring self-consistency in the [Fe/H] solution.   In the case of NGC 1718,  we are able to slightly improve on the stability of the [Fe/H] solution by allowing for sampling uncertainties.

\subsubsection{NGC 1978}
\label{sec:1978}

\begin{figure}
\centering
\includegraphics[scale=0.5]{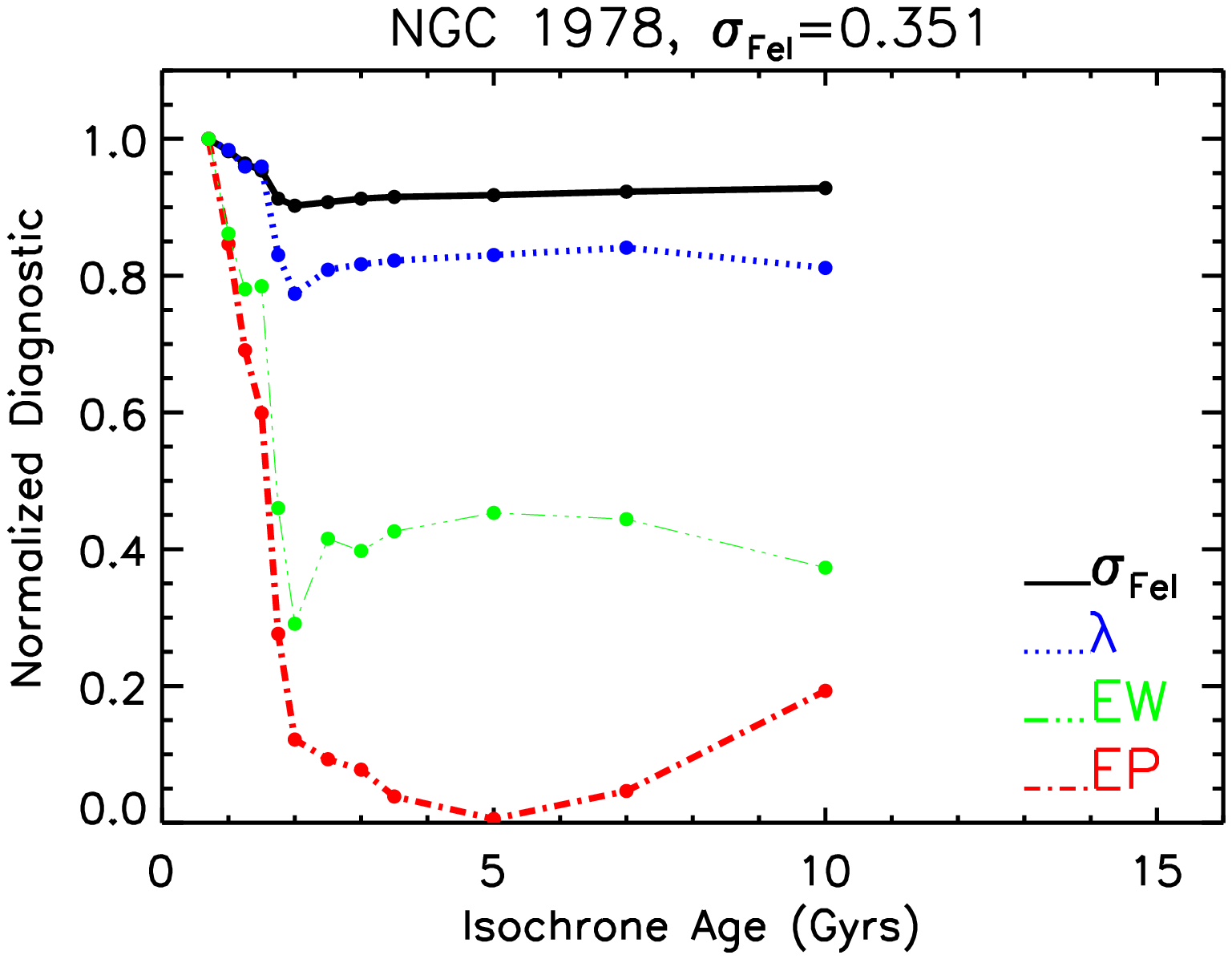}
\caption[Same as Figure~\ref{fig:2019 diag} for NGC 1978.]{Same as Figure~\ref{fig:2019 diag} for NGC 1978.  Best solutions are found for ages $>$1.5 Gyr.  }
\label{fig:CMD trends1978} 
\end{figure}

The second intermediate age cluster in the LMC training set,  NGC 1978,  has an age of $\sim$2 Gyr  and is more massive  ( \mv$\sim-7.7$) than NGC 1718.    Similar to NGC 2005, the S/N of the NGC 1978 spectrum  is lower than for most of the other clusters in our MW and LMC training sets.  We measure EWs for only 36 clean, reasonably unblended Fe I lines, whereas  we typically measure 50$-$100 Fe I lines for clusters with high S/N data.   The best solution has   $\sigma_{\rm{Fe}}$= 0.351, which is larger than the $\sigma_{\rm{Fe}}$ for NGC 1718, as well as  the $\sigma_{\rm{Fe}}$ for the old clusters in the LMC training set.  The  
low S/N makes NGC 1978 particularly difficult to analyze.

We calculate the [Fe/H] solutions for the same grid of isochrones as NGC 1718, but note that the average CMDs with ages of 0.5 Gyrs  result in solutions with [Fe/H] $>+0.5$, which is outside the range of  isochrones and stellar atmospheres used in our analysis and so are not considered further here.    We show the normalized diagnostics for the self-consistent, average CMDs of the other 7 ages in  Figure~\ref{fig:CMD trends1978}.  In the case of NGC 1978, we find that the diagnostics show that the  worst solutions have ages between  0.7 to 1.25 Gyrs.  The solutions with ages $>$1.5 Gyr are of comparable quality, although there is some indication from the EP diagnostic that solutions with ages $>$7 Gyr are also unfavorable. The solutions with ages of 1.5 to 7 Gyr show a moderate dependence of [Fe/H] with both EP and reduced EW, and a significant dependence of [Fe/H] with wavelength.    These  diagnostics  imply that the stellar populations  are a poor match to the cluster.

We estimate that we have only observed 5$-$10\% of the total flux of NGC 1978 in the scanned region, which means that incomplete sampling is a significant issue.  The synthetic CMDs for NGC 1978 have $\sim$4,000 stars, whereas a cluster with \mv$\sim-7.7$ would have $\sim$76,000 stars in the synthetic CMDs  if we had observed 100\% of the total flux.    As in the case of  NGC 1718,  the synthetic CMDs for NGC 1978 have significantly less than 30,000 stars.

\begin{figure}
\centering
\includegraphics[angle=90,scale=0.35]{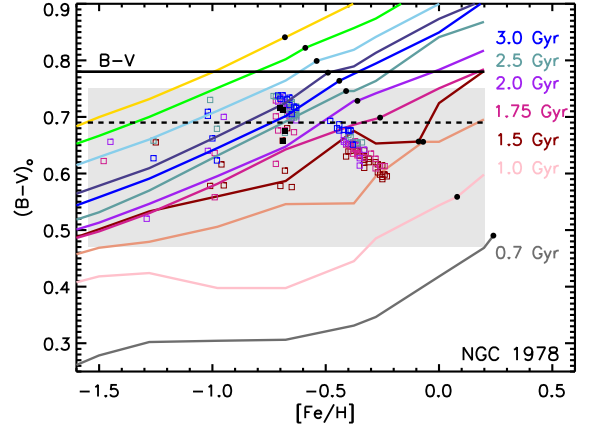}
\caption[Same as Figure~\ref{fig:2019colors} for NGC 1978.]{Same as Figure~\ref{fig:2019colors} for NGC 1978. Colors are the same as in Figure~\ref{fig:Fe plots1718}. The shaded gray region corresponds to the range in \bvo~ allowed for the subset of CMD realizations used for abundance analysis.}
\label{fig:1978colors} 
\end{figure}

We create CMD realizations for ages of 1.5 to 3 Gyr and [Fe/H]$=-1.5$ to [Fe/H]$=0$  to see if we can find a population that results in a more stable [Fe/H] solution. 
Like NGC 1718, for NGC 1978 we find that we are able to eliminate the majority of the possible CMD realizations using the self-consistency of the Fe lines.   Out of these $>$3000 solutions, we only keep a subset of $\sim$150 CMD realizations that satisfy the requirement $0.47 < $\bvo$ < 0.75$.  This is a fairly large range in \bvo, and is chosen to include both the $E(B-V)$=0.09 of \cite{2007AJ....133.2053M} and the significantly larger value of $E(B-V)$= 0.25 from the catalog of \cite{2008MNRAS.385.1535P}.   This \bvo~ requirement is highlighted by the shaded gray area in Figure~\ref{fig:1978colors}, where it can be seen that  the average CMD solutions with ages $>$3 Gyr have \bvo~ colors inconsistent with this requirement.  Moreover, from Figure~\ref{fig:1978colors} there is some justification for eliminating the average CMDs with ages $>$5 Gyr from consideration for a best-fitting solution because they have redder colors than the observed $B-V$.

\begin{figure}
\centering
\includegraphics[scale=0.5]{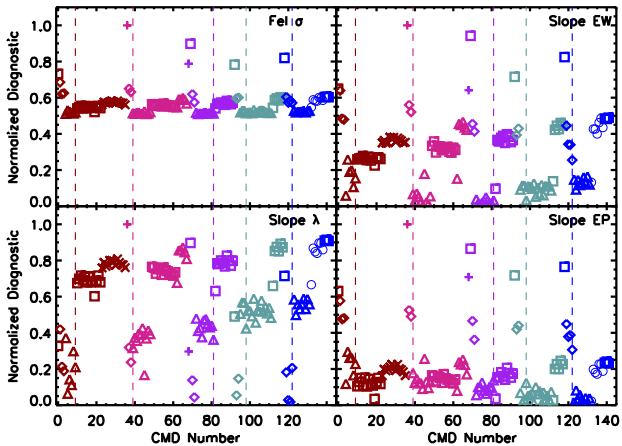}
\caption[Same as Figure~\ref{fig:2019newdiag} for NGC 1978]{Same as Figure~\ref{fig:1718newdiag} for NGC 1978. Colors are the same as in Figure~\ref{fig:Fe plots1718}. Diagnostics are strongly correlated with both age and metallicity.   }
\label{fig:1978compare} 
\end{figure}

\begin{figure}
\centering
\includegraphics[scale=0.5]{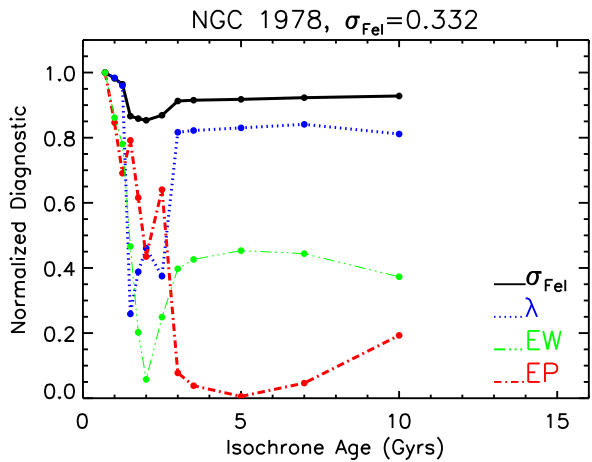}
\caption[Same as Figure~\ref{fig:CMD trends1978}, except the solutions at ages of 1.5$-$2.5 Gyr have been replaced by the solutions for the best-fitting CMD realization at these ages.]{Same as Figure~\ref{fig:CMD trends1978}, except the solutions at ages of 1.5$-$2.5 Gyr have been replaced by the solutions for the best-fitting CMD realization at these ages. The best solutions are found for ages of 1.5$-$2.5 Gyr.}
\label{fig:CMD trends1978new} 
\end{figure}

\begin{figure*}
\centering
\includegraphics[angle=90,scale=0.2]{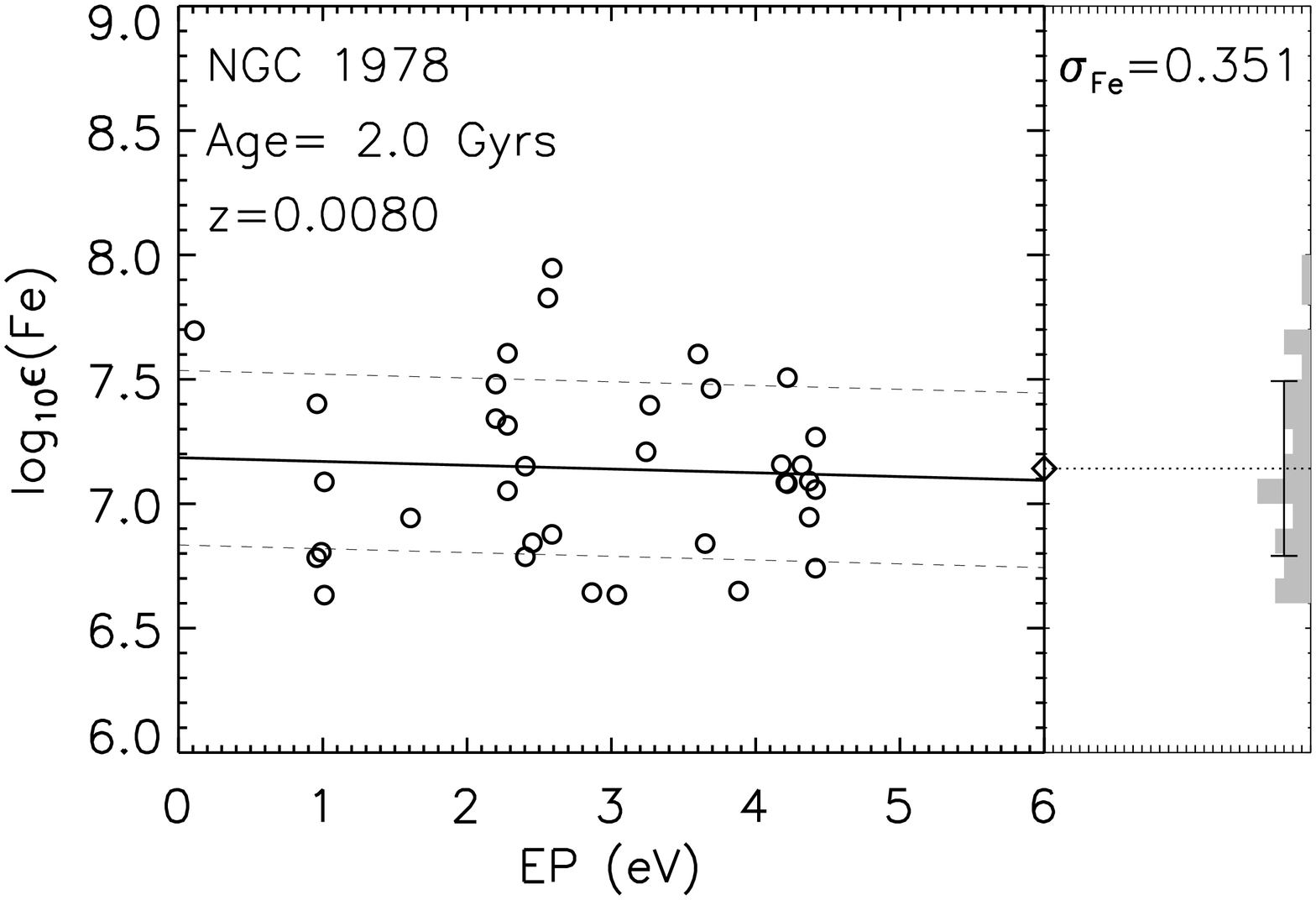}
\includegraphics[angle=90,scale=0.2]{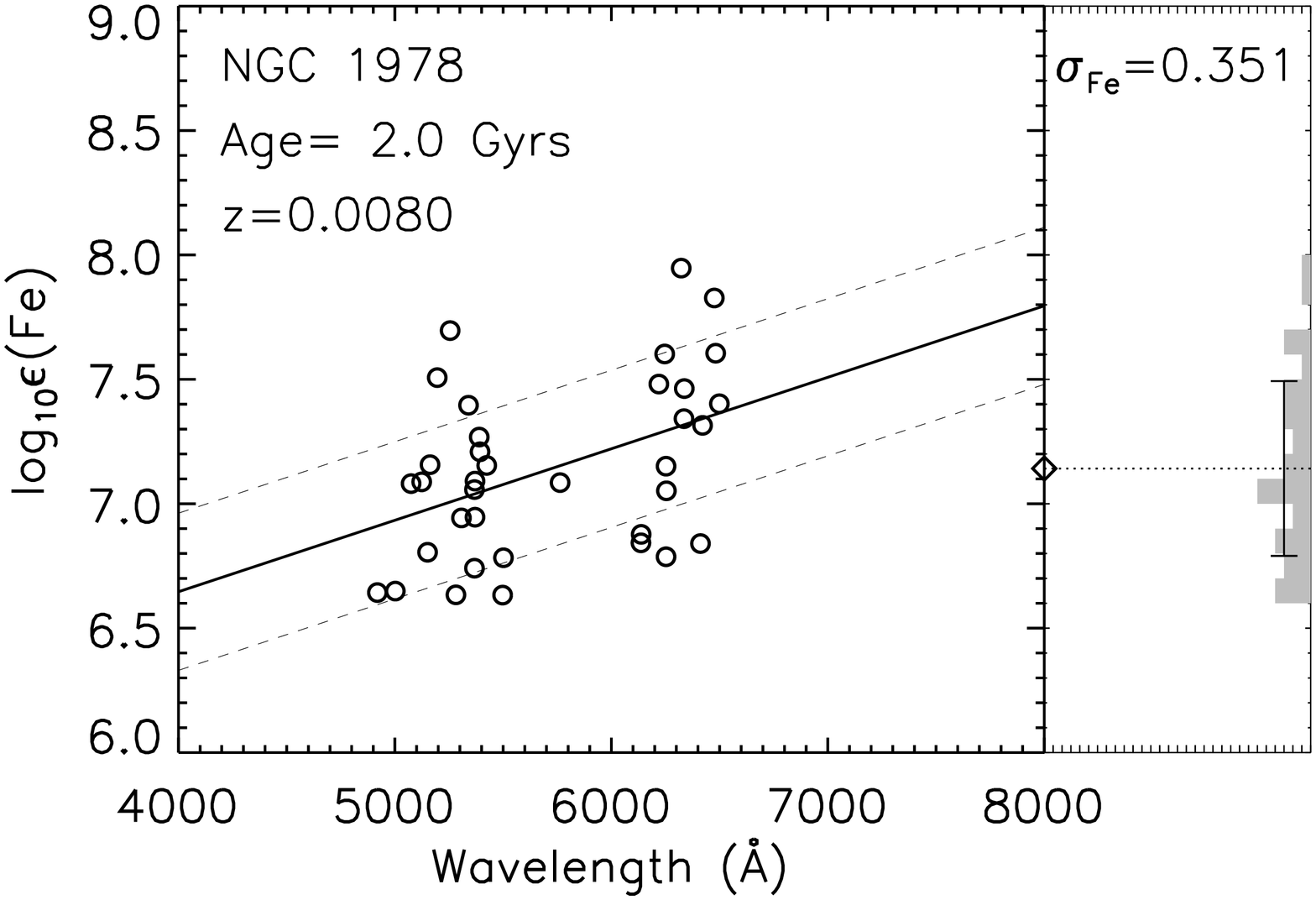}
\includegraphics[angle=90,scale=0.2]{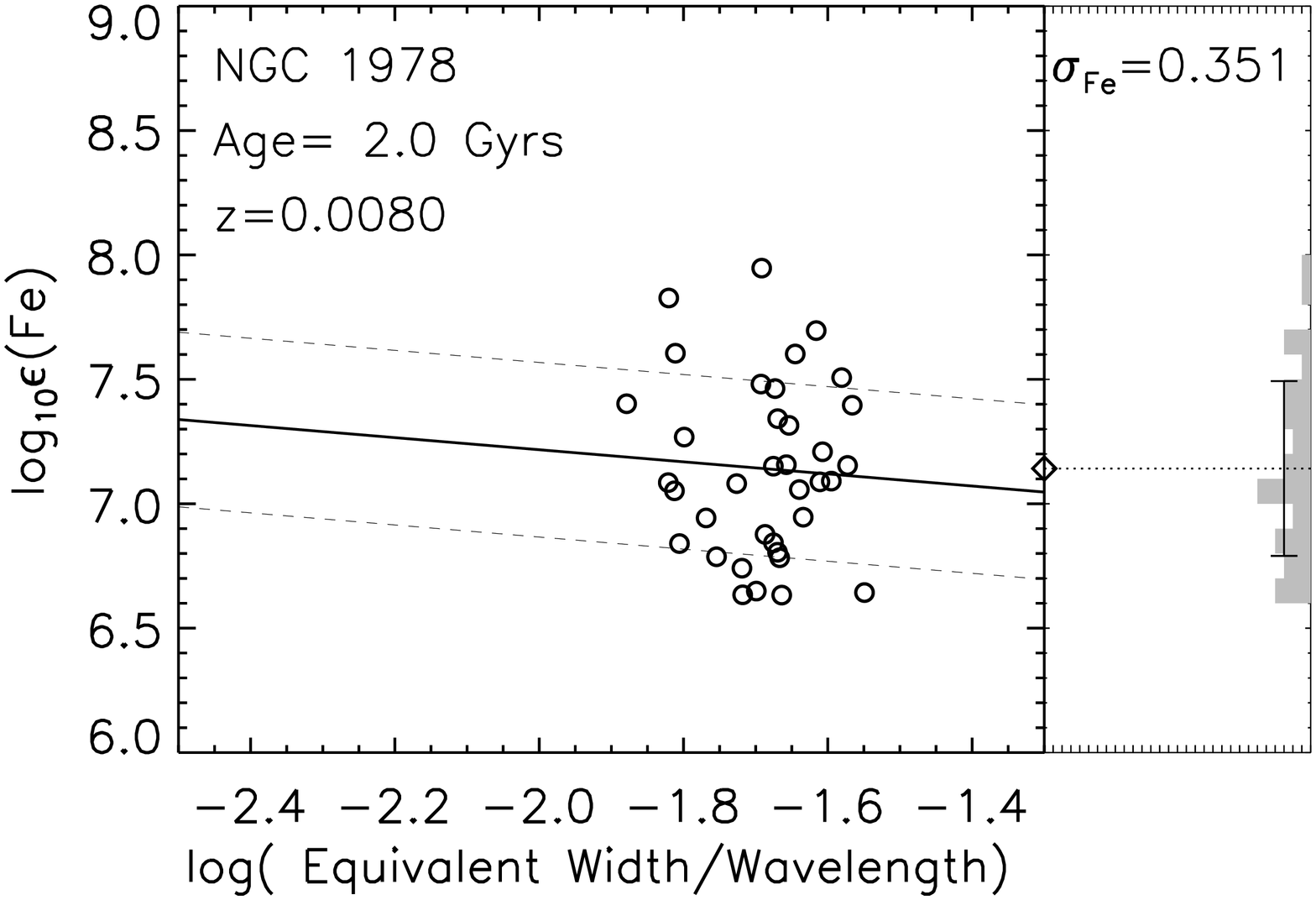}
\includegraphics[angle=90,scale=0.2]{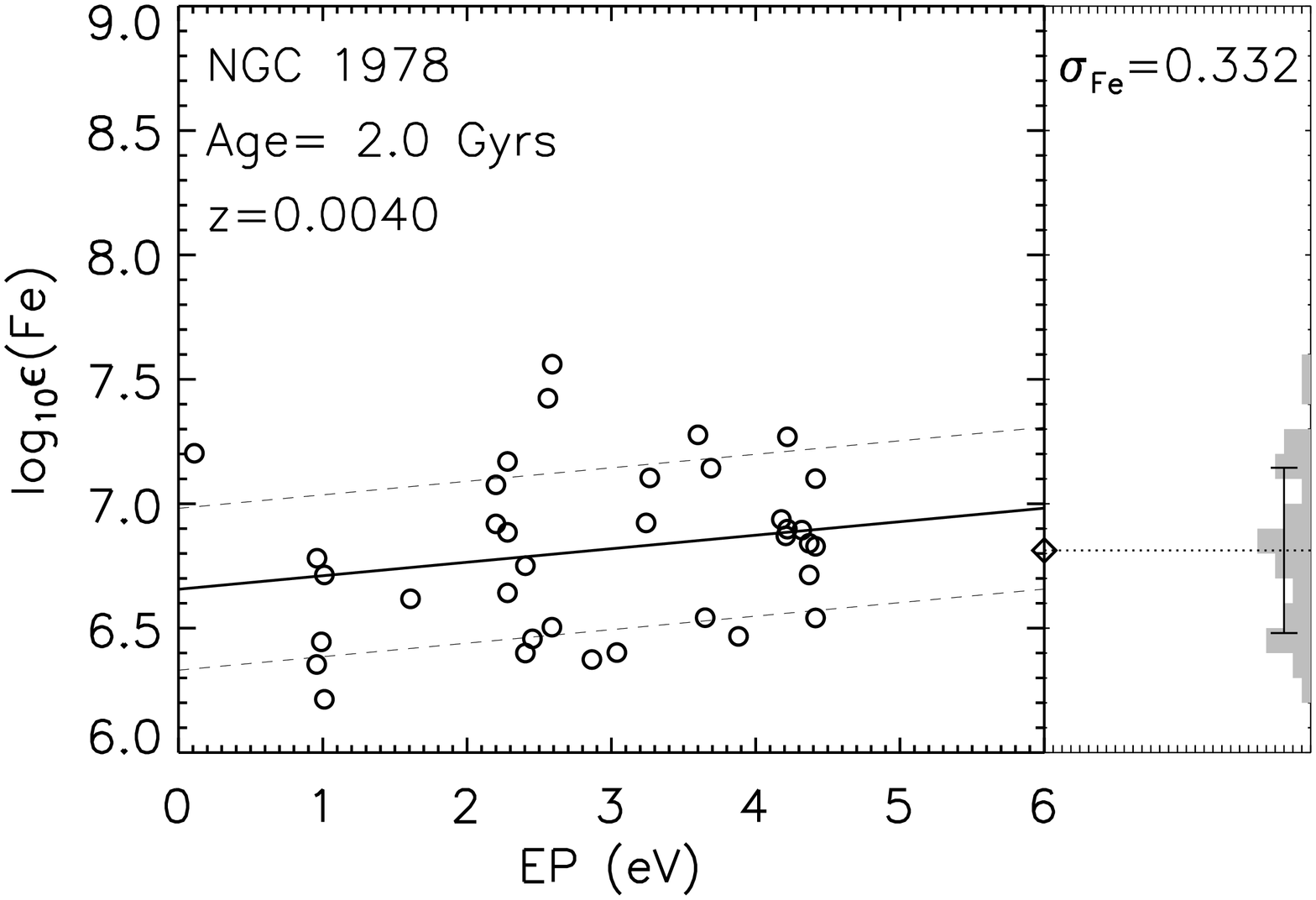}
\includegraphics[angle=90,scale=0.2]{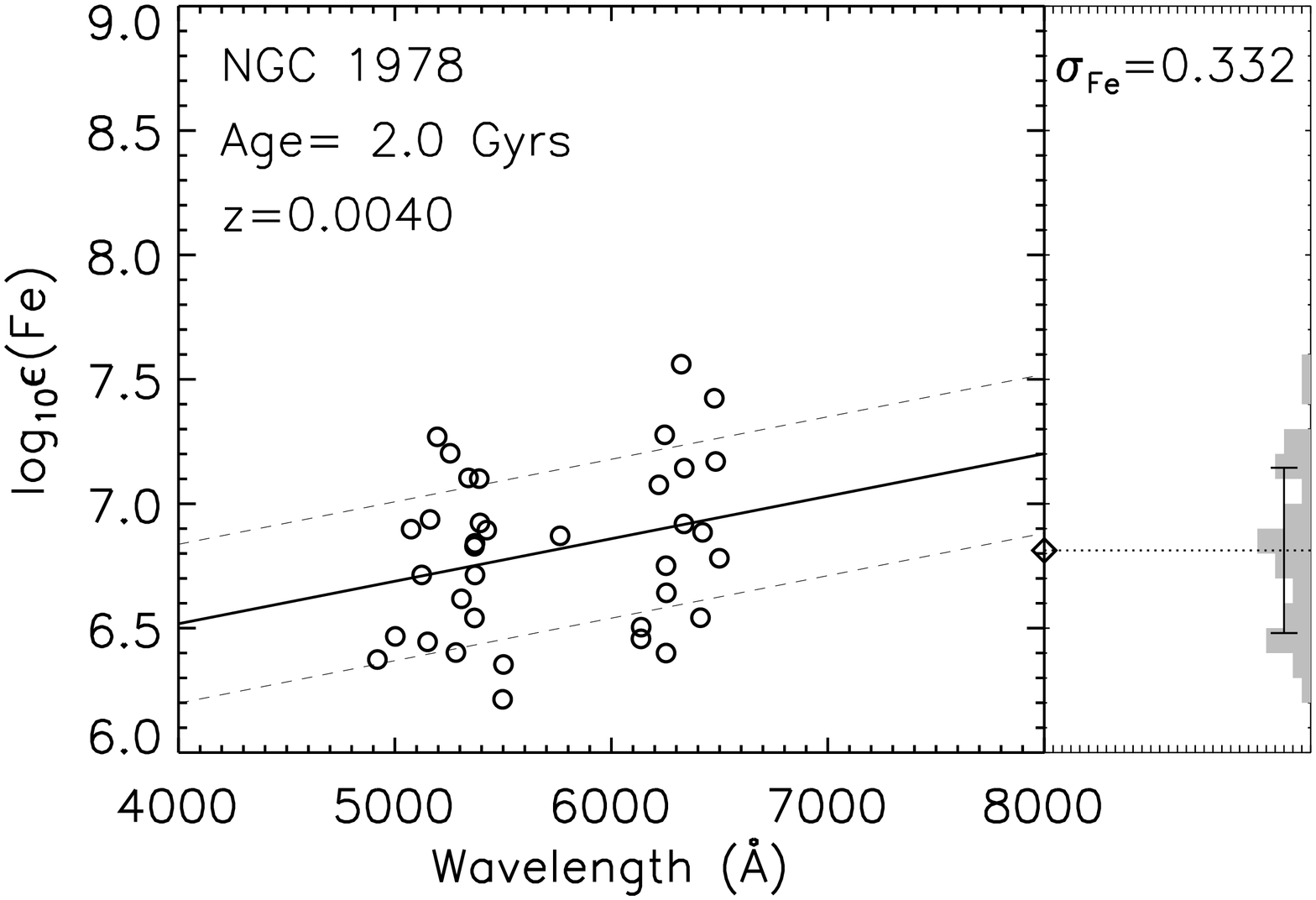}
\includegraphics[angle=90,scale=0.2]{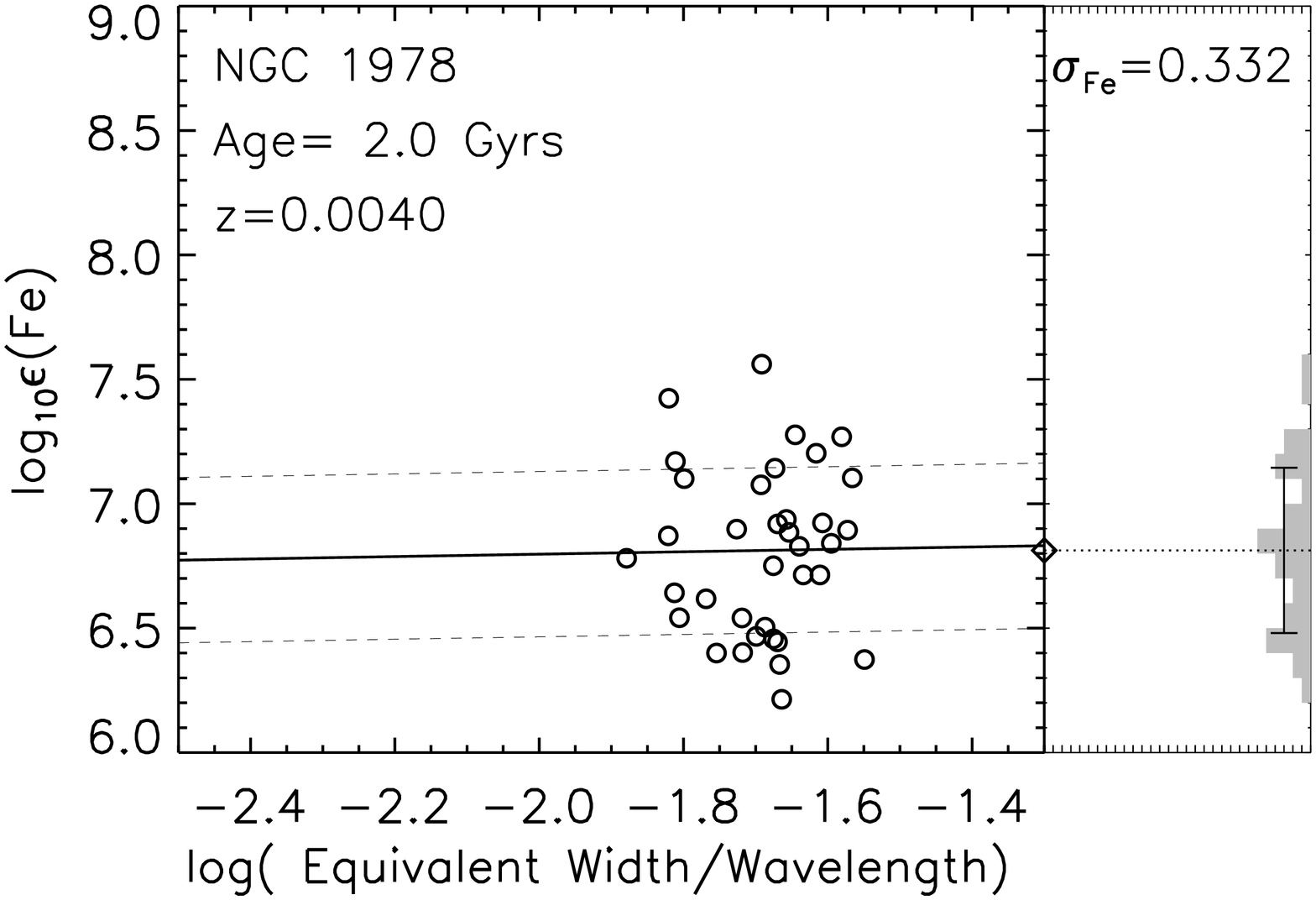}
\caption[Same as Figure~\ref{fig:2019 diagnostics} for NGC 1978]{Same as Figure~\ref{fig:2019 diagnostics} for NGC 1978.  The top panels correspond to the average CMD solution using a 2 Gyr, [Fe/H]=$-0.38$ isochrone, and the bottom panels correspond to the best-fitting CMD realization using a 2 Gyr, [Fe/H]=$-0.66$ isochrone. The solution in the bottom panels has a smaller $\sigma_{\rm{Fe}}$, and smaller [Fe/H] dependence on wavelength and reduced EW.}
\label{fig:1978 diagnostics} 
\end{figure*}

  We show the diagnostics for the subset of  $\sim$150 self-consistent solutions in Figure~\ref{fig:1978compare}, where it can be seen that all four Fe diagnostics show similar, correlated behavior with age and metallicity. We pick one best-fitting solution for each age (5 total), and compare the normalized diagnostics to the solutions for the other ages in Figure~\ref{fig:CMD trends1978new}.  From  Figure~\ref{fig:CMD trends1978new} it appears clearer that there is a preferred age for NGC 1978 between 1.5 and 2.5 Gyr.  There is significant improvement in the $\sigma_{\rm{Fe}}$, and a much smaller dependence of [Fe/H] on wavelength and reduced EW when compared to the average CMD solutions in Figure~\ref{fig:CMD trends1978}, although a dependence of [Fe/H] on EP and wavelength is still present.  The difference between the best-fitting 2.5 Gyr CMD realization and the original averaged 2.5 Gyr CMD can be seen most clearly in Figure~\ref{fig:1978 diagnostics}, where we compare the individual diagnostics.

Interestingly, we find that  all of best-fitting CMD realizations for NGC 1978 have [Fe/H]$\sim-0.7$, regardless of the age.
The solution for NGC 1978 is therefore much better constrained than for NGC 1718.  Note that this is also much more metal-poor than the solution obtained from the average CMDs. Comparing with the original solution, we find that the main difference is that the CMD realizations that allowed for partial sampling of the CMD  have flux in much cooler AGB stars than the original solutions.  This is further evidence that we are able to tell whether significant flux in cool AGB stars is needed in the synthetic CMDs by using the Fe I diagnostics.  We also find that the best-fitting CMD realizations have 0.66 $<$\bvo$<$ 0.72, which corresponds to 0.07$<$$E(B-V)$$<$ 0.12. This range is very consistent with the $E(B-V)$=0.09 derived by \cite{2007AJ....133.2053M} from their very deep HST CMD, and supports the use of this method to provide independent constraints on the $E(B-V)$ of unresolved clusters as done in \citet{m31paper}.

In summary, the mean solution for NGC 1978 using the best-fitting CMD realizations with ages of 1.5 to 2.5  Gyr is [Fe/H]=$-0.74$, with $\sigma_{\rm{age}}=\pm0.05$.  Although the line-to-line scatter for the Fe abundance is also high, it is likely that it can be explained by the low S/N of the data.  Like NGC 1718, we find that the best-fitting solutions for NGC 1978 show pretty significant dependence of [Fe/H] on EP and wavelength.  However, unlike the case of NGC 1718, for NGC 1978 we find the [Fe/H] has only a weak dependence on  age within the preferred age range, despite the low quality of the data and large standard deviation of the Fe lines.

These two clusters are very interesting illustrations of our method at intermediate ages.  However, the two clusters are very different from each other. 
  Given the data quality and that we only have two clusters in the intermediate age range, it is impossible to determine if the difficulties we encounter in finding an accurate stellar population match for these clusters is due to broader, generic problems in the isochrones at these ages.
Along those lines,  we note that there is still some debate over the presence or degree of convective overshooting  needed in the input physics of the isochrones in order to match the observed CMDs of young clusters, which we discuss further in \textsection \ref{sec:uncertainties} \citep[e.g.][]{2004ApJ...612..168P,2007AJ....133.2053M}.  There are also many open questions as to the modeling of AGB star evolution and mass loss on the RGB, both of which can have a particularly significant effect on the integrated properties of clusters in this age range \citep[e.g.][]{2007AJ....133..468C,2000A&AS..141..371G,2005ARA&A..43..387G}.  A more extensive training set of high quality integrated light spectra of clusters in this age range is clearly needed to begin to address these questions.

\subsection{Results: Young Clusters}
\label{sec:young}

There are four young clusters in the LMC training set, which have ages $<$ 1 Gyr.  Like for  the intermediate age clusters, it is immediately apparent using our original technique that these clusters are best matched by the youngest CMDs in our grid, and so we  add synthetic CMDs with ages between 30$-$300 Myr to the analysis.   

  In the following sections we present the analysis for NGC 1866, NGC 1711 and NGC 2100. 
 The remaining cluster, NGC 2002, has an estimated age of  $\sim$15 Myrs \citep{1991ApJS...76..185E}. Note that  the youngest age in the grid of isochrones we currently employ for our analysis is 30 Myr.   While we are able to measure EWs for  NGC 2002 (as reported in Table~\ref{tab:t4_stub}),  the solutions for synthetic CMDs with ages between 30 Myrs and 1 Gyr do not converge at self-consistent values of [Fe/H]. Indeed this cluster appears to be beyond the reach of our current technique, and so we do not discuss  NGC 2002 further in this work.

\subsubsection{NGC 1866}
\label{sec:1866}

\begin{figure}
\centering
\includegraphics[angle=90,scale=0.35]{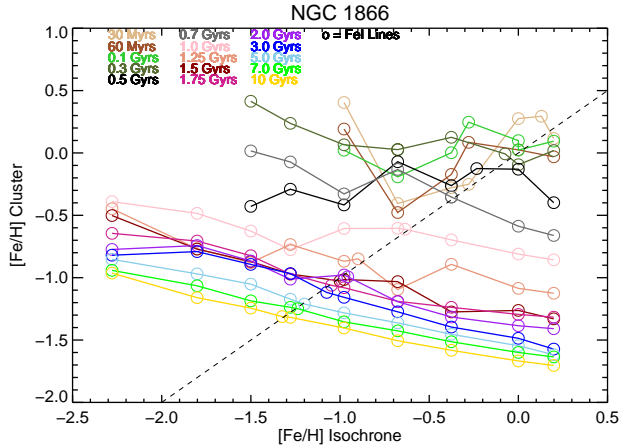}
\caption[Same as Figure~\ref{fig:2019 Fe plot} for NGC 1866.]{Same as Figure~\ref{fig:2019 Fe plot} for NGC 1866.  Additional CMDs with ages $<$0.5 Gyr are shown by the labeled colors.}
\label{fig:1866Fe plots} 
\end{figure}

\begin{figure}
\centering
\includegraphics[angle=90,scale=0.35]{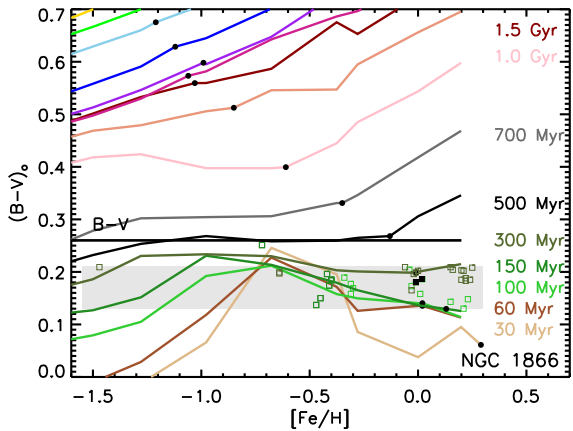}
\caption[Same as Figure~\ref{fig:2019colors} for NGC 1866.]{Same as Figure~\ref{fig:2019colors} for NGC 1866. Colors are the same as in Figure~\ref{fig:1866Fe plots}. }
\label{fig:1866color} 
\end{figure}

NGC 1866 is a relatively massive (\mv$\sim-9$), well-studied cluster, with an age of 100$-$200 Myr \citep[e.g.][]{1999AJ....118.2839T, 2003AJ....125.3111B}.   From the reasonably high S/N spectra for NGC 1866 we are able to measure 49 Fe I lines.  We find a line-to-line scatter for the Fe I abundance that is comparable to that of the old clusters ($\sigma_{\rm{Fe}}\sim$0.244), and significantly smaller than that of the intermediate age clusters, which makes NGC 1866  a valuable  test case for our abundance analysis methods.

We derive mean  [Fe/H] values for the full grid of synthetic CMDs, which are shown  in Figure~\ref{fig:1866Fe plots}. The  [Fe/H]$_{\rm{cluster}}$ solutions for synthetic CMDs with ages younger than 150 Myr  deviate from the behavior of the [Fe/H]$_{\rm{cluster}}$ of the older CMDs, which generally result in monotonically decreasing output [Fe/H]$_{\rm{cluster}}$ with increasing input [Fe/H]$_{\rm{iso}}$. 
 This behavior is due to  the stellar atmospheres over-compensating the Fe abundance to match the observed Fe EWs  when the input [Fe/H]$_{\rm{iso}}$  is lower than the real abundance, or under-compensating when the  input [Fe/H]$_{\rm{iso}}$ is too high. 
 For synthetic CMDs with old ages, the stellar populations generally change slowly and smoothly with increasing metallicity, which leads to the smooth change in derived [Fe/H]$_{\rm{cluster}}$.  For synthetic CMDs with ages $<$150 Myr, the temperatures of core He-burning,  luminous supergiants not only vary significantly on short timescales, but are also very sensitive to metallicity \citep[e.g.][]{1992ARA&A..30..235C, 1993ApJ...410...99B, 1999A&AS..136...65B}.   
For this grid of isochrones, the averaged, synthetic CMDs  with ages $<$150 Myr and [Fe/H]=$-1$ to $-0.3$ contain supergiants at cooler temperatures than CMDs at other metallicities at the same age. Because the cooler supergiants have stronger Fe I  features, the derived [Fe/H] tend to be lower, which is seen in Figure~\ref{fig:1866Fe plots}.

The temperature of the luminous supergiants also significantly affects the integrated colors of the synthetic CMDs.   In Figure~\ref{fig:1866color}, we show the behavior of the integrated \bvo~ for the grid of isochrones as a function of  [Fe/H]$_{\rm{iso}}$.  As expected, the youngest synthetic CMDs with [Fe/H]=$-1$ to $-0.3$ have redder \bvo~ colors than CMDs at other metallicities (note the ``bump''  in \bvo~ at $-0.7<$[Fe/H]$<-0.5$
for the 30$-$100 Myr age isochrones).   A more subtle point that can be seen in  Figure~\ref{fig:1866color} is that  there is a complicated \bvo-metallicity-age degeneracy for synthetic CMDs with ages $<$150 Myr.  We will return to this point below.

\begin{figure}
\centering
\includegraphics[scale=0.5]{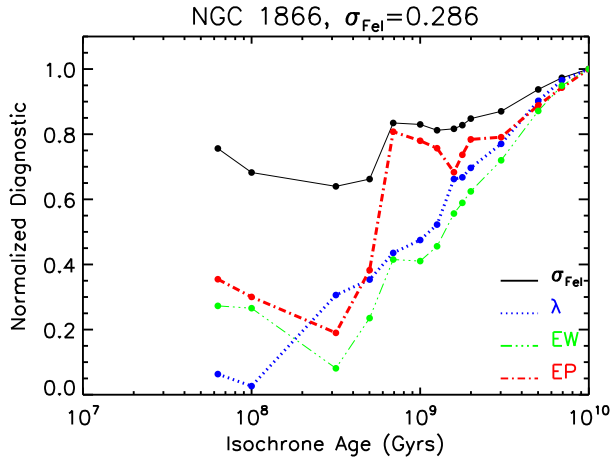}
\caption[Same as Figure~\ref{fig:2019 diag} for NGC 1866]{Same as Figure~\ref{fig:2019 diag} for NGC 1866. Best solutions are for ages of 100$-$300 Myr.}
\label{fig:1866 diag} 
\end{figure}

\begin{figure}
\centering
\includegraphics[angle=90,scale=0.35]{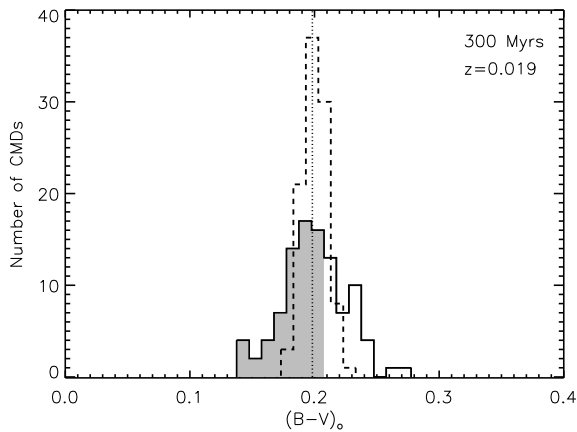}
\caption[Histogram of integrated \bvo~ color for 100 CMD realizations of a 300 Myr, z=0.019  (Fe/H=0)  isochrone]{Histogram of integrated \bvo~ color for 100 CMD realizations of a 300 Myr, z=0.019 ([Fe/H]=$0$) isochrone.  Solid black line shows the histogram for a population where the total flux in stars has been normalized to 14\% of a \mv$=-9$  cluster, which is appropriate for our integrated light spectrum of NGC 1866.  Dashed black line shows the histogram for a population normalized to 100\% of a \mv$=-9$ cluster.  CMDs with \bvo~ color consistent with the observed, reddening-corrected $B-V$ of NGC 1866 are shaded in gray. }
\label{fig:1866 histo} 
\end{figure}

The complex behavior of the [Fe/H]$_{\rm{cluster}}$ solutions in Figure~\ref{fig:1866Fe plots} hints that constraining the [Fe/H] of clusters with ages $<$1 Gyr may be difficult. 
For NGC 1866, the self-consistent synthetic CMDs with younger ages (and hotter stars),  tend to have more metal-rich [Fe/H] solutions.  
In Figure~\ref{fig:1866 diag} we show the diagnostics for the 14 self-consistent CMD solutions for the initial grid.  It is very clear  from the Fe I line diagnostics alone that a young age for NGC 1866 is preferred.  Specifically, the most stable [Fe/H] solutions for NGC 1866 are found at ages of 100$-$500 Myrs.   Note that this  independent age constraint using our abundance analysis method is very consistent with the observed $B-V$=0.26 from  \cite{2008MNRAS.385.1535P}.

\begin{figure}
\centering
\includegraphics[angle=90,scale=0.35]{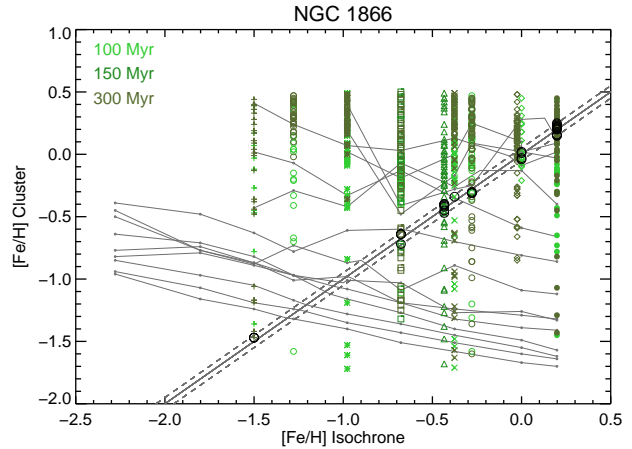}
\caption[Same as Figure~\ref{fig:1866Fe plots}, with the addition of the Fe I abundance results for the CMD realizations of NGC 1866]{Same as Figure~\ref{fig:1866Fe plots}, with the addition of the Fe I abundance results for the CMD realizations of NGC 1866.  Colors of the points denote the same ages as Figure~\ref{fig:1866Fe plots}.  Only CMD realizations consistent with the observed $B-V$ color of NGC 1866 are shown.   Black circles on the solid black line denote CMD realizations that satisfy the self-consistent criterion [Fe/H]$_{\rm{\rm{iso}}}$=[Fe/H]$_{\rm{cluster}}$.  Only a small number of the possible CMD realizations have self-consistent [Fe/H] solutions. }
\label{fig:1866Fe plots all} 
\end{figure}

In view of the earlier discussion of the importance of luminous supergiants for clusters in this age regime, one would expect sampling uncertainties to be important here.   However, in the case of NGC 1866, the original, average CMD solutions for ages of 100$-$300 Myr are very stable and show only a very slight dependence of [Fe/H] with EP and wavelength.  Because these solutions are very stable already, we 
are unlikely to find an improved CMD realization.
Nevertheless, it is interesting to ask what range in age or [Fe/H] can produce a similarly good CMD when stochastic sampling is considered.
To that end,  the dashed line in  Figure~\ref{fig:1866 histo} shows a histogram of the integrated \bvo~ of 100 CMD realizations   created for an isochrone with an age of 300 Myr, and [Fe/H]=0, which is  consistent with the properties we derive for NGC 1866.  The CMD realizations that make up the dashed line histogram have been normalized to the total flux of a  \mv$\sim-9$ cluster.  It can be seen in Figure~\ref{fig:1866 histo}  that a cluster with  \mv$\sim-9$ is massive enough that the range of \bvo~ is small ($\sim$0.05 mag) because the CMDs are not significantly impacted by statistical fluctuations  in the number and properties of the supergiant stars.  Similarly, even though we estimate that we have only observed 14\% of the total flux of NGC 1866, we find that CMD realizations normalized to 14\% of the flux of a  \mv$\sim-9$ cluster are still so well-populated  that the spread in \bvo~ is only a little larger ($\sim$0.1 mag), and  comparable to that for the better sampled old clusters in the training set.  The histogram for 100 CMD realizations normalized to the observed region of NGC 1866 is shown by the solid line in Figure~\ref{fig:1866 histo}.

\begin{figure*}
\centering
\includegraphics[angle=90,scale=0.2]{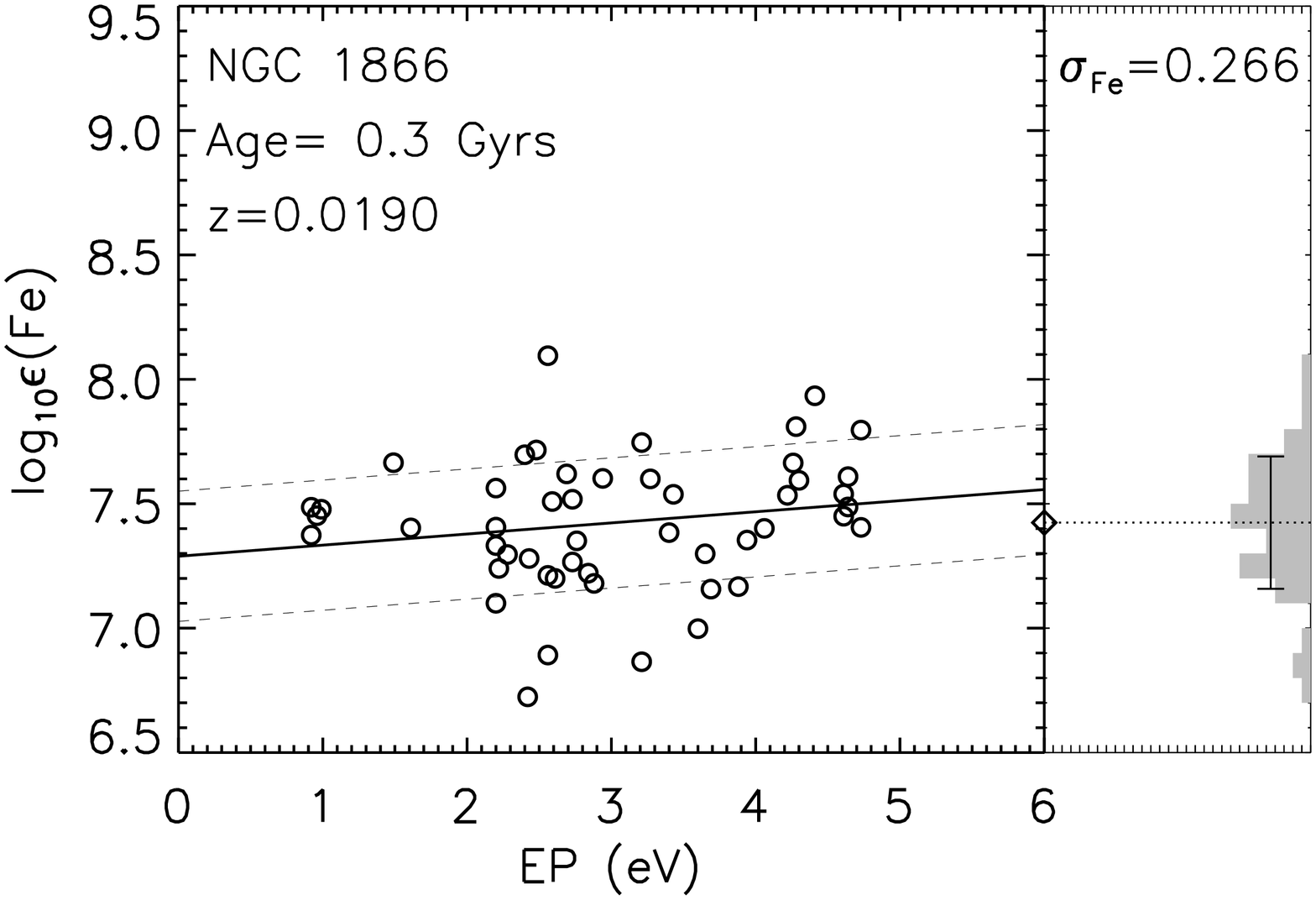}
\includegraphics[angle=90,scale=0.2]{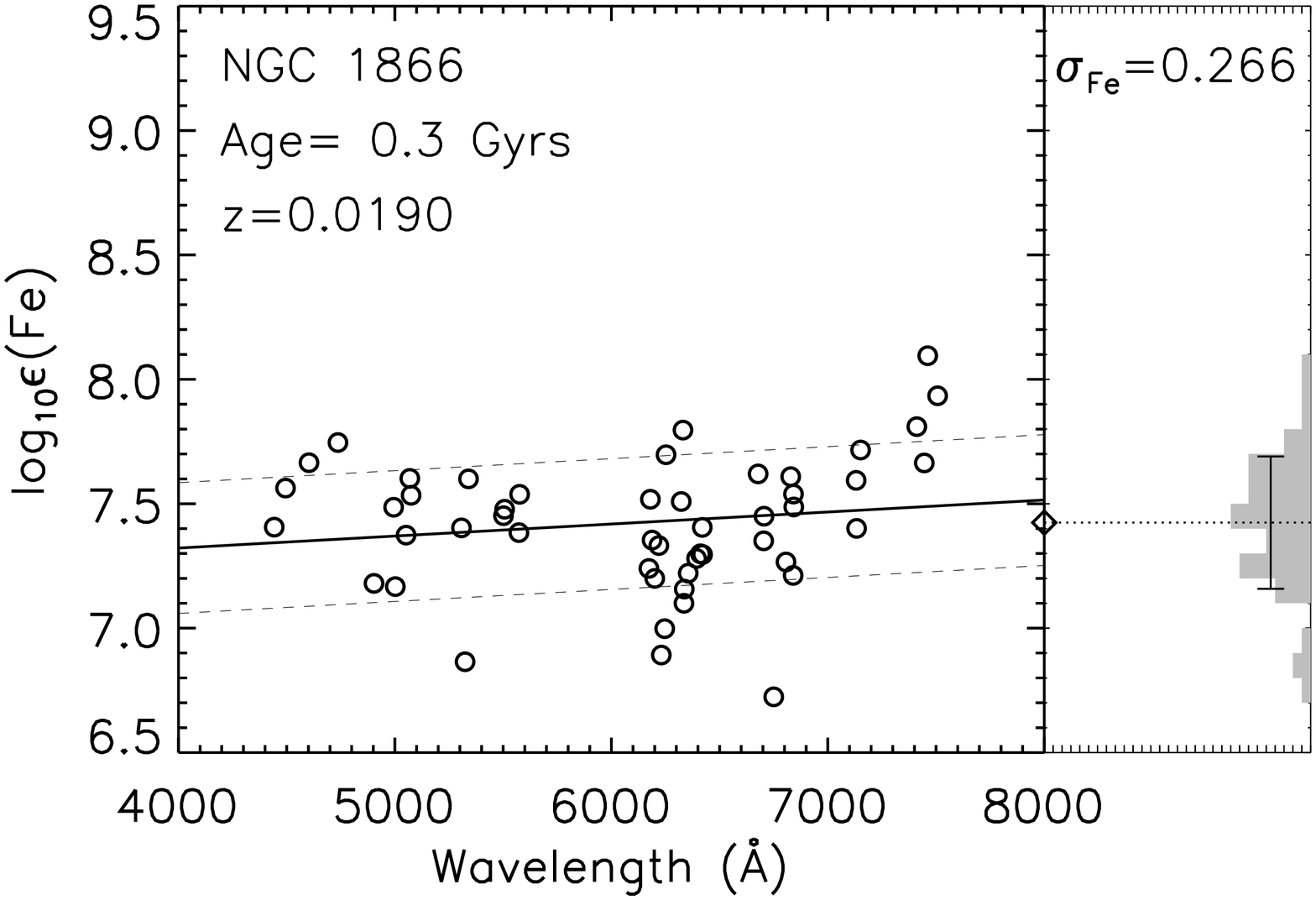}
\includegraphics[angle=90,scale=0.2]{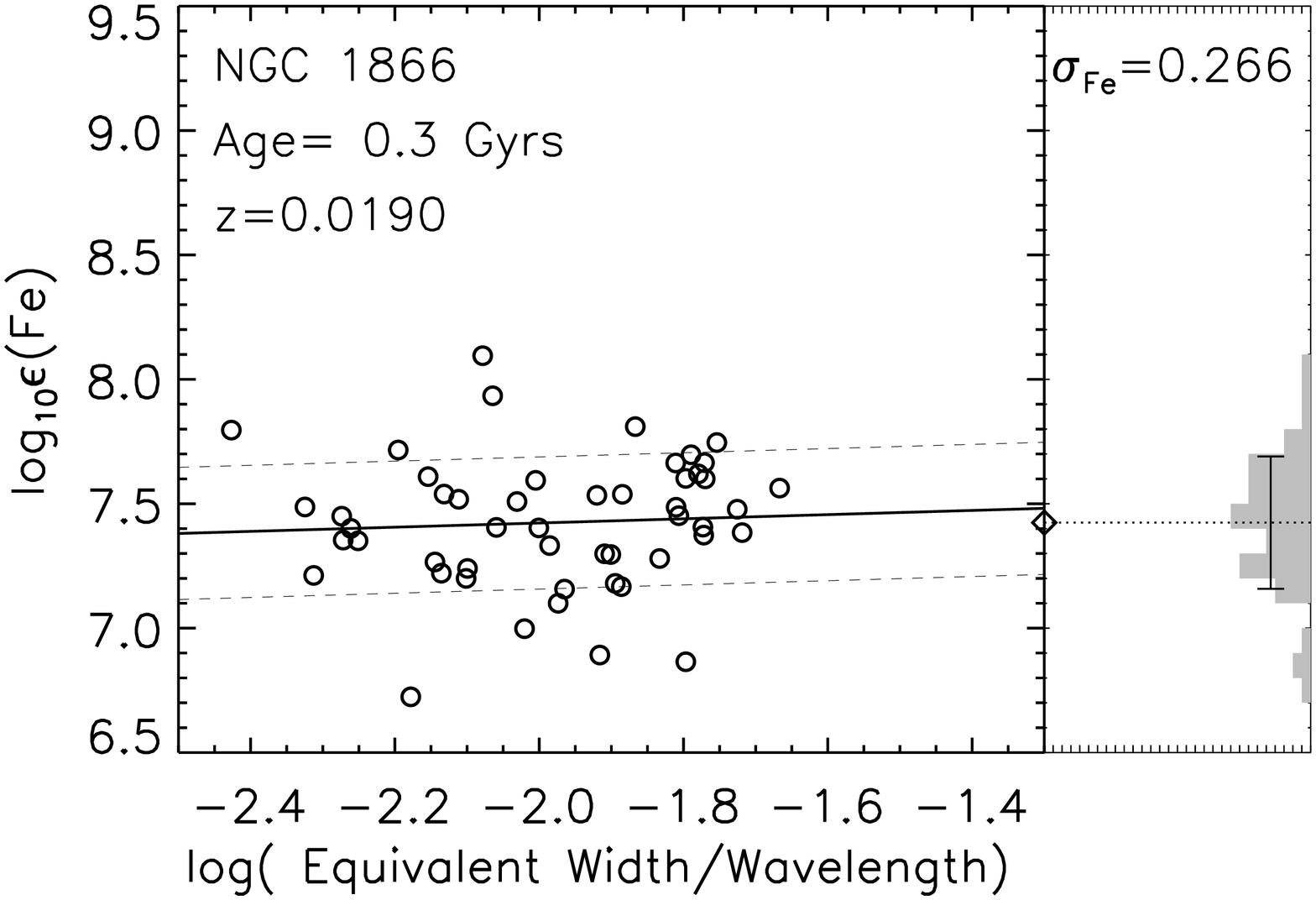}
\includegraphics[angle=90,scale=0.2]{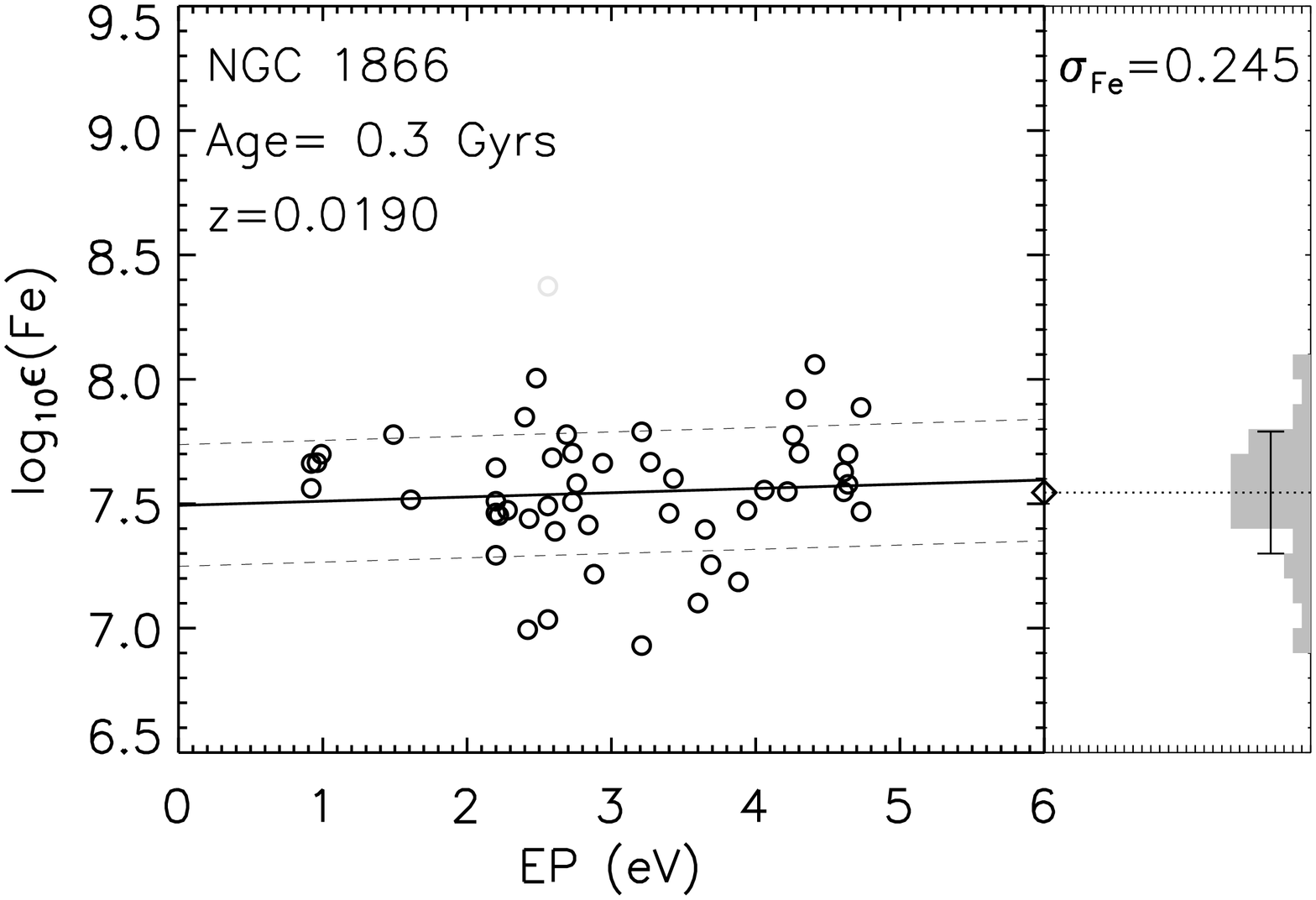}
\includegraphics[angle=90,scale=0.2]{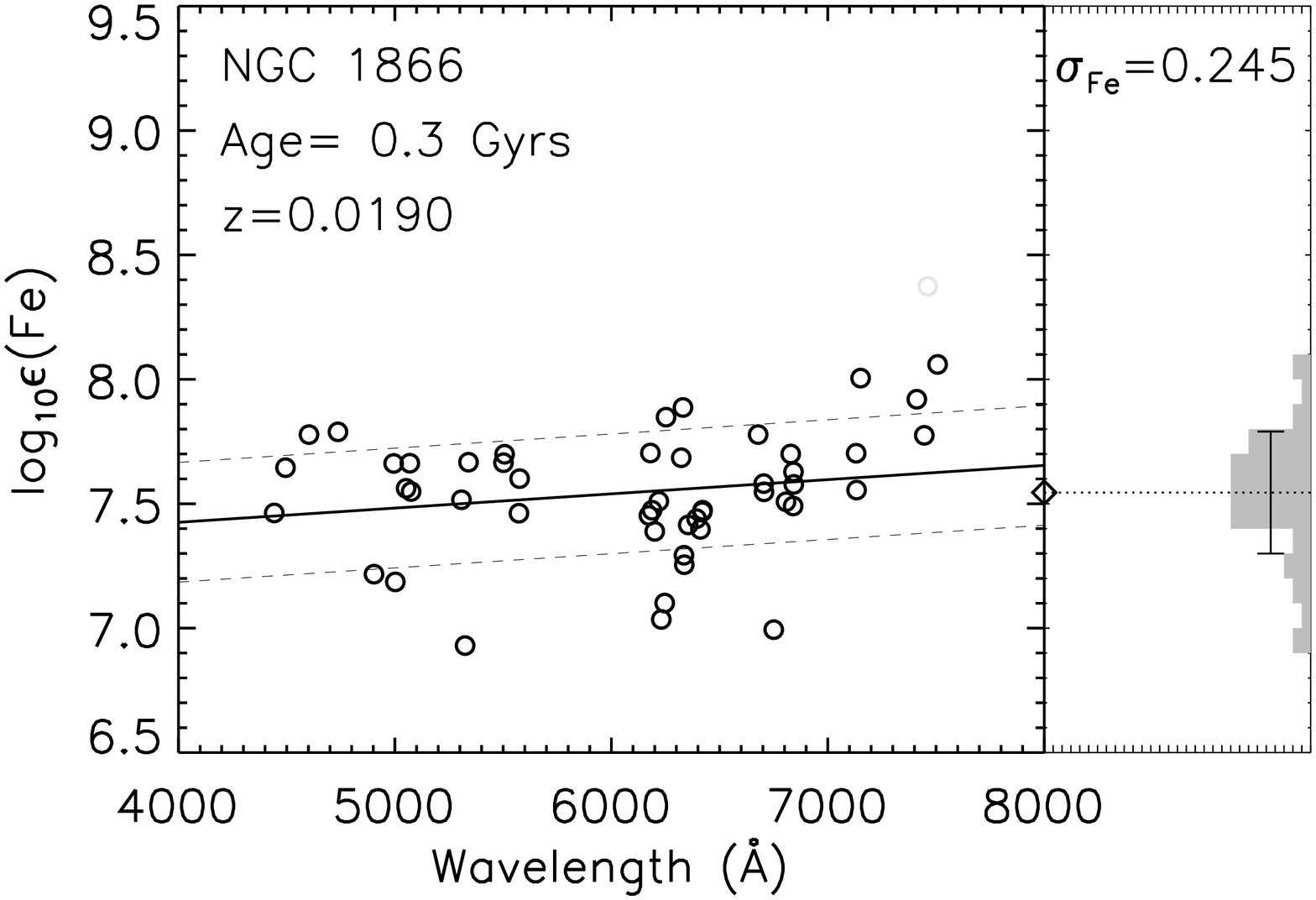}
\includegraphics[angle=90,scale=0.2]{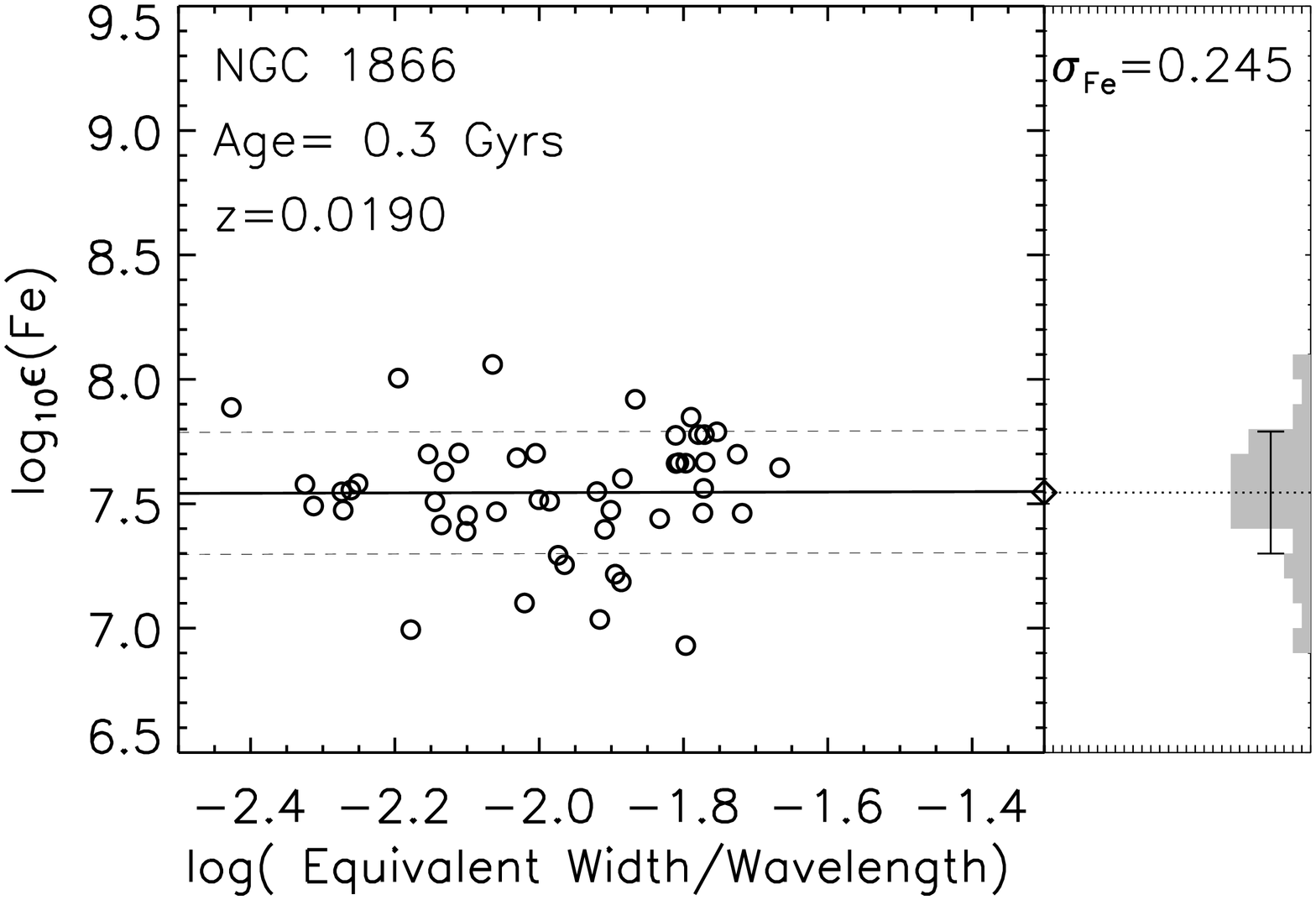}
\caption[Same as Figure~\ref{fig:2019 diagnostics} for NGC 1866]{Same as Figure~\ref{fig:2019 diagnostics} for NGC 1866.  The top panels correspond to the average CMD solution using a 300 Myr, [Fe/H]=$0$ isochrone, and the bottom panels correspond to the best-fitting CMD realization using a 300 Gyr, [Fe/H]=$0$ isochrone. The solution in the bottom panels has a smaller $\sigma_{\rm{Fe}}$. }
\label{fig:1866 diagnostics} 
\end{figure*}

\begin{figure}
\centering
\includegraphics[scale=0.5]{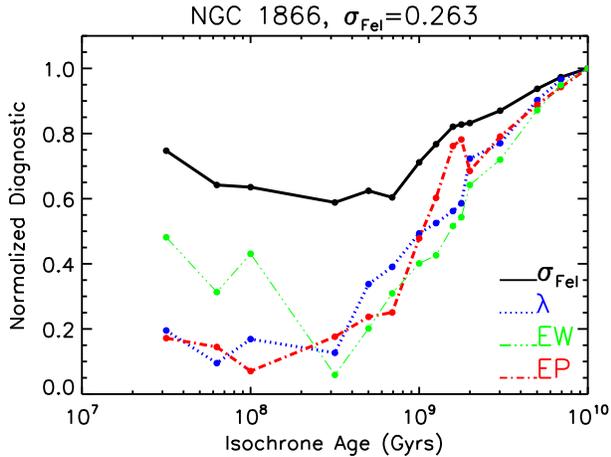}
\caption[Same as Figure~\ref{fig:1866 diag}, except the solutions at ages of 100$-$ 300 Myr have been replaced by the solutions for the best-fitting CMD realization at these ages.]{Same as Figure~\ref{fig:1866 diag}, except the solutions at ages of 100$-$300 Myr have been replaced by the solutions for the best-fitting CMD realization at these ages. Best solutions are for ages of 100-300 Myr. }\label{fig:1866 diagnew} 
\end{figure}

 Fluctuations in \bvo~ tend to be larger at lower metallicities (see Figure~\ref{fig:1866color}), especially for clusters with young ages. Although it appears that NGC 1866 is massive enough that  incomplete sampling is not a major issue,  it is important to establish if acceptable CMD realizations exist at lower metallicities and whether they result in more stable [Fe/H] solutions.   For this test, we create  CMD realizations with ages of 100$-$300 Myr, and [Fe/H]=$-1.5$ to $+0.2$.  
In determining a range in \bvo~ for the self-consistent CMD realizations, we note that NGC 1866 is well-studied and there are several estimates of the $E(B-V)$ that are consistently in the range of 0.06 to 0.13.    The \bvo~ requirement that we use includes this range of $E(B-V)$, as shown by the shaded gray region in   Figure~\ref{fig:1866color}.  

As  found for the intermediate age clusters, for NGC 1866  the spread in \bvo~ leads to a large spread in [Fe/H]$_{\rm{cluster}}$.   The solutions range from [Fe/H]=$-1.7$ to $>+0.5$.  We show the results for the CMD realizations with ages of 100, 150, and 300 Myr in Figure~\ref{fig:1866Fe plots all}; compared to the original average CMD solutions for reference.  Out of the large number of CMD realizations,  only $\sim$50 result in self-consistent solutions and satisfy the \bvo~ requirement  (see those falling on the diagonal line in Figure~\ref{fig:1866Fe plots
  all}).  Because so many of the possible CMD realizations do not satisfy the [Fe/H] self-consistency requirement, 
 we can conclude that for young clusters our  analysis method is  very successful in picking out a small range of viable  CMD realizations from a large set of CMD possibilities, just as discussed for the intermediate age clusters in \textsection~\ref{sec:int}.   This is due to the sensitivity of the Fe line EWs to the temperature of the luminous supergiants.

From the $\sim$50 possible self-consistent CMD realizations, we pick one best-fitting CMD realization for each age by minimizing the Fe I line diagnostics.   We find that solutions with [Fe/H]$\sim$0 are the most stable solutions overall, just as was found from the original average CMD solutions. These solutions have a smaller   $\sigma_{\rm{Fe}}$  than the original average CMD solutions, and small improvements in the EP, wavelength, and reduced EW diagnostics.  A comparison of the individual diagnostics for the 300 Myr average CMD solution and the best-fitting 300 Myr CMD realization is shown for reference in Figure~\ref{fig:1866 diagnostics}.   In Figure~\ref{fig:1866 diagnew} we show the normalized diagnostics for the solutions with ages of 100$-$500 Myr compared to the average CMD solutions at other ages. From Figure~\ref{fig:1866 diagnew} we derive a final age constraint for NGC 1866 of 100$-$300 Myr.  The range of [Fe/H] for the 100$-$300 Myr solutions results in a mean of  [Fe/H]=$+0.05$ and $\sigma_{\rm{age}}=0.06$, which is comparable to the standard error in the mean of the Fe I abundance.

\subsubsection{NGC 1711}
\label{sec:1711}

NGC 1711 is estimated to have an age of 50 Myr  \citep{2000A&A...360..133D}, and is less massive than NGC 1866 ( \mv$-8.3$).    Our analysis of NGC 1711 is hampered by low S/N spectra and poor sampling ($\sim$16\%).  We are only able to measure $\sim$25 reasonably clean Fe I lines, and find a  high line-to-line scatter of $\sigma_{\rm{Fe}}\sim$ 0.45.  These difficulties mean that the constraints we can make from our abundance analysis are limited, but we find the general trends are consistent with those of NGC 1866.

\begin{figure}
\centering
\includegraphics[angle=90,scale=0.35]{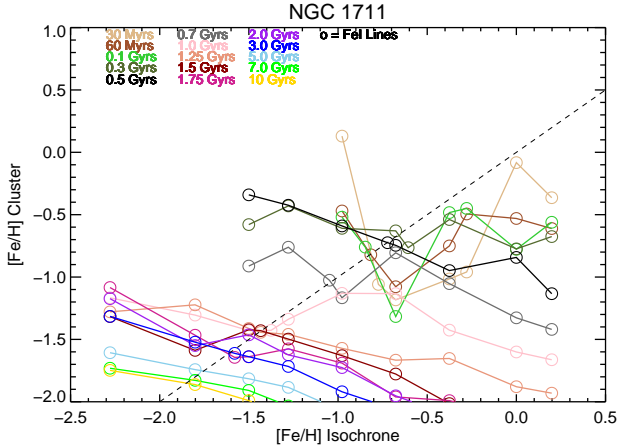}
\caption{Same as Figure~\ref{fig:2019 Fe plot} for NGC 1711.}
\label{fig:Fe 1711} 
\end{figure}

We find that the self-consistent [Fe/H] solutions appear at [Fe/H]$\sim-0.9$ for synthetic CMDs with ages $<$300 Myrs, as  shown in Figure~\ref{fig:Fe 1711}.  These solutions fall into the region of \bvo-metallicity space that is very sensitive to the evolution of cool supergiants.  These solutions also imply that NGC 1711 is reasonably metal-poor for its young age, although the high scatter in the Fe I abundance solution must be kept in mind.  However, there is also some indication from Figure~\ref{fig:Fe 1711} that the 30 Myr CMD solutions near solar metallicity are also close to a self-consistent [Fe/H].  The large jumps in output [Fe/H]  indicate that sampling uncertainties are having an impact on the  synthetic CMDs and the [Fe/H] solution, as well.

\begin{figure}
\centering
\includegraphics[scale=0.5]{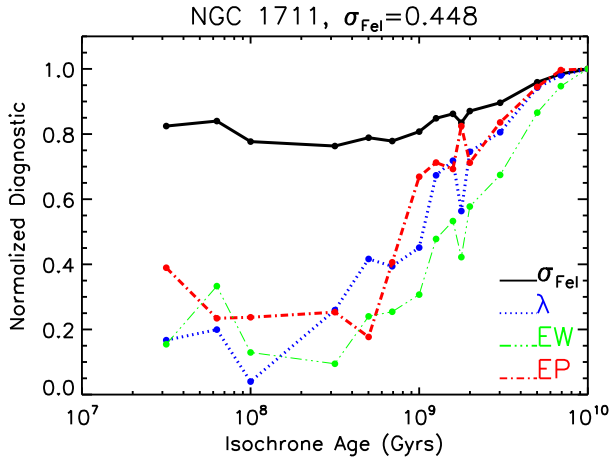}
\caption[Same as Figure~\ref{fig:2019 diag} for NGC 1711]{Same as Figure~\ref{fig:2019 diag} for NGC 1711. Best solutions are for ages of $<$300 Myr.}
\label{fig:1711trends2} 
\end{figure}

Although the [Fe/H] solutions are not ideal in terms of absolute quality, we are still able to evaluate the relative quality of the solutions for different ages.     We find that, like NGC 1866, the normalized diagnostics for the self-consistent solutions for each age  clearly suggest the best solutions are at the youngest ages, as seen in Figure~\ref{fig:1711trends2}.  For NGC 1711, we find that the solutions imply an age $<$300 Myr.     This is further evidence that the Fe lines are powerful diagnostics for determining ages of young clusters, and it suggests that with high quality data our abundance analysis method could be used to put tighter constraints on age and [Fe/H] of $\sim$50 Myr clusters that are sufficiently massive and well-sampled.

\begin{figure}
\centering
\includegraphics[angle=90,scale=0.35]{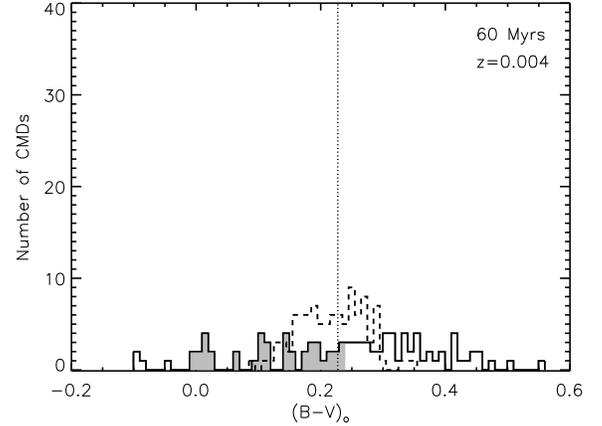}
\caption[Histogram of integrated \bvo~ color for 100 CMD realizations of a 60 Myr,z=0.004 (Fe/H=$-0.66$)  isochrone]{Histogram of integrated \bvo~ color for 100 CMD realizations of a 60 Myr, z=0.004 ([Fe/H]=$-0.66$) isochrone.  Solid black line shows the histogram for a population where the total flux in stars has been normalized to 16\% of a \mv$=-8.3$  cluster, which is appropriate for our integrated light spectrum of NGC 1711.  Dashed black line shows the histogram for a population normalized to 100\% of a \mv$=-8.3$ cluster.  CMDs with \bvo~ color consistent with the observed, reddening-corrected $B-V$ of NGC 1711 are shaded in gray. The range in \bvo~ for the CMD realizations is very large, demonstrating that sampling uncertainties are a big concern.  }\label{fig:1711histo} 
\end{figure}

We repeat the exercise of creating CMD realizations covering the range of interest in age and [Fe/H], primarily to evaluate if we can see evidence for better solutions at more metal-rich abundances.  For a comparison to the older clusters in the training set, we show the \bvo~ histogram of 100 CMD realizations created for a \mv$=-8.3$, 60 Myr cluster, like NGC 1711, by the dashed lines in  Figure~\ref{fig:1711histo}.  We note that the \bvo~ scale in Figure~\ref{fig:1711histo} is bigger than for the histograms shown for other clusters in this work, because the range in \bvo~ for the CMD realizations is  larger.  This is most apparent from the solid line histogram, which is made from CMD realizations normalized to the total flux of  16\% of a \mv$-8.3$ cluster, which corresponds to the fraction of NGC 1711 we have observed.  The range of \bvo~ for the CMD realizations in this histogram is $\sim$0.6 mag, significantly higher than the $\sim$0.1 mag discussed for NGC 1866.  Figure~\ref{fig:1711histo} underscores the fact that it is crucial  that young clusters are  very luminous and well-sampled when observed for integrated light analyses \citep[e.g.][]{1999A&AS..136...65B}.

\begin{figure*}
\centering
\includegraphics[angle=90,scale=0.2]{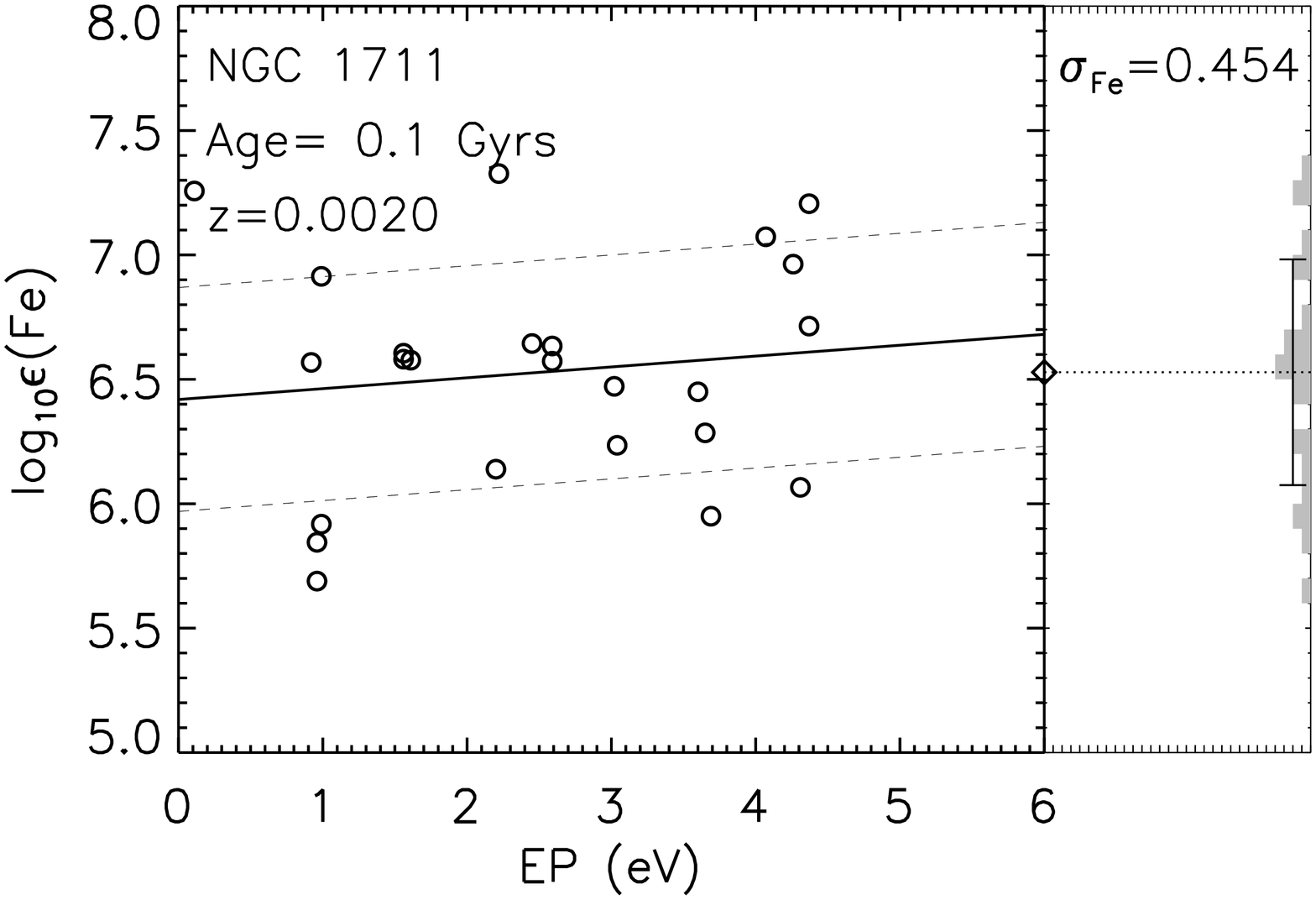}
\includegraphics[angle=90,scale=0.2]{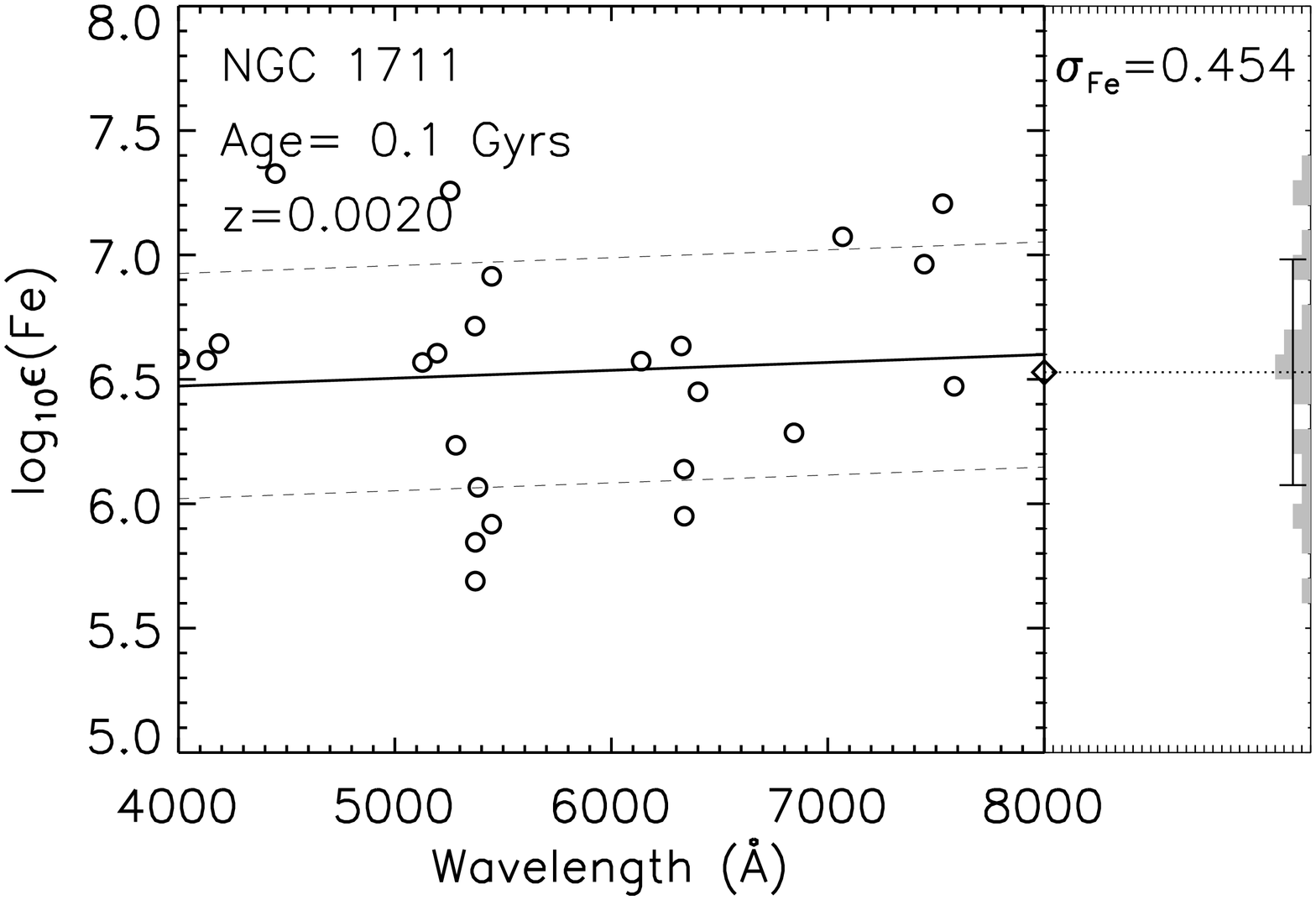}
\includegraphics[angle=90,scale=0.2]{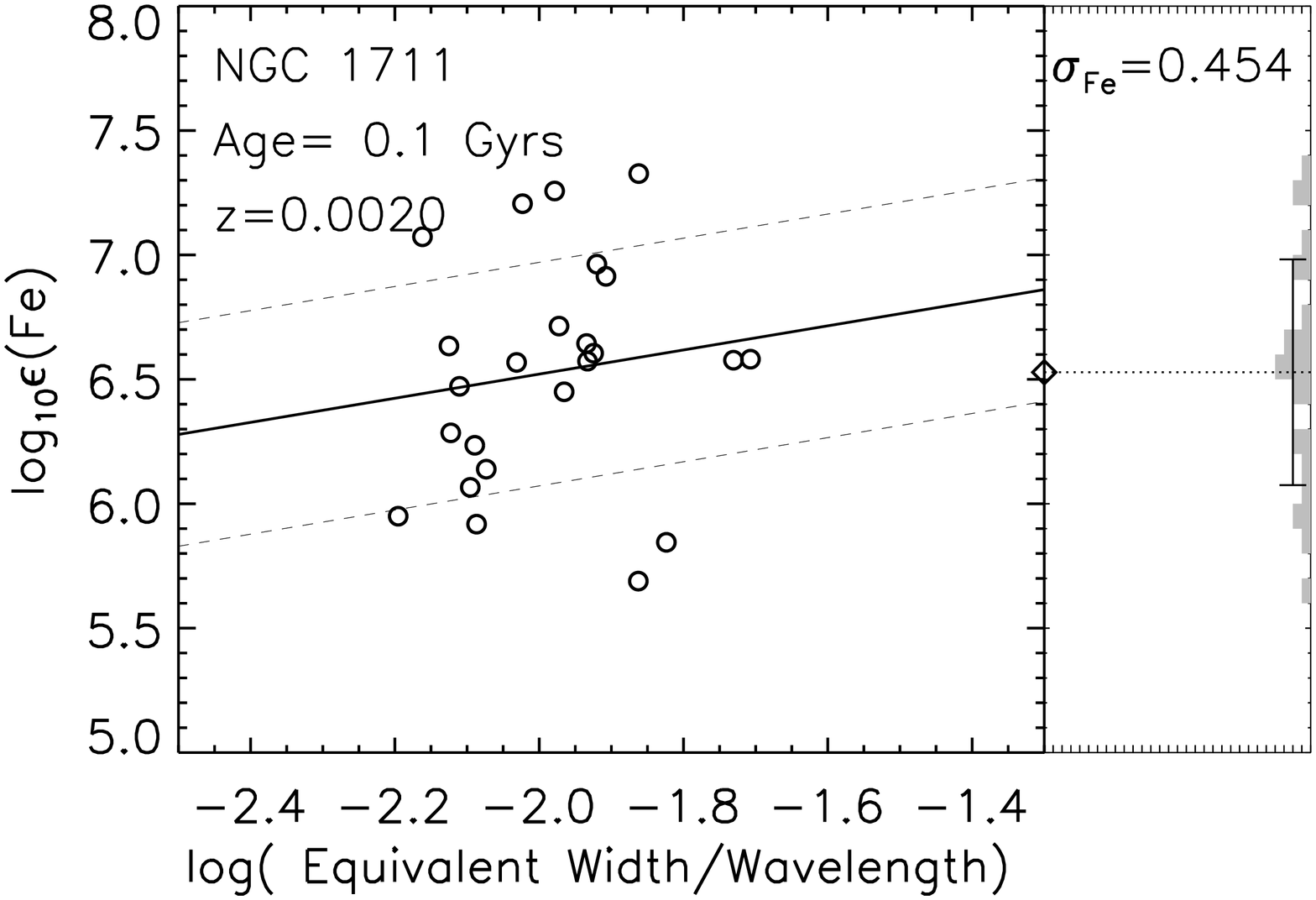}
\caption{Same as Figure~\ref{fig:2019 diagnostics} for NGC 1711.  The best CMD realization using a 60 Myr, [Fe/H]=$-0.96$ isochrone.  }
\label{fig:1711 diagnostics} 
\end{figure*}

 We next explore the range of solutions that can be obtained from the CMD realizations for NGC 1711.   We allow the subset of CMD realizations used for abundance analysis to have a wide range in color ( 0.0$<$\bvo$<$ 0.3), due to large uncertainties in the observed $B-V$ and $E(B-V$) of  \cite{2008MNRAS.385.1535P} and \cite{2000A&A...360..133D}.  Because the stellar populations change rapidly at the youngest ages,  we also refine the age grid for the  CMD realizations to include additional ages of 40, 50, and 80 Myr.
While in general we do not find that the self-consistent CMD realizations offer significant improvements over the average CMD solutions, it is clearer that the best solutions are for ages of 60-300 Myr.  
The solutions at the best fit ages cover a range in [Fe/H] of $-0.66$ to $-0.97$, so we quote an abundance of [Fe/H]=$-0.82$ with an uncertainty due to the assumed age of $\sigma_{\rm{age}}$= 0.15 dex.  The individual diagnostics for the best-fitting 60 Myr CMD realization are shown in Figure~\ref{fig:1711 diagnostics}, where the large line-to-line scatter can be fully appreciated.

 In conclusion, although the [Fe/H] solution for NGC 1711 has significant uncertainties, the general trends from the analysis are  very similar to that of NGC 1866.  This suggests that in the future it may be possible to obtain even tighter constraints for $\sim$50 Myr clusters that are more massive and better sampled than NGC 1711.

\subsubsection{NGC 2100}
\label{sec:2100}

\begin{figure}
\centering
\includegraphics[angle=90,scale=0.35]{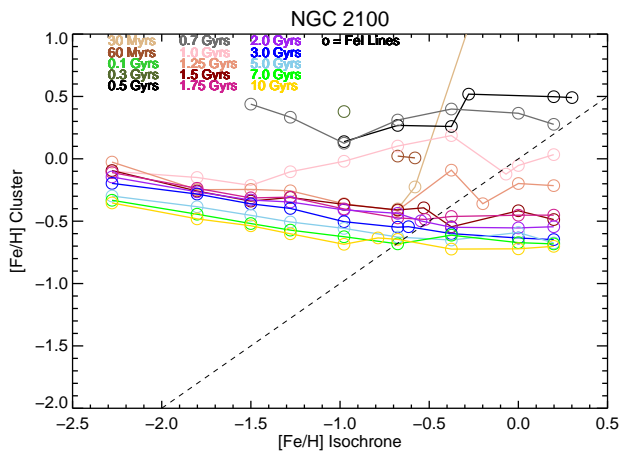}
\caption{Same as Figure~\ref{fig:2019 Fe plot} for NGC 2100.}
\label{fig:Fe 2100} 
\end{figure}

The youngest cluster in the LMC training set for which we can constrain abundances is NGC 2100, which is estimated to have an age of 15 Myr  \citep{1991ApJS...76..185E}, and \mv$\sim-8.8$.  Our best solution is from 33 clean Fe I lines and has an uncertainty of  $\sigma_{\rm{Fe}}$=0.32, which is lower than the line-to-line scatter for NGC 1711 but larger than for NGC 1866.

NGC 2100 is sufficiently young and metal-rich that we are unable  to find solutions that converge at self-consistent [Fe/H] for the average CMDs with ages between 30 Myr$-$300 Myr.  The average CMD solutions for NGC 2100 are shown in Figure~\ref{fig:Fe 2100}.  Although the average CMD solutions do not converge at self-consistent [Fe/H],  there is some indication that there may be possible solutions for an age of 30 Myr and [Fe/H]$>-0.7$, given the spread we expect from incomplete sampling effects.  To determine if self-consistent solutions exist, we create CMD realizations for  ages of 30, 40, and 100 Myr and $-0.66 <$[Fe/H]$<+0.0$ in the sampling uncertainty tests.  We use an additional  color constraint of $-0.13 <$ \bvo $<0.22$, which covers the range in observed $B-V$ and $E(B-V)$ from Table~\ref{tab:lmctable}.    

\begin{figure}
\centering
\includegraphics[scale=0.5]{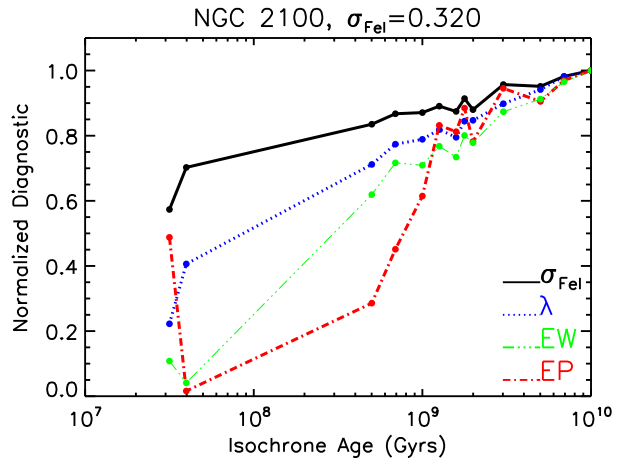}
\caption{Same as Figure~\ref{fig:2019 diag} for NGC 2100. Best solutions are for ages of $<$40 Myr.}
\label{fig:2100trends} 
\end{figure}

\begin{figure*}
\centering
\includegraphics[angle=90,scale=0.2]{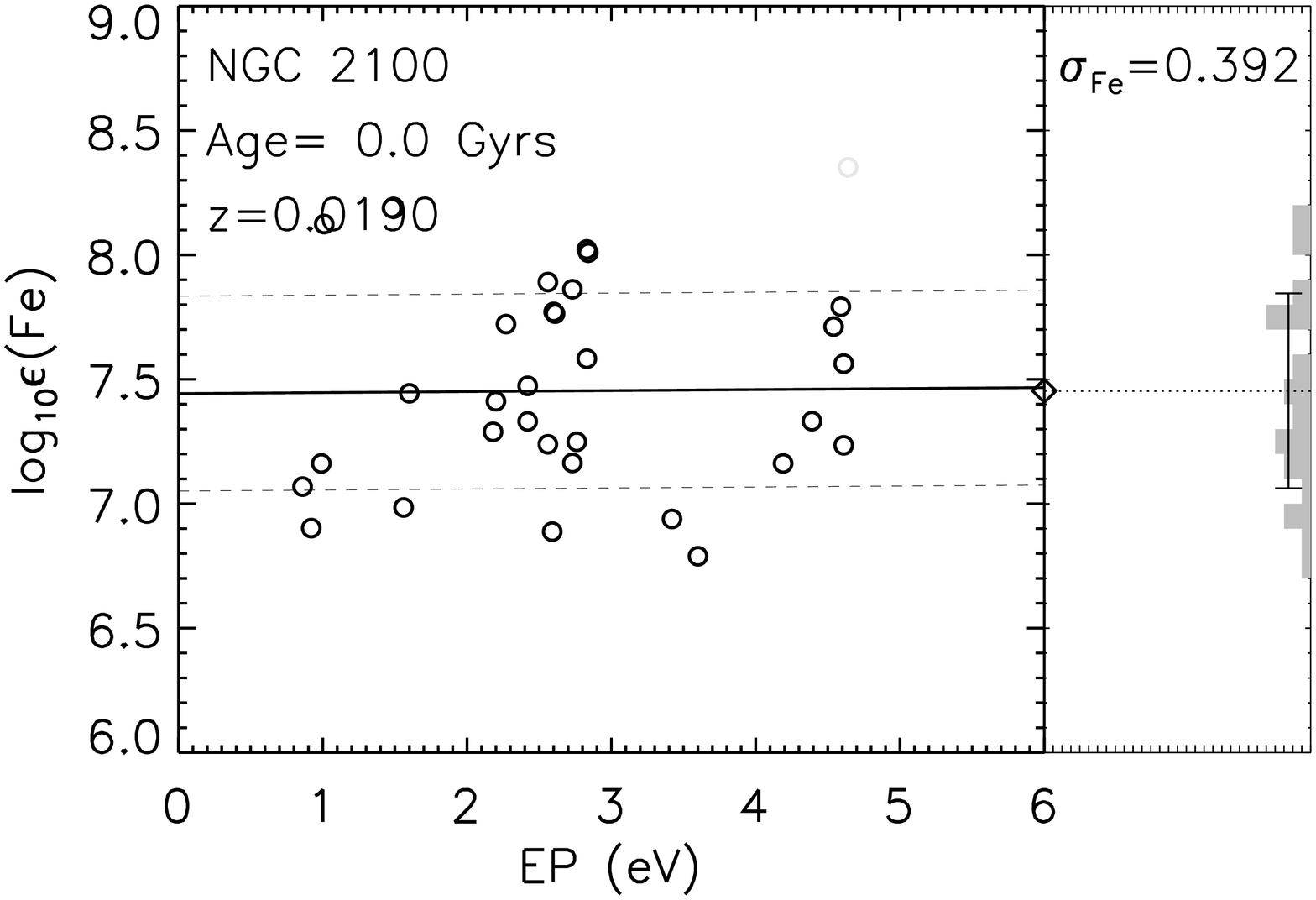}
\includegraphics[angle=90,scale=0.2]{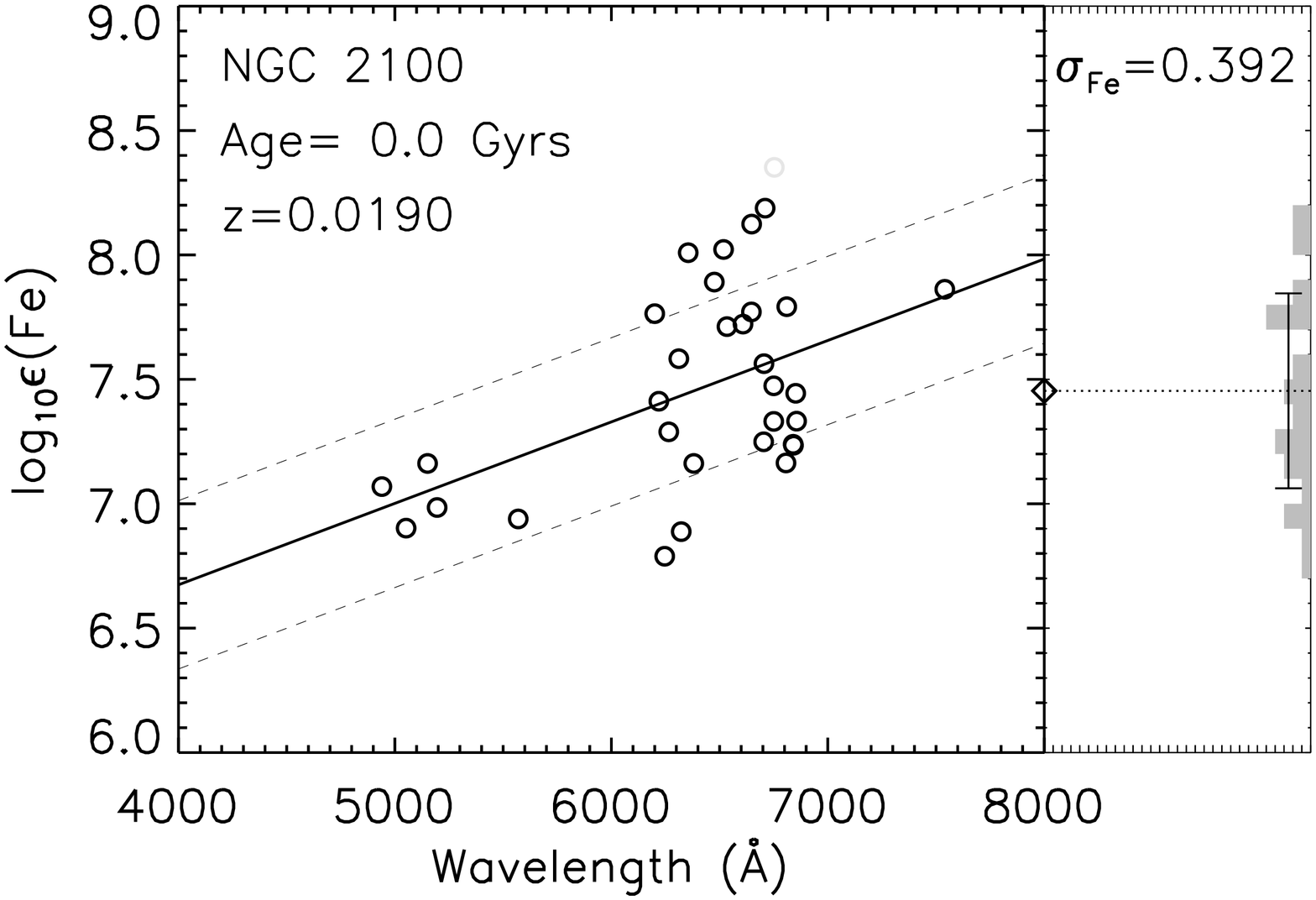}
\includegraphics[angle=90,scale=0.2]{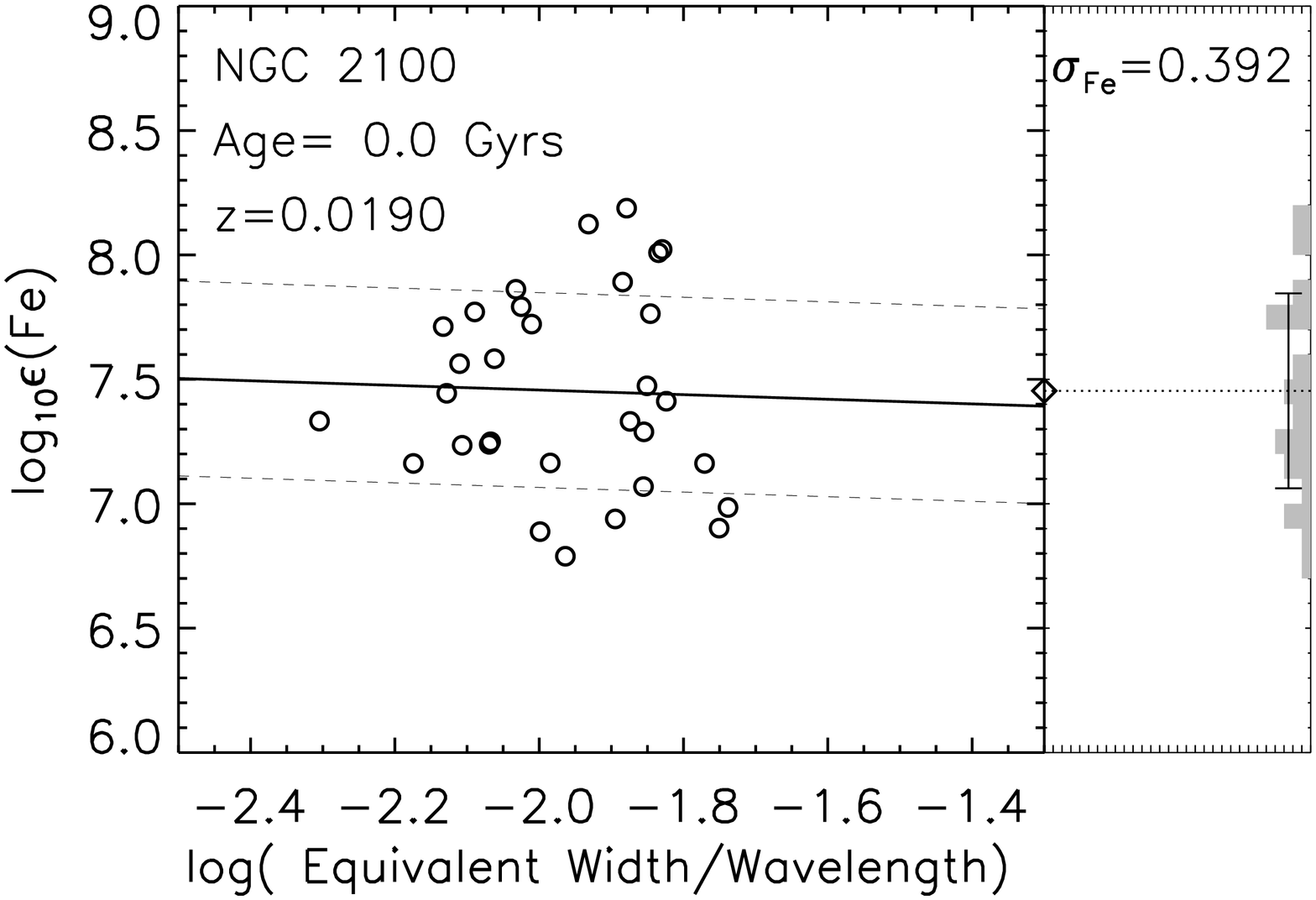}
\caption{Same as Figure~\ref{fig:2019 diagnostics} for NGC 2100.  The best CMD realization using a 40 Myr, [Fe/H]=$0$ isochrone.  }
\label{fig:2100 diagnostics} 
\end{figure*}

We find that a handful of CMD realizations with [Fe/H]$\sim0$ result in self-consistent solutions, and that all of the self-consistent solutions have  ages $<$40 Myr.   These solutions have diagnostics that are significantly better than any  self-consistent solutions with ages $>$500 Myr, as can be seen in Figure~\ref{fig:2100trends}.    The individual diagnostics for the best-fitting  40 Myr, [Fe/H]=$0$ CMD realization are shown in Figure~\ref{fig:2100 diagnostics}. This solution has a negligible dependence of [Fe/H] with EP, a small dependence of [Fe/H] with reduced EW, and a significant dependence of [Fe/H] with wavelength.

 In summary, even though our grid of synthetic CMDs does not probe ages as young as that determined for NGC 2100 from resolved photometry,  we are in fact able to put an upper limit on the age of NGC 2100 of 40 Myr using the Fe lines and our abundance analysis method.    Our final abundance constraint for NGC 2100 is [Fe/H]$=-0.03$; this abundance can be considered a lower limit given our age constraint.  

The analysis of all three young clusters in the LMC training set demonstrates that using our abundance analysis method, we can obtain very tight constraints on the age and abundance of  massive clusters with ages of  of $\sim50$ to 500 Myr, and limits on the age and abundance for clusters as young as 10 Myr.   We see evidence for the significant impact of luminous supergiants on both the integrated \bvo~ colors and the derived [Fe/H] abundances.   As was the case for the intermediate age clusters, for the young clusters  it is the sensitivity of the [Fe/H] solution to luminous, cool stars that  facilitates strong constraints on age and metallicity from high resolution integrated light spectra. By allowing for statistical fluctuations in the synthetic CMDs, we are able to reduce the $\sigma_{\rm{Fe}}$ of the original solutions and improve the [Fe/H] stability with EP, wavelength, and reduced EW.  Moreover, we find that  with high quality data, the self-consistency and stability of the [Fe/H] solution can break the \bvo-metallicity degeneracy seen in Figure~\ref{fig:1866color}.

\subsection{Summary of Results }
\label{sec:results}

We have measured the ages and determine [Fe/H] abundances  for eight of the nine clusters in the LMC training set (NGC 2019, 2005, 1916, 1718, 1978, 1866, 1711, and 2100), as listed in Table~\ref{tab:lmcresults}. With this sample of clusters, we have demonstrated the ability to measure both a wide range in metallicity ($-1.5 < $[Fe/H]$ < +0.0$) and age (40 Myr to 12 Gyr)  using a single abundance analysis technique.    Four of these clusters  (NGC 1916, 1718, 1711, and 2100) have no prior measurements of   [Fe/H] from high S/N, high resolution spectroscopy.

In the analysis of each cluster, we have discussed two sources of  uncertainty in our measurement of [Fe/H].  The first is the line-to-line statistical scatter in the abundance of individual Fe lines, $\sigma_{\rm{Fe}}$, and the second is the uncertainty in [Fe/H] due to the assumed age, $\sigma_{\rm{age}}$.    For a final uncertainty for each cluster, $\sigma_{\rm{total}}$, we add the standard error in the mean of the Fe lines and the age uncertainty in quadrature, which we have listed in column 7 of Table \ref{tab:lmcresults}.  The standard error in the mean of the Fe lines is defined as $\sigma_{\rm{Lines}}$= $\sigma_{\rm{Fe}}$/(N$_{\rm{Lines}}$-1), where N$_{\rm{Lines}}$, the number of measured Fe I lines for each cluster, is listed in column 3 of Table \ref{tab:lmcresults}.  An evaluation of systematic offsets is presented in \textsection \ref{sec:uncertainties}.

\begin{deluxetable*}{rrrrcccc}
\tablecolumns{8}
\tablewidth{0pc}
\tablecaption{IL spectra Age and Abundance Results \label{tab:lmcresults}}
\tablehead{
\colhead{Cluster}  &\colhead{Age (Gyrs)} & \colhead{[Fe/H]\tablenotemark{a}} &
\colhead{N$_{\rm{Lines}}$\tablenotemark{b}} &
\colhead{$\sigma_{\rm{Lines}}$\tablenotemark{c}} 
& \colhead{$\sigma_{\rm{Age}}$}& \colhead{$\sigma_{\rm{
{total}}}$\tablenotemark{d}} & \colhead{$\alpha_{\rm{offset}}$\tablenotemark{e}}}
\startdata

NGC 2019 &     $>$7 &     $-$1.61	& 49   & 0.03	& 0.04	& 0.05  	& $-$0.03	\\
NGC 2005  &   $>$5   &     $-$1.50 &  34	&  0.04	& 0.05	& 0.06  	& $-$0.02\\
NGC 1916  &   $>$5    &      $-$1.48&  50	  & 0.04	& 0.07	& 0.08  	& $<$0.01\\
NGC 1718 &    1.0-2.5   &   $-$0.64&  69	   &0.04	& 0.25	& 0.25  	& $+$0.15\\
NGC 1978 &    1.5-2.5  &    $-$0.74& 36	  & 0.06	& 0.05	& 0.08  	& $+$0.03\\

NGC 1866  &  0.10-0.30  &   $+$0.04&49	    & 0.04   & 0.02	& 0.04  	& $-$0.04\\
NGC 1711  &  0.06-0.30  &   $-$0.82 &	25    &    0.09	& 0.15	& 0.17  	& $+$0.04 \\
NGC 2100  &  $<$0.04  &     $>$$-$0.03& 33	 &   0.06	& \nodata	& 0.06  	& \nodata \\
\enddata

\tablecomments{ a.  The [Fe/H] value is the average value of the solutions with ages listed in column 2, and uncorrected for the $\alpha_{
\rm{offset}}$ in column 8.   b. N$_{
m{Lines}}$ corresponds to the number of Fe I lines used in the abundance analysis for each cluster.  c. The standard error in the mean of the [Fe/H]
  measurement is defined as $\sigma_{
\rm{Lines}}$ =
  $\sigma_{\rm{Fe}}$/(N$_{
\rm{Lines}}$-1).    d.  The total statistical error is
  defined as $\sigma_{
\rm{
{total}}}$= $\sqrt{\sigma_{
\rm{Lines}}^2 +
    \sigma_{
\rm{Age}}^2 }$.   e.  $\alpha_{
\rm{offset}}$ is a systematic error due
  to a difference in the $\alpha$-enhancement of the Teramo isochrones
and the true $\alpha$-enhancement of the cluster, as discussed in
\textsection \ref{sec:uncertainties}.}

\end{deluxetable*}

\section{Discussion}
\label{sec:discussion}

In the first part of this section,  we discuss additional sources of uncertainty in our analysis due to the isochrones used to construct synthetic CMDs.  Next, we compare the results of the IL spectra analysis method to cluster 
 properties  previously determined with other methods.  In \textsection~\ref{sec:age and met}, we discuss the age-metallicity relationship derived for our training set LMC clusters.

\subsection{Other Sources of Uncertainty: Choice of Isochrones}
\label{sec:uncertainties}

An important additional source of uncertainty is, of course, the physics and assumptions that went into the isochrones.  For example, we determine whether to use scaled-solar or $\alpha$-enhanced Teramo isochrones on a cluster-by-cluster basis, after evaluating each clusters' Ca, Si and Ti abundances (presented in \citetalias{paper4}).  In principle, the level of   $\alpha$-enhancement can have an effect on the derived age and [Fe/H], because $\alpha$-enhanced stars have higher opacities, resulting in slightly stronger metal lines \citep{1993ApJ...414..580S,2006ApJ...642..797P}.
Because the Teramo isochrones only come in two extremes --- [$\alpha$/Fe]=$+0.4$ or $+0.0$ --- a potential concern is that our [Fe/H] and age solutions could be affected by a mismatch in the  $\alpha$-enhancement of the isochrone and the true $\alpha$-enhancement of the cluster. 

Fortunately, because we explicitly measure the [$\alpha$/Fe] ratio of each cluster,  we can estimate the systematic offset in [Fe/H]  that is due to a mismatch between the Teramo isochrones and the true [$\alpha$/Fe].  To estimate the systematic offset we derive solutions for our clusters with both of the values of the Teramo isochrones, and then interpolate an offset for each cluster based on the measured [$\alpha$/Fe].     The final systematic error for each cluster, $\alpha_{\rm{offset}}$,  is listed in the last column of Table \ref{tab:lmcresults}.

 We find that for the old clusters, the systematic offset is small, and is $\leq0.03$ dex in all three cases. This offset is always smaller than the statistical uncertainty in [Fe/H] due to the line-to-line scatter or assumed age of the older clusters.   For the younger clusters,  we also derive small errors of $\sim$0.04 dex for all of the clusters with the exception of NGC 1718, for which we find an offset of $+0.15$ dex.   This is not unexpected because this cluster already has a large uncertainty in the [Fe/H] value due to the assumed age.   We are unable to derive an offset for the case of NGC 2100, because the $\alpha$-enhanced  isochrone solutions do not converge, as in most cases for the solar isochrones for this young cluster.

 We can also evaluate whether the choice of scaled-solar or $\alpha$-enhanced isochrones adds a significant uncertainty to our age constraints.  In all cases, we find that our original age constraints are conservative enough that the choice of isochrones does not change our final solutions.  This is, in fact, consistent with previous work on  the effect of $\alpha$-enhancement on age measurements using  low resolution IL line index techniques.  In this case, indexes with Balmer line absorption are typically used for age determinations,  and at the same age and [Fe/H] an $\alpha$-enhanced star  has weaker Balmer lines than a scaled-solar star.  This leads to younger inferred ages, which are typically different by    $\sim$3 Gyr for old populations and $\sim$1 Gyr for intermediate age populations \citep[e.g.][]{2007ApJS..171..146S}.      $1-3$ Gyr is well within our original constraints, moreover, metal lines are not nearly as sensitive to $\alpha$-enhancement  as the Balmer lines.  Therefore, an additional strength in our analysis when compared to low resolution line index methods is that  we can independently determine the age of a cluster with metal lines, without the additional complications of Balmer line age dating.

Another potential source of uncertainty in our choice of isochrones involves the treatment of convective overshooting in the cores of massive ($>1.1\msol$) stars.   This work is the first high resolution integrated light  abundance analysis where the treatment of convective overshooting is relevant, because for the first time we have a sample of young clusters containing massive stars.
In this paper we chose to use the non-canonical Teramo isochrones, which include convective overshooting, because they were found to best  reproduce the  high quality photometry of NGC 1978 \citep{2007AJ....133.2053M}.    However, after this paper was submitted, \cite{mucc1866} measured detailed abundances of individual stars in NGC 1866 (age $\sim$150 Myr), and found a lower [Fe/H] than what we have found in our IL analysis. Because the non-canonical isochrones we have used have higher temperatures than the canonical isochrones, in the case of NGC 1866 we would in fact derive higher [Fe/H] solutions at ages of 100 to 300 Myrs with the non-canonical tracks than with the canonical tracks.    If this is indeed the reason that our [Fe/H] for NGC 1866 is higher than that measured by   \cite{mucc1866}, then our analysis has extremely interesting implications for the long-standing convective overshooting debate in the literature \citep[e.g.][]{1999AJ....118.2839T,2002A&A...385..847B}.  Therefore, we have begun  preliminary tests to evaluate the effect of convective overshooting  on our [Fe/H] solutions, and present [Fe/H]  estimates using the canonical isochrones for the intermediate age and young clusters in Table \ref{tab:canvsnoncan}.   We will present the full analysis of this interesting topic and conclusions on the most appropriate level of convective overshooting in an upcoming paper.

\subsection{Previous Estimates of Cluster Properties}
\label{sec:photo}

 \begin{deluxetable}{lrr}
\centering
\tablecolumns{3}
\tablewidth{0pc}
\tablecaption{Comparison of Canonical and Non-Canonical Isochrones \label{tab:canvsnoncan}}
\tablehead{  &  \colhead{Non-Canonical} & \colhead{Canonical}  \\
\colhead{Cluster}  & \colhead{[Fe/H]}& \colhead{[Fe/H]}  }
\startdata

NGC 1718 &     $-$0.64 &$-$0.70\\
NGC 1978 &     $-$0.74 &$-$0.60\\
NGC 1866  &   $+$0.04  & $-$0.19\\
NGC 1711  &      $-$0.82 &  $-$0.54\\
NGC 2100  &         $>$$-$0.03 & $>+$0.03\\
\enddata
\end{deluxetable}

\begin{figure}
\centering
\includegraphics[angle=90,scale=0.5]{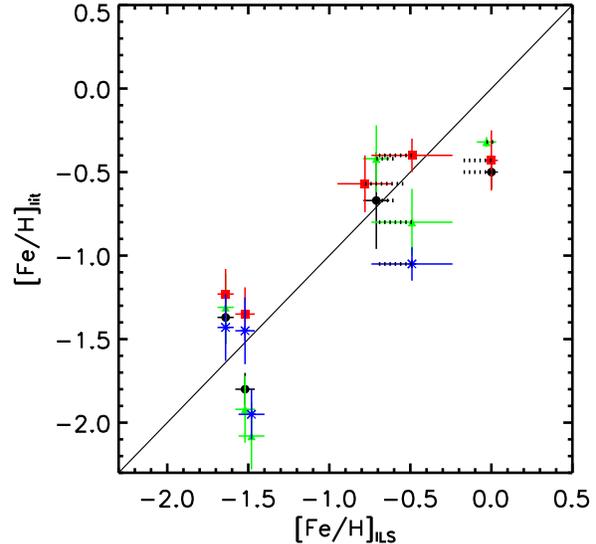}
\caption{Comparison of previous metallicity estimates to IL abundance analysis results.  Results from the literature for high resolution spectroscopy of individual stars, for low resolution CaT measurements, for photometric estimates, and for low resolution line index methods are shown as black circles, green triangles, red squares, and blue stars, respectively.  References for metallicities from the literature are discussed in \textsection~\ref{sec:photo}. Black dotted lines show the anticipated change in [Fe/H] when isochrones with more appropriate treatment of convective overshooting are used.}
\label{fig:fecomparison} 
\end{figure}

  Our goal in this section is to look for broad consistencies or inconsistencies with previous results from a variety of techniques: high resolution spectra of individual stars,  low resolution  Ca II triplet (Ca T) spectroscopy, CMDs, Str\"{o}mgren photometry, and low resolution line index methods. 
  We note that, unlike for the MW training set clusters, there is only a subset of clusters in the LMC training set (NGC 2019, NGC 2005, NGC 1978, and NGC 1866) for which comparisons can be made to abundances obtained from  high resolution analysis of individual stars.  However, even for these clusters  it is important to keep in mind that many systematic uncertainties can arise due to choice of analysis methods, and that
 detailed abundances of stars determined by different authors typically do not agree to
 better than $\pm0.1$ dex \citep[see discussions in][]{2004ARA&A..42..385G, 2003PASP..115..143K}. 
In addition to a handful of detailed abundances from individual stars, CMDs for all clusters are available in the literature, albeit of variable depth and quality.   These CMDs have been used to constrain ages and estimate metallicities for the clusters, just as is done for clusters in the MW.   Metallicity estimates  from low resolution Ca T spectroscopy or line index methods exist for most of the clusters as well.  Below, we summarize the previous estimates for ages and metallicities of the training set clusters compared to what we measure with high resolution abundance analysis.  In addition, we compare our high resolution IL [Fe/H] results with previous measurements in Figure \ref{fig:fecomparison}. Note that our IL abundances in Figure \ref{fig:fecomparison} have been corrected with the $\alpha_{\rm{offset}}$ derived for each cluster.  In addition, for the younger clusters, we show the anticipated shift in IL abundances  when an analysis  with a more appropriate treatment of convective overshooting is used.

{\it NGC 2019}. We obtain a measurement of [Fe/H]=$-1.64 \pm 0.05$, which includes the $\alpha_{\rm{offset}}$ as well as uncertainties due to line-to-line scatter and age.   We derive an age of $>$7 Gyr.    \cite{2006ApJ...640..801J} measured detailed abundances for a sample of 3 RGB stars in NGC 2019 and found   [FeI/H]=$-1.37 \pm 0.07$ and a significantly higher [FeII/H]$=-1.10 \pm 0.16$ from Fe II lines.  Reasons for the  offset in Fe I and Fe II abundances  have been discussed profusely in \cite{2003PASP..115..143K} and Paper I and II.
 \cite{1998MNRAS.300..665O}  estimate [Fe/H]=$-1.23 \pm 0.15$ from an HST WFPC2 CMD.  Using low resolution Ca T spectroscopy of individual stars, \citet{1991AJ....101..515O} estimate [Fe/H]=$-1.8 \pm 0.2$, while \cite{2006AJ....132.1630G} find  a more metal-rich [Fe/H]=$-1.31$.     
 \cite{2002MNRAS.336..168B}  find [Fe/H]=$-1.43 \pm 0.2$ for NGC 2019 using low resolution IL spectra. This wide range in measurements from $-1.8 < $[Fe/H]$ < -1.10$ is consistent with our measurement.  \cite{1998MNRAS.300..665O}  also derive an age of $>$12 Gyr  by comparison the HST CMD  to MW clusters and the location of the HB, consistent with our age constraint. 

{\it NGC 2005}.   We obtain a measurement of  [Fe/H]$=-1.52 \pm 0.06$, with an age constraint of $>$ 5 Gyr.  As was the case for NGC 2019, \cite{2006ApJ...640..801J} find significantly different abundances for Fe I and Fe II lines of  [FeI/H]$=-1.78 \pm 0.09$ and [FeII/H]$=-1.31 \pm 0.03$.  \cite{1998MNRAS.300..665O} find  [Fe/H]=$-1.35 \pm 0.16$ using an HST CMD.  The metallicity estimate of  \citet{1991AJ....101..515O} from low resolution Ca T spectroscopy is more metal-poor than the CMD estimate, at [Fe/H]=$-1.92 \pm 0.2$.   
The line index metallicity estimate  is [Fe/H]=$-1.45 \pm 0.2$ \citep{2002MNRAS.336..168B}.  The range of these measurements is $-1.92 <$ [Fe/H] $< -1.31$, and is consistent with our result.  \cite{1998MNRAS.300..665O} also  find  an old  age of $>$12 Gyr for NGC 2005.

{\it NGC 1916}.  We measure an abundance of   [Fe/H]=$-1.48 \pm 0.08$ and we are able to constrain the age  to $>$5 Gyr. \citet{1991AJ....101..515O} measured an abundance of [Fe/H]=$-2.08 \pm 0.2$ using Ca T spectroscopy.  \cite{2002MNRAS.336..168B}  estimate [Fe/H]=$-1.8$ to $-2.1$ using various  line indexes, which highlights the  importance of  high resolution abundances for GCs with moderate to low metallicities. 
 These estimates of the metallicity of NGC 1916  
 are significantly lower than our measurement, but the greater uncertainties in low resolution techniques at lower metallicities must be kept in mind  \citep[see discussion in][]{m31paper}.
 While an HST CMD for NGC 1916 exists in \cite{1998MNRAS.300..665O}, no detailed analysis of cluster properties was performed due to differential reddening.  However, \cite{1998MNRAS.300..665O} note there is a blue horizontal branch present in the CMD, indicating that NGC 1916 is old ($>$10 Gyr), similar to our age  constraint.

{\it NGC 1718}. We find  [Fe/H]=  $-0.49 \pm 0.25$ and constrain the age of NGC 1718 to 1$-$2.5 Gyr.   \cite{2007A&A...462..139K} estimate [Fe/H]=$-0.40\pm0.10$ using  an HST  CMD. With Ca T spectroscopy, \cite{2006AJ....132.1630G} find [Fe/H]=$-0.80$, and with line indexes, \cite{2002MNRAS.336..168B} find [Fe/H]$=-0.98$ to $-1.12$.  This range of $-1.12 <$ [Fe/H] $< -0.40$  and the age of 2$\pm0.15$ of  Gyr \cite{2007A&A...462..139K} is  consistent with our result   that NGC 1718 is a relatively metal-poor intermediate age cluster.   Our preliminary analysis shows that the  canonical isochrones have a minor effect on the solution, and only lower the abundance  by 0.06 dex.

{\it NGC 1978}. We obtain a measurement of  [Fe/H]=$-0.71 \pm 0.08$, with an age constraint of 1.5$-$2.5 Gyr.  There are two groups that have measured detailed abundances from individual RGB stars in NGC 1978.   \cite{2006ApJ...645L..33F} find [Fe/H]$=-0.38 \pm 0.07$ from a sample of 11 stars, while  \cite{2000A&A...364L..19H} find [Fe/H]=$-0.96 \pm 0.2$ from 2 stars.  The mean of these two studies is in good agreement with our result.  Using low resolution Ca T, \citet{1991AJ....101..515O} determine [Fe/H]=$-0.42 \pm 0.2$, which is consistent with our result within the uncertainties.  \cite{2007AJ....133.2053M} present a very deep HST ACS CMD of NGC 1978, and derive a  best fit age of $1.9 \pm 0.1$ Gyr,  but with an assumption about the metallicity based on high resolution spectroscopy of individual stars.   We find that canonical isochrones may have a significant effect on the abundance we derive for NGC 1978, with an average offset of $+0.14$ dex.  The positive offset would put our solution more in line with that of         \cite{2006ApJ...645L..33F} and \citet{1991AJ....101..515O}, and will be investigated further in a future paper.    

{\it NGC 1866}. Our measurement   is [Fe/H]=$+0.00 \pm 0.04$, and an age of  100 to 300 Myrs. \cite{2000A&A...364L..19H} find  [Fe/H]=$-0.50 \pm 0.1$ from a sample of 3 giant stars, and \cite{mucc1866} find [Fe/H]$=-0.43 \pm 0.04$ from a sample of 14 stars. Using Str\"{o}mgren photometry, \cite{1995A&A...294..648H} derive a metallicity for NGC 1866 of [Fe/H]=$-0.43 \pm 0.18$, with the caveat that the calibration assumes that MW and LMC clusters have similar metallicities, which is not generally the case.    Our result is more metal-rich than these other studies, however in \textsection \ref{sec:uncertainties} we discussed how the treatment of convective overshooting in the isochrones could be influencing our results.  As shown in Table \ref{tab:canvsnoncan}, preliminary results suggest that using the canonical Teramo isochrones in our analysis would lower the abundance to [Fe/H]=$-0.19$, closer to previous studies of individual stars.      \cite{1999AJ....118.2839T} find an age of 100$-$200 Myr for NGC 1866 from a deep, ground-based CMD, while from a HST CMD, \cite{2003AJ....125.3111B} derive an age of 140$-$250 Myr depending on assumptions about binary fractions and convective overshooting.  Our age estimate  is consistent with these results.

{\it NGC 1711}. We  measure [Fe/H]$=-0.78 \pm 0.17$, and constrain the age to a range of  60 to 300 Myr.     \cite{2000A&A...360..133D}  derive an age of 50 Myr from a CMD, which is close to our constraint.    \cite{2000A&A...360..133D} find [Fe/H]$=-0.57 \pm 0.17$ using Str\"{o}mgren photometry, which is consistent with our result given the large uncertainties.  We note that, like NGC 1866, we find preliminary evidence that the treatment of convective overshooting may   impact  our solution, and raise the abundance to [Fe/H]=$-0.54$, which would bring it closer to the result of   \cite{2000A&A...360..133D}.

{\it NGC 2100}. We are able to put a lower limit on the metallicity of [Fe/H]=$-0.03$, and an upper limit on the age  of 40 Myr.    From medium resolution spectroscopy  and spectral synthesis matching, \cite{1994A&A...282..717J} find a more metal-poor result of  [Fe/H]$=-0.32$.  While our measurement is significantly more metal-rich,   it is impossible to evaluate systematic uncertainties between  our technique and that of  \cite{1994A&A...282..717J}.   We do not find evidence that our limit will change substantially if canonical isochrones are used in the analysis.  To investigate this further,  we will report follow-up high resolution spectroscopy of individual stars in NGC 2100  to look for  possible  systematic offsets from  our technique.  \cite{1994A&A...282..717J} find an age of 9 Myr, and  \cite{1991ApJS...76..185E} find 15 Myr, which are both consistent with our upper limit.

\subsection{Age-Metallicity Relationship}
\label{sec:age and met}

\begin{figure}
\centering
\includegraphics[angle=90,scale=0.5]{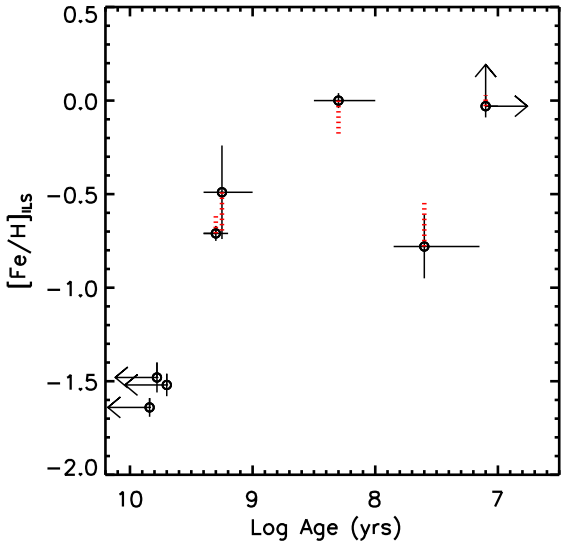}
\caption{Age-metallicity relationship from IL abundance analysis for LMC training set clusters. Red dotted lines show the anticipated change in [Fe/H] when isochrones with more appropriate treatment of convective overshooting are used.}
\label{fig:agemet} 
\end{figure}

In Figure~\ref{fig:agemet} we show the age-metallicity relationship from our high resolution abundance analysis  for the LMC clusters in our training set.  
The [Fe/H] values have been corrected with $\alpha_{\rm{offset}}$, and the possible shift due to isochrone choice is shown by red dotted lines for each cluster.
 In general, the more recently formed (younger) clusters have higher [Fe/H], as one would expect, and our results agree well with the age-metallicity relationship determined by \cite{harzar09} and \cite{2008AJ....135..836C}.  However, even though this sample of LMC clusters is small, we find a small spread in [Fe/H] at constant age for clusters with ages of  $>$10 Gyr and a larger spread for clusters with ages of $\sim$150 Myr, both of which are  larger than the abundance uncertainties of the individual clusters. This suggests that chemical enrichment across the LMC was inhomogeneous or that some clusters or gas may have been accreted from other merging satellites. The largest metallicity difference at a fixed age occurs for NGC 1866 and NGC 1711, at an age of $\sim$100 Myr.   These clusters are on the outskirts of the LMC and 
separated from each other by several degrees on the sky, so this may be a result of accretion,  poor mixing of the ISM, and/or infall of low metallicity gas.   This is discussed further in  \citetalias{paper4} of this series, where we present $\sim$20 more element abundances for these LMC clusters to  investigate the chemical enrichment and star formation history of the LMC in greater detail.

\section{Summary}
\label{sec:summary}

We have analyzed a training set of clusters in the LMC using the high resolution   IL abundance analysis method originally developed and tested on MW clusters   in  \citetalias{2008ApJ...684..326M} and  \citetalias{milkyway}.    In this work we have tested the method on clusters with ages $<$10 Gyr for the first time, and developed a strategy to accommodate the effect of incomplete statistical sampling of  CMD stellar populations.  While this strategy is motivated by incomplete sampling in the training set clusters in particular, for which we have only observed a fraction of the stellar population due to the large spatial extent of the clusters on the sky, it is applicable without any a priori knowledge of a cluster's resolved stellar  populations and can also be applied to distant extragalactic clusters.   This is primarily important for low luminosity and young clusters, as their integrated properties are more sensitive to statistical fluctuations in the number of stars in short-lived, cool, luminous stages of stellar evolution.

Our  IL abundance analysis of clusters with ages of 1$-$5 Gyr and $<$1 Gyr has demonstrated that clusters of these ages can be distinguished from old clusters using constraints derived from the Fe lines in the  high resolution spectra alone.  For intermediate age clusters we find that we can constrain ages to a $\leq$2.5 Gyr range,  but that this range can result in a larger error on the derived abundances than typically found for old clusters.    For the two intermediate age clusters in the training set we have found abundance solutions with higher scatter and less stability in [Fe/H] than generally found in the analysis of old clusters. This is likely due to low S/N data and poor statistical sampling,  but it is possible that it is a systematic problem in the synthetic CMD populations at these ages. 
For the young clusters in the training set we have found that we can constrain ages to a 200 Myr range for a well sampled cluster with high S/N spectra, and in some cases can determine upper limits to ages and abundances for clusters with ages $<$30 Myr.  Preliminary evidence suggests that the abundance solutions for young clusters could be sensitive to the input physics of the isochrones used to construct our synthetic CMDs.

The IL abundance analysis of $>$10 Gyr LMC clusters has demonstrated that even with low S/N data ($\sim40$), 20 to 30  Fe lines are generally available for analysis,  and that good constraints on the ages and abundances can be obtained with  even a small number of lines.  The analysis of NGC 1916 has shown that precise ages and abundances  can be obtained for clusters with significant differential reddening.

In conclusion, we have demonstrated the utility of the new high resolution IL spectra abundance analysis method to obtain tight age,  abundance, and stellar population  constraints on clusters spanning  a large range in age from $\sim$50 Myrs to $>$10 Gyrs. Using this method, we have measured properties of LMC clusters spanning both a large range in age and a large range in [Fe/H] using a single, consistent technique.  The age-metallicity relationship for this small set of LMC clusters suggests that the ISM of the LMC was not well mixed during all epics of star formation. Chemical abundances for an additional $\sim$20 elements will be presented in \citetalias{paper4}, where we will investigate the chemical evolution history of the LMC in greater detail.

\acknowledgements
This research was supported by NSF grant AST-0507350. The authors thank the anonymous referee for his or her comments.   J.E.C. thanks N. Calvet, P. GuhaThakurta, M. Mateo, M. S. Oey, and G. Evrard for reading of an earlier version of this  manuscript and insightful comments. The authors thank D. Osip for help with the scanning algorithm for MIKE on the Magellan Clay Telescope.

\clearpage
\LongTables
\scriptsize

\begin{deluxetable}{rrrrccccccccc}
\scriptsize
\tablecolumns{13}
\tablewidth{0pc}
\tablecaption{Line Parameters and Integrated Light Equivalent Widths for LMC GCs \label{tab:linetable_fe}}
\tablehead{\colhead{Species} & \colhead{$\lambda$}   & \colhead{E.P.} & \colhead{log gf}   & \colhead{EW} & \colhead{EW} & \colhead{EW} & \colhead{EW} & \colhead{EW}& \colhead{EW}& \colhead{EW}& \colhead{EW}& \colhead{EW}\\ 
\colhead{} & \colhead{(\AA)} & \colhead{(eV)} &\colhead{} & \colhead{(m\AA)} &\colhead{(m\AA)} &\colhead{(m\AA)} &\colhead{(m\AA)} &\colhead{(m\AA)} &\colhead{(m\AA)} &\colhead{(m\AA)} &\colhead{(m\AA)} &\colhead{(m\AA)} 
\\ & & & & \colhead{1916}&\colhead{2005}&\colhead{2019}&\colhead{1978}&\colhead{1718}&\colhead{1866}&\colhead{1711}&\colhead{2100}&\colhead{2002}}

\startdata

Fe  I &4005.254 &  1.557 & -0.583 & \nodata & \nodata & \nodata & \nodata & \nodata & 78.6 & \nodata & \nodata & \nodata\\
 Fe  I &4076.637 &  3.211 & -0.528 & \nodata & \nodata & \nodata & \nodata & \nodata & 41.8 & \nodata & \nodata & \nodata\\
 Fe  I &4132.067 &  1.608 & -0.675 & \nodata & \nodata & \nodata & \nodata & \nodata & 76.8 & \nodata & \nodata & \nodata\\
 Fe  I &4187.047 &  2.449 & -0.514 & \nodata & \nodata & \nodata & \nodata & \nodata & 48.7 & \nodata & \nodata & \nodata\\
 Fe  I &4199.105 &  3.047 &  0.156 & 85.1 & \nodata & \nodata & \nodata & \nodata & \nodata & \nodata & \nodata & \nodata\\
 Fe  I &4206.702 &  0.052 & -3.960 & 65.6 & \nodata & \nodata & \nodata & \nodata & \nodata & \nodata & \nodata & \nodata\\
 Fe  I &4250.130 &  2.469 & -0.380 & \nodata & \nodata &145.7 & \nodata & \nodata & \nodata & \nodata & \nodata & \nodata\\
 Fe  I &4250.797 &  1.557 & -0.713 &133.0 & \nodata & \nodata & \nodata & \nodata & \nodata & \nodata & \nodata & \nodata\\
 Fe  I &4282.412 &  2.176 & -0.779 & \nodata & \nodata & 96.3 & \nodata & \nodata & \nodata & \nodata & \nodata & \nodata\\
 Fe  I &4442.349 &  2.198 & -1.228 & \nodata & \nodata & \nodata & \nodata & \nodata & 74.9 & \nodata & \nodata & \nodata\\
 Fe  I &4447.728 &  2.223 & -1.339 & \nodata & \nodata & \nodata & \nodata & \nodata & 61.1 & \nodata & \nodata & \nodata\\
 Fe  I &4461.660 &  0.087 & -3.194 & \nodata & \nodata &128.6 & \nodata & \nodata & \nodata & \nodata & \nodata & \nodata\\
 Fe  I &4494.573 &  2.198 & -1.143 & \nodata & \nodata & 90.7 & \nodata & 98.5 & 96.8 & \nodata & \nodata & \nodata\\
 Fe  I &4602.949 &  1.485 & -2.208 & \nodata & \nodata & 84.2 &157.0 & 53.4 & 78.0 & \nodata & \nodata & \nodata\\
 Fe  I &4632.918 &  1.608 & -2.901 & \nodata & \nodata & \nodata & \nodata & 51.7 & \nodata & \nodata & \nodata & \nodata\\
 Fe  I &4654.504 &  1.557 & -2.721 & \nodata & \nodata & \nodata & \nodata & 63.3 & \nodata & \nodata & \nodata & \nodata\\
 Fe  I &4691.420 &  2.990 & -1.523 & \nodata & \nodata & \nodata & \nodata & 54.1 & \nodata & \nodata & \nodata & \nodata\\
 Fe  I &4736.783 &  3.211 & -0.752 & \nodata & \nodata & \nodata & \nodata & 74.0 & 83.4 & \nodata & \nodata &107.8\\
 Fe  I &4871.325 &  2.865 & -0.362 & \nodata &128.9 & \nodata & \nodata & \nodata & \nodata & \nodata & \nodata & \nodata\\
 Fe  I &4872.144 &  2.882 & -0.567 &121.0 & 77.7 & \nodata & \nodata &107.1 & \nodata & \nodata & \nodata & \nodata\\
 Fe  I &4890.763 &  2.875 & -0.394 &107.8 & 84.9 & 84.8 & \nodata &125.7 & \nodata & \nodata & \nodata & \nodata\\
Fe  I &4891.502 &  2.851 & -0.111 & \nodata &131.6 &127.6 & \nodata & \nodata & \nodata & \nodata & \nodata & \nodata\\
 Fe  I &4903.316 &  2.882 & -0.926 & 83.4 & 84.9 & \nodata & \nodata & \nodata & 62.5 & \nodata & \nodata &128.4\\
 Fe  I &4918.998 &  2.865 & -0.342 & 98.4 & \nodata & \nodata &138.9 & \nodata & \nodata & \nodata & \nodata & \nodata\\
 Fe  I &4920.514 &  2.832 &  0.068 & \nodata & \nodata &155.7 & \nodata & \nodata & \nodata & \nodata & \nodata & \nodata\\
 Fe  I &4938.820 &  2.875 & -1.077 & \nodata & \nodata & \nodata &152.0 & 64.2 & \nodata & \nodata & \nodata & \nodata\\
 Fe  I &4939.694 &  0.859 & -3.252 & 78.9 & \nodata & \nodata & \nodata & 90.7 & \nodata & \nodata & 68.9 & \nodata\\
 Fe  I &4966.095 &  3.332 & -0.871 & \nodata & \nodata & \nodata & \nodata & 57.2 & \nodata & \nodata & \nodata &136.4\\
 Fe  I &4994.138 &  0.915 & -2.969 & \nodata & 68.4 & \nodata & \nodata & \nodata & 77.3 & \nodata & \nodata & \nodata\\
 Fe  I &5001.870 &  3.881 &  0.050 & \nodata & \nodata & \nodata &100.1 & \nodata & 65.1 & \nodata & \nodata & \nodata\\
 Fe  I &5006.120 &  2.832 & -0.615 & \nodata &108.4 & \nodata & \nodata & \nodata & \nodata & \nodata & \nodata & \nodata\\
 Fe  I &5014.951 &  3.943 & -0.303 & 50.0 & 74.8 & \nodata & \nodata & 70.9 & \nodata & \nodata & \nodata & \nodata\\
 Fe  I &5041.763 &  1.485 & -2.203 & \nodata & \nodata &104.8 & \nodata & \nodata & \nodata & \nodata & \nodata & \nodata\\
 Fe  I &5049.827 &  2.279 & -1.355 & 93.6 & 79.6 & 91.4 & \nodata &121.0 & \nodata & \nodata & \nodata &140.6\\
 Fe  I &5051.640 &  0.915 & -2.764 & 80.3 & \nodata &105.7 &153.3 &122.4 & 85.4 & \nodata & 89.7 & \nodata\\
 Fe  I &5051.640 &  0.915 & -2.764 &103.6 & \nodata &\nodata &\nodata &\nodata & \nodata & \nodata & \nodata & \nodata\\
 Fe  I &5068.771 &  2.940 & -1.041 & \nodata & \nodata & \nodata & \nodata & \nodata & 80.9 & \nodata & \nodata & \nodata\\
 Fe  I &5074.753 &  4.220 & -0.160 & 57.1 & \nodata & \nodata & 95.4 & \nodata & 61.0 & \nodata & \nodata & 73.7\\
 Fe  I &5079.745 &  0.990 & -3.245 & \nodata & \nodata & \nodata & \nodata & 84.1 & \nodata & \nodata & \nodata & \nodata\\
 Fe  I &5083.345 &  0.958 & -2.842 & \nodata & 74.4 & \nodata & \nodata & \nodata & \nodata & \nodata & \nodata & \nodata\\
 Fe  I &5123.730 &  1.011 & -3.058 & \nodata & \nodata & \nodata &125.5 & \nodata & \nodata & \nodata & \nodata & \nodata\\
 Fe  I &5127.368 &  0.915 & -3.249 & \nodata & \nodata & 73.9 & \nodata & \nodata & \nodata & 47.7 & \nodata & \nodata\\
 Fe  I &5150.852 &  0.990 & -3.037 & \nodata & \nodata & 87.4 &110.2 & 93.2 & \nodata & \nodata & 87.3 & \nodata\\
 Fe  I &5151.917 &  1.011 & -3.321 & \nodata & \nodata & \nodata &150.9 & \nodata & \nodata & \nodata & \nodata & \nodata\\
 Fe  I &5162.281 &  4.178 &  0.020 & 52.8 & \nodata & \nodata &113.6 &142.3 & \nodata & \nodata & \nodata & \nodata\\
 Fe  I &5166.284 &  0.000 & -4.123 & \nodata & \nodata &103.3 & \nodata &107.0 & \nodata & \nodata & \nodata & \nodata\\
 Fe  I &5171.610 &  0.000 & -1.721 & \nodata & \nodata & \nodata &151.9 & \nodata & \nodata & \nodata & \nodata & \nodata\\
 Fe  I &5191.465 &  3.038 & -0.551 &100.2 & \nodata & 95.3 & \nodata &128.1 & \nodata & \nodata & \nodata & \nodata\\
 Fe  I &5192.353 &  2.998 & -0.421 &118.3 & \nodata & \nodata & \nodata &133.4 & \nodata & \nodata & \nodata & \nodata\\
 Fe  I &5194.949 &  1.557 & -2.021 & \nodata & \nodata & \nodata &151.3 & \nodata & \nodata & 61.8 & 94.9 & \nodata\\
 Fe  I &5195.480 &  4.220 & -0.002 & \nodata & \nodata & \nodata &136.5 & \nodata & \nodata & \nodata & \nodata & \nodata\\
 Fe  I &5216.283 &  1.608 & -2.082 & 86.3 & \nodata & 91.9 & \nodata & 69.8 & \nodata & \nodata & \nodata & \nodata\\
 Fe  I &5232.952 &  2.940 & -0.057 &114.3 & \nodata &118.9 & \nodata &149.9 & \nodata & \nodata & \nodata & \nodata\\
 Fe  I &5254.953 &  0.110 & -4.764 & \nodata & 68.0 & \nodata &127.2 & \nodata & \nodata & 55.2 & \nodata & \nodata\\
 Fe  I &5266.563 &  2.998 & -0.385 & \nodata & 96.0 &106.3 & \nodata & \nodata & \nodata & \nodata & \nodata & \nodata\\
 Fe  I &5281.798 &  3.038 & -0.833 & \nodata & 75.9 & \nodata &101.2 & 98.1 & \nodata & 43.0 & \nodata & \nodata\\
 Fe  I &5283.629 &  3.241 & -0.524 &110.5 & \nodata & \nodata &152.9 & 97.4 & \nodata & \nodata & \nodata & \nodata\\
 Fe  I &5302.307 &  3.283 & -0.720 & \nodata & \nodata & 69.5 & \nodata &124.3 & \nodata & \nodata & \nodata & \nodata\\
 Fe  I &5307.369 &  1.608 & -2.912 & 62.2 & \nodata & 57.8 & 90.4 & 72.5 & 53.0 & \nodata & \nodata & \nodata\\
 Fe  I &5324.191 &  3.211 & -0.103 &110.2 &119.4 &102.6 & \nodata & \nodata & 85.0 & \nodata & \nodata & \nodata\\
 Fe  I &5339.937 &  3.266 & -0.720 & \nodata & 69.2 & 63.8 &145.0 & 61.3 & 90.7 & \nodata & \nodata & \nodata\\
 Fe  I &5367.476 &  4.415 &  0.443 & \nodata & \nodata & \nodata &102.6 & \nodata & \nodata & \nodata & \nodata &104.7\\
 Fe  I &5367.476 &  4.415 &  0.443 & \nodata & \nodata & \nodata &123.2 & \nodata & \nodata & \nodata & \nodata &\nodata\\
 Fe  I &5369.974 &  4.371 &  0.536 & \nodata & \nodata & \nodata &124.8 &101.8 & \nodata & 57.2 & \nodata & \nodata\\
 Fe  I &5369.974 &  4.371 &  0.536 & \nodata & \nodata & \nodata &136.5 &\nodata& \nodata & \nodata & \nodata & \nodata\\
 Fe  I &5371.501 &  0.958 & -1.644 & \nodata & \nodata &168.6 & \nodata &150.7 & \nodata & 73.7 & \nodata & \nodata\\
 Fe  I &5383.380 &  4.312 &  0.645 & 70.6 & 97.1 & 87.2 & \nodata & 93.6 & \nodata & 43.2 & \nodata &135.8\\
 Fe  I &5389.486 &  4.415 & -0.410 & 34.9 & \nodata & \nodata & 85.7 & \nodata & \nodata & \nodata & \nodata & \nodata\\
 Fe  I &5393.176 &  3.241 & -0.715 & 82.3 & 63.5 & 71.6 &133.2 &114.9 & \nodata & \nodata & \nodata & \nodata\\
 Fe  I &5397.141 &  0.915 & -1.982 & \nodata & \nodata &136.7 & \nodata &153.7 & \nodata & 45.6 & \nodata & \nodata\\
 Fe  I &5405.785 &  0.990 & -1.852 & \nodata & \nodata &139.0 & \nodata & \nodata & \nodata & \nodata & \nodata & \nodata\\
 Fe  I &5424.080 &  4.320 &  0.520 & 76.8 & \nodata & \nodata &145.2 & \nodata & \nodata & \nodata & \nodata & \nodata\\
 Fe  I &5434.534 &  1.011 & -2.126 & \nodata & 69.0 &121.2 & \nodata &116.2 & \nodata & \nodata & \nodata & \nodata\\
 Fe  I &5446.924 &  0.990 & -3.109 & \nodata &155.9 & \nodata & \nodata & \nodata & \nodata & 67.4 & \nodata & \nodata\\
 Fe  I &5455.624 &  0.000 & -2.091 & \nodata & \nodata &155.9 & \nodata & \nodata & \nodata & \nodata & \nodata & \nodata\\
 Fe  I &5455.624 &  0.000 & -2.091 & \nodata & \nodata &169.5 & \nodata & \nodata & \nodata & \nodata & \nodata & \nodata\\
 Fe  I &5497.526 &  1.011 & -2.825 & \nodata & 93.0 &102.4 &119.2 &123.7 & \nodata & \nodata & \nodata & \nodata\\
 Fe  I &5501.477 &  0.958 & -3.046 &109.8 & \nodata &116.9 &118.6 & 89.9 & 85.9 & \nodata & \nodata & \nodata\\
 Fe  I &5506.791 &  0.990 & -2.789 &142.4 &108.7 &112.3 & \nodata & 83.7 &103.6 & \nodata & \nodata & \nodata\\
 Fe  I &5569.631 &  3.417 & -0.500 & 75.9 & \nodata & 65.8 & \nodata &123.0 & \nodata & \nodata & 71.0 &155.3\\
 Fe  I &5572.851 &  3.396 & -0.275 & \nodata & \nodata & \nodata & \nodata & \nodata &106.5 & \nodata & \nodata & \nodata\\
 Fe  I &5576.099 &  3.430 & -0.900 & \nodata & \nodata & 46.6 & \nodata & \nodata & 72.7 & \nodata & \nodata &125.3\\
 Fe  I &5586.771 &  4.260 & -0.096 & \nodata & \nodata & \nodata & \nodata &155.3 & \nodata & \nodata & \nodata &157.4\\
 Fe  I &5586.771 &  4.260 & -0.096 & \nodata & \nodata & \nodata & \nodata &127.1 & \nodata & \nodata & \nodata &\nodata\\
 Fe  I &5763.002 &  4.209 & -0.450 & \nodata & \nodata & 57.9 & 87.1 & 81.1 & \nodata & \nodata & \nodata &122.2\\
 Fe  I &6136.624 &  2.453 & -1.410 &129.1 &106.6 & \nodata &129.7 & \nodata & \nodata & \nodata & \nodata & \nodata\\
 Fe  I &6137.702 &  2.588 & -1.346 &108.7 & \nodata & 85.9 &126.3 & \nodata & \nodata & 71.6 & \nodata & \nodata\\
 Fe  I &6151.623 &  2.180 & -3.330 & \nodata & \nodata & \nodata & \nodata & \nodata &112.7 & \nodata & \nodata & \nodata\\
 Fe  I &6173.341 &  2.220 & -2.863 & 40.0 & \nodata & \nodata & \nodata & 62.5 & 49.1 & \nodata & \nodata & \nodata\\
 Fe  I &6180.209 &  2.730 & -2.628 & \nodata & \nodata & \nodata & \nodata & 61.2 & 47.8 & \nodata & \nodata & \nodata\\
 Fe  I &6187.995 &  3.940 & -1.673 & \nodata & \nodata & \nodata & \nodata & \nodata & 33.1 & \nodata & \nodata & \nodata\\
 Fe  I &6200.321 &  2.610 & -2.386 & 56.2 & 52.3 & \nodata & \nodata & 72.3 & 49.1 & \nodata & 88.4 &105.4\\
 Fe  I &6213.437 &  2.220 & -2.490 & \nodata & 54.7 & \nodata & \nodata & 89.2 & \nodata & \nodata & \nodata & \nodata\\
 Fe  I &6219.287 &  2.200 & -2.428 & 76.6 & 56.5 & \nodata &126.3 & 73.5 & 64.3 & \nodata & 93.3 &152.4\\
 Fe  I &6229.232 &  2.830 & -2.821 & 31.5 & \nodata & \nodata & \nodata & \nodata & \nodata & \nodata & \nodata & 82.9\\
 Fe  I &6230.736 &  2.559 & -1.276 &101.3 & \nodata & \nodata & \nodata & \nodata & 75.6 & \nodata & \nodata & \nodata\\
 Fe  I &6246.327 &  3.600 & -0.796 & \nodata & \nodata & \nodata &141.4 & 65.0 & 59.6 & \nodata & 67.9 & \nodata\\
 Fe  I &6252.565 &  2.404 & -1.767 & 71.9 & \nodata & \nodata &132.1 &116.5 &101.5 & \nodata & \nodata & \nodata\\
 Fe I &6254.253 &  2.280 & -2.435 & 55.6 & \nodata & 60.5 & 96.3 & 72.3 & \nodata & \nodata & \nodata & \nodata\\
 Fe  I &6265.141 &  2.180 & -2.532 & \nodata & \nodata & \nodata & \nodata & 86.3 & \nodata & \nodata & 87.5 & \nodata\\
 Fe  I &6311.504 &  2.830 & -3.153 & \nodata & \nodata & \nodata & \nodata & \nodata & 54.7 & \nodata & \nodata & \nodata\\
 Fe  I &6322.694 &  2.590 & -2.438 & \nodata & 36.3 & 47.7 &128.7 & \nodata & 58.9 & 47.4 & 63.4 & \nodata\\
 Fe  I &6330.852 &  4.730 & -1.640 & \nodata & \nodata & \nodata & \nodata & \nodata & 23.7 & \nodata & \nodata & 33.0\\
 Fe  I &6335.337 &  2.200 & -2.175 & 74.2 & 58.6 & 67.4 &135.6 & 74.1 & 67.3 & 53.5 & \nodata & \nodata\\
 Fe  I &6336.830 &  3.690 & -0.667 & 59.5 & \nodata & \nodata &134.6 & 65.8 & 68.7 & 40.4 & \nodata & \nodata\\
 Fe  I &6353.849 &  0.910 & -6.360 & \nodata & \nodata & \nodata & \nodata & \nodata & 86.5 & \nodata & \nodata & \nodata\\
 Fe  I &6355.035 &  2.840 & -2.328 & 60.7 & \nodata & 49.6 & \nodata & 68.0 & 46.5 & \nodata & 93.1 & \nodata\\
 Fe  I &6380.750 &  4.190 & -1.366 & \nodata & \nodata & \nodata & \nodata & \nodata & 42.7 & \nodata & \nodata & \nodata\\
 Fe  I &6392.538 &  2.280 & -3.957 & \nodata & \nodata & \nodata & \nodata & 30.2 & \nodata & \nodata & \nodata & \nodata\\
 Fe  I &6393.612 &  2.430 & -1.505 & 95.0 & \nodata & 91.7 & \nodata &129.9 & 93.9 & \nodata & \nodata & \nodata\\
 Fe  I &6400.009 &  3.602 & -0.290 & \nodata & \nodata & \nodata & \nodata &143.3 & \nodata & 69.3 & \nodata & \nodata\\
 Fe  I &6411.658 &  3.650 & -0.646 & 73.2 & 79.8 & 67.9 &100.4 & \nodata & 78.9 & \nodata & \nodata &158.7\\
 Fe  I &6419.956 &  4.730 & -0.183 & \nodata & \nodata & \nodata & \nodata & \nodata & 56.0 & \nodata & \nodata & \nodata\\
 Fe  I &6421.360 &  2.280 & -1.979 & 83.9 & 90.2 & 88.0 &142.5 & 92.0 & 80.7 & \nodata & \nodata & \nodata\\
 Fe  I &6430.856 &  2.180 & -1.954 & \nodata & 74.3 & 76.7 & \nodata &118.9 & \nodata & \nodata & \nodata & \nodata\\
 Fe  I &6475.632 &  2.560 & -2.929 & \nodata & \nodata & 36.7 & 97.9 & 40.9 & \nodata & \nodata & 84.5 & \nodata\\
 Fe  I &6481.878 &  2.280 & -2.985 & \nodata & \nodata & \nodata &100.1 & \nodata & \nodata & \nodata & \nodata & \nodata\\
 Fe  I &6494.994 &  2.400 & -1.246 &103.3 & 87.8 &124.5 & \nodata &143.7 & \nodata & \nodata & \nodata & \nodata\\
 Fe  I &6498.945 &  0.960 & -4.675 & \nodata & \nodata & \nodata & 86.0 & \nodata & \nodata & \nodata & \nodata & \nodata\\
 Fe  I &6518.373 &  2.830 & -2.397 & \nodata & \nodata & \nodata & \nodata & \nodata & 96.5 & \nodata & \nodata &132.9\\
 Fe  I &6533.940 &  4.540 & -1.360 & \nodata & \nodata & \nodata & 58.3 & \nodata & \nodata & \nodata & 48.1 & 36.4\\
 Fe I &6546.252 &  2.750 & -1.536 & \nodata & \nodata & \nodata & \nodata &143.1 & \nodata & \nodata & 95.2 &134.0\\
 Fe  I &6593.874 &  2.430 & -2.377 & \nodata & 58.4 & \nodata & \nodata & \nodata & \nodata & \nodata & \nodata & \nodata\\
 Fe  I &6597.571 &  4.770 & -0.970 & \nodata & \nodata & \nodata & \nodata & \nodata & 37.5 & \nodata & \nodata & \nodata\\
 Fe  I &6608.044 &  2.270 & -3.939 & \nodata & \nodata & \nodata & \nodata & 44.9 & \nodata & \nodata & 64.5 & \nodata\\
 Fe  I &6646.966 &  2.600 & -3.917 & \nodata & \nodata & \nodata & \nodata & \nodata & 54.1 & \nodata & \nodata & 52.3\\
 Fe  I &6648.121 &  1.010 & -5.730 & \nodata & \nodata & \nodata & \nodata & \nodata & 77.8 & \nodata & \nodata &107.0\\
 Fe  I &6677.997 &  2.690 & -1.395 & 93.9 & \nodata & 91.4 & \nodata &103.3 &111.0 & \nodata & \nodata & \nodata\\
 Fe  I &6703.576 &  2.760 & -3.059 & 31.5 & \nodata & 23.6 & 79.0 & 40.1 & 37.6 & \nodata & 57.4 & \nodata\\
 Fe  I &6705.105 &  4.610 & -1.060 & \nodata & \nodata & \nodata & \nodata & 34.0 & 35.7 & \nodata & 52.0 & \nodata\\
 Fe  I &6710.323 &  1.480 & -4.807 & \nodata & \nodata & \nodata & \nodata & \nodata & 88.7 & \nodata & \nodata & \nodata\\
 Fe  I &6725.364 &  4.100 & -2.227 & \nodata & \nodata & \nodata & \nodata & \nodata & 60.7 & \nodata & \nodata & \nodata\\
 Fe  I &6739.524 &  1.560 & -4.801 & \nodata & \nodata & \nodata & 67.4 & 44.9 & \nodata & \nodata & \nodata & \nodata\\
 Fe  I &6750.164 &  2.420 & -2.592 & \nodata & \nodata & \nodata & \nodata & 66.0 & 44.8 & \nodata & 90.2 &151.1\\
 Fe  I &6752.716 &  4.640 & -1.263 & \nodata & \nodata & \nodata & \nodata & \nodata & 70.8 & \nodata & \nodata & \nodata\\
 Fe  I &6806.856 &  2.730 & -2.633 & 27.1 & \nodata & \nodata & \nodata & 38.7 & 48.8 & \nodata & 70.5 & 91.3\\
 Fe  I &6810.267 &  4.590 & -0.992 & \nodata & \nodata & \nodata & \nodata & \nodata & 64.3 & \nodata & \nodata & 67.6\\
 Fe I &6828.596 &  4.640 & -0.843 & \nodata & \nodata & \nodata & \nodata & \nodata & 47.9 & \nodata & \nodata & 82.3\\
 Fe  I &6839.835 &  2.560 & -3.378 & \nodata & \nodata & \nodata & 77.4 & 56.5 & 33.3 & \nodata & 58.3 &100.0\\
 Fe  I &6841.341 &  4.610 & -0.733 & 39.3 & \nodata & \nodata & \nodata & 72.7 & 50.5 & \nodata & 53.5 & 83.7\\
 Fe  I &6842.689 &  4.640 & -1.224 & \nodata & \nodata & \nodata & \nodata & \nodata & 32.4 & \nodata & \nodata & 50.2\\
 Fe  I &6843.655 &  3.650 & -0.863 & \nodata & \nodata & \nodata & \nodata & \nodata & 51.6 & \nodata & \nodata & 52.6\\
 Fe  I &6851.652 &  1.600 & -5.247 & \nodata & \nodata & \nodata & \nodata & \nodata & 51.0 & \nodata & \nodata & 73.4\\
 Fe  I &6855.723 &  4.390 & -1.747 & \nodata & \nodata & \nodata & \nodata & \nodata & 34.0 & \nodata & \nodata & \nodata\\
 Fe  I &6916.686 &  4.150 & -1.359 & \nodata & \nodata & \nodata & \nodata & \nodata &133.7 & \nodata & \nodata & \nodata\\
 Fe  I &6960.330 &  4.570 & -1.907 & \nodata & \nodata & \nodata & \nodata & \nodata & 37.8 & \nodata & \nodata & \nodata\\
 Fe  I &7007.976 &  4.180 & -1.929 & \nodata & \nodata & \nodata & \nodata & \nodata &105.2 & \nodata & \nodata & \nodata\\
 Fe  I &7022.957 &  4.190 & -1.148 & \nodata & \nodata & \nodata & \nodata & \nodata &134.8 & \nodata & \nodata & \nodata\\
 Fe  I &7038.220 &  4.220 & -1.214 & \nodata & \nodata & \nodata & \nodata & 65.5 & \nodata & \nodata & \nodata &139.6\\
 Fe  I &7068.423 &  4.070 & -1.319 & \nodata & \nodata & 26.3 & \nodata & \nodata & \nodata & 48.7 & \nodata &143.1\\
 Fe  I &7071.866 &  4.610 & -1.627 & \nodata & \nodata & \nodata & \nodata & \nodata & 34.9 & \nodata & \nodata & \nodata\\
 Fe  I &7072.800 &  4.070 & -2.767 & \nodata & \nodata & \nodata & \nodata & \nodata & 23.0 & \nodata & \nodata & \nodata\\
 Fe  I &7090.390 &  4.230 & -1.109 & \nodata & \nodata & \nodata & \nodata & \nodata &135.1 & \nodata & \nodata & \nodata\\
 Fe  I &7127.573 &  4.990 & -1.177 & \nodata & \nodata & \nodata & \nodata & \nodata & 55.7 & \nodata & \nodata & \nodata\\
 Fe  I &7130.925 &  4.300 & -0.708 & \nodata & \nodata & \nodata & \nodata & 91.7 & 70.5 & \nodata & \nodata &146.2\\
 Fe  I &7132.985 &  4.060 & -1.635 & \nodata & \nodata & \nodata & \nodata & 64.0 & 39.1 & \nodata & \nodata & 94.2\\
 Fe  I &7142.517 &  4.930 & -1.017 & \nodata & \nodata & \nodata & \nodata & \nodata & 41.7 & \nodata & \nodata & \nodata\\
 Fe  I &7145.312 &  4.610 & -1.240 & \nodata & \nodata & \nodata & \nodata & 41.5 & \nodata & \nodata & \nodata & 80.1\\
 Fe  I &7151.464 &  2.480 & -3.657 & \nodata & \nodata & \nodata & \nodata & \nodata & 45.6 & \nodata & \nodata &123.0\\
 Fe  I &7155.634 &  4.990 & -1.017 & \nodata & \nodata & \nodata & \nodata & \nodata & 81.4 & \nodata & \nodata & \nodata\\
 Fe  I &7411.162 &  4.280 & -0.287 & 73.6 & \nodata & \nodata & \nodata & \nodata &100.7 & \nodata & \nodata & \nodata\\
 Fe  I &7445.758 &  4.260 &  0.053 & 82.3 & 69.4 & \nodata & \nodata & 99.8 &115.1 & 89.5 & \nodata & \nodata\\
 Fe  I &7454.004 &  4.190 & -2.337 & \nodata & \nodata & \nodata & \nodata & \nodata & 47.1 & \nodata & \nodata & \nodata\\
 Fe  I &7461.527 &  2.560 & -3.507 & \nodata & \nodata & \nodata & \nodata & \nodata & 62.3 & \nodata & \nodata & \nodata\\
 Fe  I &7491.652 &  4.280 & -1.067 & \nodata & \nodata & \nodata & \nodata & 89.4 & \nodata & \nodata & \nodata & \nodata\\
 Fe  I &7507.273 &  4.410 & -1.107 & \nodata & \nodata & \nodata & \nodata & 49.7 & 64.7 & \nodata & \nodata & \nodata\\
 Fe  I &7531.153 &  4.370 & -0.557 & \nodata & \nodata & \nodata & \nodata & 86.6 & \nodata & 71.4 & \nodata & \nodata\\
 Fe  I &7540.444 &  2.730 & -3.777 & \nodata & \nodata & \nodata & 60.2 & \nodata & \nodata & \nodata & 70.0 &124.2\\
 Fe  I &7583.790 &  3.018 & -1.885 & \nodata & \nodata & 57.2 & \nodata & \nodata & \nodata & 58.8 & \nodata & \nodata\\

\enddata

\tablecomments{~ Lines listed twice correspond to those measured in adjacent orders with overlapping wavelength coverage.}

\end{deluxetable}

\end{document}